\documentclass[11pt]{article}
\usepackage{graphics,amsmath,amssymb,stmaryrd,color}
\usepackage{subfigure,amsfonts,array}
\providecommand{\U}[1]{\protect\rule{.1in}{.1in}}
\usepackage{graphicx}
\usepackage{amsmath, amsthm, amsfonts}
\usepackage{bigints}
\usepackage{flafter}
\usepackage{bbm}
\usepackage{lscape}
\usepackage{pdflscape}
\usepackage{varwidth}
\usepackage{caption}
\usepackage{booktabs, topcapt}
\usepackage{threeparttable}
\usepackage{lscape}
\usepackage{pdflscape}
\usepackage{natbib}
\usepackage{xcolor}
\usepackage{xfrac}
\usepackage{mathtools}
\usepackage{titlesec}
\usepackage{indentfirst}
\usepackage{setspace}
\usepackage{ulem}
\titleformat*{\section}{\large\bfseries}
\titleformat*{\subsection}{\large\bfseries}
\newcommand{\indep}{\rotatebox[origin=c]{90}{$\models$}}
\allowdisplaybreaks

\theoremstyle{plain}

\newtheorem{assumption}{Assumption}

\newtheorem{lemma}{Lemma}

\newtheorem{proposition}{Proposition}
\newtheorem{remark}{Remark}

\DeclareMathOperator*{\argmin}{\arg\!\min}

\usepackage[utf8]{inputenc}
\usepackage{newunicodechar}
\def\texteqcolon{≕} \newunicodechar{≕}{\ifmmode \mathrel{{=}{\vcentcolon}} \else \texteqcolon \fi}

\newcommand\independent{\protect\mathpalette{\protect\independenT}{\perp}}
\def\independenT#1#2{\mathrel{\rlap{$#1#2$}\mkern2mu{#1#2}}}
\numberwithin{equation}{section}

\newcommand{\prob}{\mathbb{P}}

\newcounter{bean}

\usepackage{hyperref} 
\hypersetup{
	colorlinks=true, breaklinks=true, bookmarksnumbered,
	urlcolor=blue, linkcolor=blue, citecolor=blue, 
	pdftitle={}, 
	pdfauthor={\textcopyright}, 
	pdfsubject={}, 
	pdfkeywords={}, 
	pdfcreator={pdfLaTeX}, 
	pdfproducer={LaTeX with hyperref and ClassicThesis} 
}

\begin{document}
	\nonstopmode
	\title{{\huge Identifying Marginal Treatment Effects in the Presence of Sample Selection}\thanks{The present version is of \today. First draft: August, 2018. Email addresses: Ot\'avio Bartalotti - bartalot@iastate.edu; D\'esir\'e K\'edagni (corresponding author) dkedagni@iastate.edu; Vitor Possebom - vitoraugusto.possebom@yale.edu. This paper replaces Bartalotti and K\'edagni (2019) under the same title and Possebom (2019):``Sharp Bounds for the MTE with Sample Selection.''}}

	\author{ \begin{tabular}{ccc}
			Ot\'avio Bartalotti & D\'esir\'e K\'edagni & Vitor Possebom \\
			\footnotesize Iowa State University and IZA & \footnotesize Iowa State University & \footnotesize Yale University
		\end{tabular}
	}
	\date{~}

	\maketitle

	\newsavebox{\tablebox} \newlength{\tableboxwidth}

	\begin{center}
		%

		\normalsize
		\

		\textbf{Abstract}
	\end{center}

	This article presents identification results for the marginal treatment effect (MTE) when there is sample selection.  We show that the MTE is partially identified for individuals who are always observed regardless of treatment, and derive uniformly sharp bounds on this parameter under three increasingly restrictive sets of assumptions. The first result imposes standard MTE assumptions with an unrestricted sample selection mechanism. The second set of conditions imposes monotonicity of the sample selection variable with respect to treatment, considerably shrinking the identified set. Finally, we incorporate a stochastic dominance assumption which tightens the lower bound for the MTE. Our analysis extends to discrete instruments. The results rely on a mixture reformulation of the problem where the mixture weights are identified, extending \citeauthor{lee2009training}'s (\citeyear{lee2009training}) trimming procedure to the MTE context. We propose estimators for the bounds derived and use data made available by \cite{Deb2006} to empirically illustrate the usefulness of our approach.

	\




	\

	\newpage

	\doublespacing

	\section{Introduction}

	Many interesting applications in the treatment effects literature involve two simultaneous identification challenges: endogenous selection into treatment and sample selection. For instance, in labor economics, when a researcher wishes to evaluate the effect of a job training program on wages, she has to consider the individuals' decision to enroll in the training program as well as their decision to participate in the labor market. In the health sciences, the same identification challenges appear when analyzing the effect of a drug on well-being as the outcome of interest --- health status --- is observed only for those who survive. Moreover, in randomized control trials (RCTs), non-compliance and endogenous attrition in treated and control groups lead to the same identification concerns. This double selection problem is also present when analyzing the effect of attending college on wages, the effect of an educational intervention on short- and long-term outcomes, and the effect of procedural laws on litigation outcomes.\footnote{Training programs are studied by \cite{Heckman1999a}, \cite{lee2009training} and \cite{chen2015bounds}. The college wage premium is analyzed by \cite{Altonji1993}, \cite{Card1999} and \cite{carneiro2011estimating}. Some education interventions are studied by \cite{Krueger2001}, \cite{Angrist2006}, \cite{Angrist2009}, \cite{Chetty2011} and \cite{Dobbie2015}. Medical treatments are analyzed by \cite{CASS1984}, \cite{Sexton1984} and \cite{Health2004}. Litigation outcomes are discussed by \cite{Helland2017}. RCTs with attrition are illustrated by \cite{DeMel2013} and \cite{Angelucci2015}.}

	In this paper, we derive novel uniformly sharp bounds on the marginal treatment effect (MTE) for individuals who would self-select into the sample regardless of their treatment status ($MTE^{OO}$). To do so, we propose identification strategies under increasingly restrictive sets of assumptions that extend the MTE identification to scenarios with endogenous sample selection. Furthermore, the choice of treatment is allowed to be endogenous, and can be related to the sample selection mechanism. To address both identification challenges described above, we analyze a generalized sample selection model in which the realized outcome (e.g., wages)  is observed only if the individual self-selects into the sample (employment status), and the treatment choice (training program participation) is observed for all individuals in the data being analyzed.

	The $MTE$ and $MTE^{OO}$ provide important and intuitive measures of the treatment effect and its heterogeneity across the population. For example, consider a job training program. Training influences both the likelihood of employment and wages, which are observed only for individuals who are employed. The $MTE$ reflects the returns to training for individuals with different levels of the (latent) cost of attending the program. As a consequence, this parameter sheds light on the heterogeneity of the training program's effects, i.e., to understand who would benefit from taking extra training. This knowledge can be used to design policies focusing on targeting of the program, affordability and services offered. Common parameters evaluated in the literature --- such as the average treatment effect ($ATE$), the average treatment effect on the treated ($ATT$), the average treatment effect on the untreated ($ATU$), and the local average treatment effect ($LATE$) --- may not adequately capture this heterogeneity. For example, they could be positive even when most people are affected adversely by a policy, masking its effects.\footnote{The empirical relevance of the MTE is recently illustrated by \cite{carneiro2011estimating}, \cite{Kline2016_head_start}, \cite{Arnold2018}, \cite{Cornelissen2018}, \cite{Bhuller2019}, \cite{Humphries2019} and \cite{Mountjoy2019} in different contexts.} Similarly, the $MTE^{OO}$ reflects the training program's effects at the intensive margin, i.e., to the group of individuals that are more attached to the labor force, and would be employed even if they had not attended the training program.

	Our strategy to partially identify the $MTE^{OO}$ relies on standard assumptions regarding selection into treatment. Those are the same conditions usually imposed for identification of the MTE when there is no sample selection, i.e., there is an exogenous instrument excluded from the outcome determination, treatment choice is monotone in the instrument, and the propensity score is continuous.\footnote{See \cite{bjorklund1987}, \cite{heckman1999, heckman2001, heckman2005structural} and \cite{Andresen2018} for a detailed discussion.} We add the requirement that the instrument is also excluded from the sample selection mechanism. In the job training literature, a similar assumption is frequently imposed when analyzing RCTs with imperfect compliance, as the interest frequently lies on the effect on labor earnings and employment.

	In this paper's first contribution, the partial identification result for the $MTE^{OO}$ leaves sample selection unrestricted, providing very general uniformly sharp bounds on that parameter.
	Our second result tightens the bounds around the $MTE^{OO}$ by exploiting a ``monotonicity in selection'' assumption. This condition is similar to the usual LATE monotonicity assumption and requires individuals to be at least as likely to be observed in the sample if they are treated. In the job training program example, this additional assumption imposes that the treatment can induce workers to join the labor force, but not the opposite.

	The identified set is further reduced by imposing a stochastic dominance assumption to our third and final set of assumptions. This condition mandates that the subpopulation who self-selects into the sample regardless of the treatment status has higher potential outcomes if treated than the subpopulation who self-selects into the sample only when treated. Intuitively, workers with high attachment to the labor force (that would be employed regardless of receiving job training),  would earn higher wages after the extra training than those that would choose to participate in the labor force only if they participated in the training program.

	Identification relies on a reformulation of the potential outcomes' conditional probabilities as a mixture between the latent groups of individuals who are ``always observed''  and ``observed only when treated.'' This reformulation extends to the MTE case the trimming procedure proposed by \cite{Imai2008}, \cite{lee2009training} and \cite{chen2015bounds} in the context of identifying the ATE and LATE. Crucially, since we are interested in the MTE, the trimming is based on the distribution of the potential outcome conditional on unobserved individual characteristics related to treatment receipt.

	The results can be used to construct bounds for any treatment effect parameter that can be written as a weighted average of the $MTE^{OO}$. We derive new weights to obtain sharp bounds on the ATE, the ATT, the ATU, any LATE (\citealp{imbensangrist94}) and any policy-relevant treatment effect (PRTE, \citealp{Heckman2001a}) within the always-observed subpopulation. Differently from the weights for the case without sample selection (\citealp{heckman2005structural}, \citealp{carneiro2009estimating}, and \citealp{carneiro2011estimating}), the new weights must be integrated over the distribution of the latent heterogeneity for the always-observed subpopulation instead of its unconditional distribution. Furthermore, we show these weights are identified under the ``monotonicity in selection'' assumption if the support of the propensity score is the full unit interval.

	Moreover, we propose and discuss nonparametric and parametric estimators for the bounds around the $MTE^{OO}$. The usefulness and feasibility of the parametric procedure is illustrated in an empirical example using the data set organized by \cite{Deb2006}. We find that the effect of insurance plan choice on ambulatory expenditures depends negatively on the agents' relative cost of choosing managed care plans over fee for service plans. This corroborates the results found by \cite{Deb2006}, indicating that agents endogenously self-select into managed care plans.

	The main identification results are extended to the case where the researcher only has access to multi-valued discrete instruments. In this context, we derive nonparametric sharp bounds on a weighted average of the MTE. This result is conceptually similar to \cite{chen2015bounds}, who provide an outer set for the LATE parameter when the instrument is binary.\footnote{An additional extension is presented in Appendix \ref{DMTE}, where we derive sharp bounds around a more general object of interest: the distributional marginal treatment effect (DMTE), which captures the effect of the treatment on the outcome's distribution for individuals at the margin for participation. The DMTE can then be used to derive bounds on the quantile version of the marginal treatment effect and many other parameters.}

	This paper is organized as follows. Section \ref{anaF} presents the structural model and sample selection mechanism considered, followed by a discussion of the identifying assumptions. In Section \ref{Ident}, we provide the identification results for the MTE bounds in the case of a continuous instrument under each set of assumptions while Section \ref{Inf} discusses general aspects of estimation for the bounds. Section \ref{empirical} illustrates an estimation procedure and the identifying power of each set of assumptions using data made available by \cite{Deb2006}. Section \ref{extensions} presents identification results for the case of discrete instruments. Section \ref{Con} concludes. The proofs, sharp testable implications of the model, extensions to DMTE, numerical example, proposed estimation methods' details, Monte Carlo simulations for the estimator's performance, economic models illustrating our identifying assumptions, and an alternative specification for our empirical example are presented in the appendix.

	\section{Analytical Framework}\label{anaF}

	Following \cite{lee2009training}, and \cite{chen2015bounds}, we consider the generalized sample selection model, described in the potential outcomes framework:
	\begin{eqnarray}\label{seq1}
		\left\{ \begin{array}{lcl}
			Y^*&=&Y^{*}_{1}D+Y^{*}_{0}(1-D)\\ \\
			D&=&\mathbbm{1}\left\{V\leq P(Z)\right\} \\ \\
			S&=&S_1D+S_0(1-D)\\ \\
			Y&=&
			Y^{*} S 
		\end{array} \right.
	\end{eqnarray}
	where $Z$ is a vector of observable instrumental variables (e.g., random assignment of cash incentives for participation in a job training program) with support given by $\mathcal Z\subset \mathbb R^{d_{z}}$, $D$ is the treatment status indicator (job training program enrollment). The variable $Y^{*}$ is the possibly censored realized outcome variable (wages) with support $\mathcal Y \subset \mathbb R$, while $Y_{0}^{*}$ and $Y_{1}^{*}$ are the possibly censored potential outcomes when the person is untreated and treated, respectively. Similarly, $S$ is the realized sample selection indicator (employment status), and $S_{0}$ and $S_{1}$ are potential sample selection indicators when individuals are untreated and treated. Finally, $Y$ is the uncensored observed outcome (labor earnings), and $V$ represents unobserved individual characteristics (cognitive or social costs of attending the job training program). The researcher observes only the vector $\left(Y,D,S,Z\right)$, while $Y^{*}_{1}$, $Y^{*}_{0}$, $S_{1}$, $S_{0}$ and $V$ are latent variables.\footnote{For simplicity, we drop exogenous covariates from the model. All results derived in the paper hold conditionally on covariates.} This model is a generalization of that considered in \cite{heckman1999, heckman2001, heckman2005structural} to the sample selection setting.

	The treatment status $D$ is connected to the instrument $Z$ and the unobserved characteristics $V$ through the unknown function $P: \mathcal{Z} \rightarrow \mathbb{R}$. In this model, we assume that the individual receives treatment when its idiosyncratic cost $V$ is less than or equal to a threshold $P(Z)$. This assumption is equivalent to imposing monotonicity of the treatment in the instrument $Z$ \citep{imbensangrist94} as shown by \cite{vytlacil2002}.
	This setup is similar to the one proposed by \cite{heckman2005structural} and leads to the definition of the MTE,
	\begin{align*}
		MTE(p)=\mathbb E[Y^{*}_{1}-Y^{*}_{0} | V=p]
	\end{align*}
	for any $p \in \left[0, 1\right]$.

	In the setting analyzed here, the task of learning about the MTE is further complicated by the potential for nonrandom sample selection. As pointed out by \cite{lee2009training}, even in the simpler case of the ATE, point identification is no longer possible even if the treatment is randomly assigned, leading him to derive bounds for the ATE. This paper combines the insights of these literatures to develop sharp bounds for the MTE under sample selection while allowing for treatment to be endogenously determined.

	Similarly to the compliance groups defined by \cite{imbensangrist94}, we define four latent groups based on the potential sample selection indicators. The subpopulations are defined as: always-observed ($S_{0} = 1, S_{1} = 1$), observed-only-when-treated ($S_{0} = 0, S_{1} = 1$), observed-only-when-untreated ($S_{0} = 1, S_{1} = 0$), and never-observed ($S_{0} = 0, S_{1} = 0$).\footnote{Since the conditioning subpopulation is determined by post-treatment outcomes, our work is also connected to the statistical literature known as principal stratification \citep{frangakis2002principal}, in which the four latent groups would be called strata.} Those subgroups are summarized in Table~\ref{Table:subpop2}.

	\begin{table}[htbp]
		\caption{Subgroups based on the sample selection status}
		\begin{center}
			\begin{tabular}{@{} lcccccc @{}}
				\hline
				\hline\\
				subgroups & $S_{0}$ & $S_{1}$ &  Designation \\
				\hline \\

				$OO$  & 1 & 1 & Always-observed \\
				$ON$  & 1 &  0 & Observed-only-when-untreated\\
				$NO$  & 0 & 1& Observed-only-when-treated \\
				$NN$  & 0 &  0& Never-observed\\
				\hline\\
			\end{tabular}
		\end{center}
		\label{Table:subpop2}
	\end{table}

	Following \cite{Zhang2008} and \cite{lee2009training}, we focus on the always-observed subpopulation $\left(S_{0} = 1, S_{1} = 1\right)$. Importantly, this subpopulation is the only group whose censored potential outcomes are observed in both treatment arms. For the other three subpopulations, treatment effect parameters are not point-identified or bounded in a non-trivial way without further parametric assumptions, since at least one of the potential outcomes ($Y_{0}^{*}$ or $Y_{1}^{*}$) is never observed.\footnote{In some applications, the potential censored outcome $Y_{d}^{*}$ is not even properly defined when $S_{d} = 0$ for $d \in \left\lbrace 0, 1 \right\rbrace$, e.g., analyzing the impact of a medical treatment on a health quality measure where selection is given by whether the patient is alive} Since our focus is on a fully non-parametric identification strategy, we do not discuss parametric identification of unconditional treatment effect parameters or for treatment effect parameters associated with the latent groups $ON$, $NO$ and~$NN$.

	Our target parameter is the MTE function for the subpopulation who is always observed ($MTE^{OO} \colon \left[0,1\right] \rightarrow \mathbb{R}$):
	\begin{equation}\label{target}
		MTE^{OO}\left(p\right) \coloneqq \mathbb{E}\left[Y_{1}^{*} - Y_{0}^{*} \left\vert V = p, S_{0} = 1, S_{1} = 1 \right.\right],
	\end{equation}
	for any $p \in \left[0, 1\right]$. Note that this parameter captures the intensive margin of the treatment effect.\footnote{If the researcher is interested in the extensive margin of the treatment effect, captured by the MTE on the observed outcome ($\mathbb{E}\left[Y_{1} - Y_{0} \left\vert V = p\right.\right]$) and by the MTE on the selection indicator ($\mathbb{E}\left[S_{1} - S_{0} \left\vert V = p\right.\right]$), she can apply the identification strategies described by \cite{Heckman2006}, \cite{Brinch2017}, \cite{Mogstad2017} and \cite{Andresen2018}.} For example, when evaluating the effect of a job training program on wages, this parameter captures the effect of the training program on the wage of a worker who is employed regardless of her treatment status.\footnote{This conditional parameter is interesting for a policy maker and can offer meaningful information regarding treatment and its targeting. For example, the effect of a training program on individuals' wages reflects their productivity and the overall economic welfare. If a training program only affects overall earnings by attracting more participants to the labor market without any effect on productivity, the program's targeting and overall benefit to society should consider this aspect.} In medical applications where selection is due to the death of a patient, this parameter is the effect on health quality for the subpopulation who survives regardless of treatment status. In the education literature where sample selection is due to students quitting school, this parameter is the effect on test scores for the subpopulation who does not drop out of school in any case. Moreover, the definition of the $MTE^{OO}$ does not depend on the instrument being used, implying that our target parameter is policy-invariant. This feature is an advantage in comparison with the $LATE^{OO}$ \citep{chen2015bounds}, which is not policy-invariant. Although the $MTE^{OO}$ function shares the policy invariance property with the usual $MTE$ function, it has one important drawback: the definition of $MTE^{OO}$ conditions on a latent group $\left(S_{0} = 1, S_{1} = 1 \right)$ to simultaneously address the selection-into-treatment and sample selection problems.

	Analogously to \cite{lee2009training}, identification of $MTE^{OO}$ is complex because sample selection is nonrandom and possibly impacted by the treatment. To address this issue, we consider three sets of increasingly restrictive assumptions that allow partial identification of the target parameter, shrinking the identified sets as the assumptions are strengthened.\footnote{According to \citet[p. 167]{Tamer2010}, this approach to identification ``characterizes the informational content of various assumptions by providing a menu of estimates, each based on different sets of assumptions, some of which are plausible and some of which are not.'' Empirically, this approach is also illustrated by \cite{Kline2016}.}

	Assumptions \ref{RA}-\ref{Dist} are sufficient to partially identify the $MTE^{OO}$ function.

	\begin{assumption}[Random Assignment]\label{RA}
		The vector of instruments $Z$ is independent of all latent variables, i.e., $Z\ \indep\ (Y^{*}_{0},Y^{*}_{1},S_0,S_1,V)$.
	\end{assumption}
	\begin{assumption}[Propensity Score is Continuous]\label{continuity}
		$P(z)$ is a nontrivial function of $z$ and the random variable $P\left(Z\right)$ is absolutely continuous with support given by an interval $\mathcal{P} \coloneqq \left[\underline{p}, \overline{p}\right] \subseteq \left[0,1\right]$.\footnote{The assumption that $\mathcal{P}$ is a interval is made for notational simplicity. All the proofs can be easily extended to case when $\mathcal{P}$ is a set with a non-empty interior.}
	\end{assumption}
	\begin{assumption}[Positive Mass]\label{positive}
		Both treatment groups and the always-observed subpopulation exist, i.e., $0 < \mathbb{P}\left[D = 1\right] < 1$ and $\mathbb{P}\left[\left. S_{0} = 1, S_{1} = 1\right\vert V = p\right] > 0$ for any $p \in \mathcal{P}$.
	\end{assumption}
	\begin{assumption}[Finite Moments]\label{finite}
		The first population moment of the  potential outcomes for the always-observed subpopulation conditional on $V$  is finite, i.e., $\mathbb{E}\left[\left. \left\vert Y_{d}^{*} \right\vert \hspace{2pt} \right\vert S_{0} = 1, S_{1} = 1, V=p \right] < + \infty$ for any $p\in \mathcal P$ and $d \in \left\lbrace 0, 1 \right\rbrace$.
	\end{assumption}
	\begin{assumption}[Uniform Distribution of $V$]\label{Dist}
		The unconditional distribution of $V$ is uniform over $[0,1]$, i.e., $V \sim \mathcal U_{[0,1]}$.
	\end{assumption}

	Assumption \ref{RA} is a modification of the IV independence assumption to account for sample selection. Instead of assuming that the instrument is independent of the latent heterogeneity and of the potential outcomes only \citep{heckman2005structural}, we also assume independence of the potential sample selection indicators. Intuitively, we rely on changes in $Z$ shifting treatment status and, hence, sample participation to identify the marginal treatment effect bounds. Such an assumption is common in the empirical literature. A researcher interested in analyzing the impact of a job training program on wages and employment status, usually imposes a similar restriction on the marginal distributions of labor earnings and employment.

	Assumption \ref{continuity} is important for our bounding strategy of the MTE across values of the latent variable, $V$. In Section \ref{extensions}, we relax this assumption to allow for discrete instruments, and we show how our methodology can be used to bound the LATE, instead of the MTE.

	Assumptions \ref{positive}-\ref{Dist} are technical assumptions to ensure that the objects of interest are well-defined and are common in the literature about MTE \citep{Heckman2006}. Assumption \ref{positive} is crucial for the identification results and requires that there are always-observed individuals for all possible values of the unobserved heterogeneity $V$. This can be restrictive in practice, ruling out the calculation of the MTE bounds for ranges of $V$ in which receipt of treatment determines sample participation heavily. Assumption \ref{Dist} can be seen as a normalization if one assumes that the latent variable $V$ is absolutely continuous. Under the same normalization, the image of the function $P: \mathcal{Z} \rightarrow \mathbb{R}$ is contained in the unit interval.

	Assumptions \ref{RA}-\ref{Dist} form our first set of assumptions required for partial identification of the MTE for the always-observed individuals. Under those assumptions, the function $P(z)$ is identified and is equal to the propensity score $\mathbb P\left[D=1|Z=z\right]$ \cite[p. 677]{heckman2005structural}. Indeed, $\mathbb P\left[D=1|Z=z\right]=\mathbb P\left[V\leq P(z)|Z=z\right]=\mathbb P\left[V\leq P(z)\right]=P(z)$, where the second equality holds under Assumption \ref{RA} and the last holds under Assumption \ref{Dist}. This first set of restrictions partially identifies $MTE^{OO}$, as presented in Section \ref{noassumption}.

	We also stress that the identified set can be substantially tightened by imposing that the sample selection mechanism is monotone in the treatment.

	\begin{assumption}[Monotone Sample Selection]\label{MONS}
		Treatment has a non-negative effect on the sample selection indicator for all individuals, i.e., $S_{1} \geq S_{0}$.
	\end{assumption}

	This monotonicity assumption rules out the existence of the observed-only-when-untreated subpopulation and is commonly used in the sample selection literature (\citealp{lee2009training}, \citealp{chen2015bounds}).\footnote{As in \cite{lee2009training}, this assumption can be stated as $S_1\geq S_0$ with probability 1. For the sake of simplicity, we assume it to hold for all individuals. \cite{Manski1997} and \cite{Manski2000} refer to this assumption as the ``monotone treatment response'' assumption. All results can be stated with some straightforward changes if the inequality in Assumption \ref{MONS} holds in the opposite direction.} To obtain some intuition on the mechanisms behind this assumption, consider the job training program example. An individual is employed when her job search skills $\vartheta(D)$, a function of training take-up, are above a threshold $U_{S}$ so that $S=\mathbbm{1}\left\{\vartheta(D) \geq U_S\right\}.$ Additionally, suppose that attending the job training program does not decrease someone's job search skills, i.e., $\vartheta (1) \geq \vartheta (0)$, making it more likely that program's trainees would be observed in the data. In such a case, Assumption~\ref{MONS} holds. However, if attending the job training program raises the agents' reservation wages or if the lost labor market experience is very costly in terms of job finding, this assumption may not hold.

	Assumptions \ref{RA}-\ref{MONS} form our second set of identification assumptions and lead to the bounds for $MTE^{OO}$ that are the main result of this paper, presented in Proposition \ref{thm1}. This second set of assumptions has a testable implication: the treatment positively affects sample selection, i.e., $E[S_{1}-S_{0}|V=p] \geq 0$, implying
	\begin{eqnarray}
		\frac{\partial \mathbb P\left[S=1|P(Z)=p\right]}{\partial p} \geq 0\ \text{for all}\ p \in \mathcal{P}. \label{test3}
	\end{eqnarray}
	In other words, the share of the population for which the outcome is observed rises with $p$. We discuss further testable implications in Section \ref{Ident}, and formally characterize sharp testable implications arising from those assumptions in Appendix \ref{testable}. In the job training example, this testable implication means that the likelihood of employment increases with the probability of attending the training program.

	We can further shrink the identified set around the $MTE^{OO}$, by adding Assumption \ref{DOM} and completing the third set of identifying assumptions.
	\begin{assumption}[Stochastic Dominance]\label{DOM}
		The distribution of the potential outcome when treated for the always-observed subpopulation first-order stochastically dominates the distribution of the same random variable for the observed-only-when-treated subpopulation, i.e.,
		\begin{align*}
			\prob\left[\left.Y_{1}^{*} \leq y \right\vert V = p, S_0 = 1, S_1 = 1 \right] \leq \prob\left[\left.Y_{1}^{*} \leq y \right\vert V = p, S_0 = 0, S_1 = 1 \right]
		\end{align*}
		for any $y \in \mathcal{Y}$ and any $p \in \mathcal{P}$.
	\end{assumption}
	This dominance assumption imposes that the always-observed subpopulation has higher potential censored outcomes than the observed-only-when-treated group conditional on $V$. This type of assumption is common in the literature (\citealp{Imai2008}, \citealp{Blanco2013}, \citealp{Huber2015}, and \citealp{Huber2017}) and is intuitively based on the argument that some population sub-groups have more favorable underlying characteristics than others.\footnote{All of our results can be stated if the inequality in Assumption \ref{DOM} holds in the opposite direction, as it is the case if larger values of the outcome harms the agent. For example, the researcher might be interested on the effect of a drug on cholesterol levels and the selection is based on whether the patient is alive.} Naturally, the plausibility of Assumption \ref{DOM} depends on the empirical context. In some cases this stochastic dominance assumption could be hard to interpret and motivate empirically. However, this challenge can be confronted constructively in a layered policy analysis \citep{Manski2011}, since we can offer a menu of estimates based on different assumptions, allowing the researcher to understand the continuum of information that we can gather about a specific economic parameter, as advocated by \cite{Tamer2010}. In Appendix \ref{econmodel}, we provide a simple economic model related to our job training example to illustrate that Assumption \ref{DOM} may hold under plausible economic restrictions.

\begin{remark}\label{unitmass}Point identification of $MTE^{OO}$ is achieved if, in addition to assumptions \ref{RA}-\ref{Dist}, we assume that the always-observed and never-observed subpopulations are the only existing groups, i.e., $S_{0}=S_{1}$. This ``no selection effect'' assumption imposes that the treatment has no impact on sample selection. This set of assumptions has a testable implication: $\frac{\partial \mathbb P\left[S=1|P(Z)=p\right]}{\partial p} = 0\ \text{for all}\ p \in \mathcal{P}.$  In the job training program context, ``no selection'' implies that workers' employment status would not be affected by program take-up and the likelihood of employment does not depend on the probability of attending the training program.
\end{remark}

	\section{Identification Results}\label{Ident}

	This section presents the main results of this paper, the identification for $MTE^{OO}(p)$ under the three different sets of assumptions described in Section \ref{anaF}. As stepping stones, Subsection \ref{joint} shows identification of the conditional joint distribution of $\left. \left(Y_{d}^{*}, S_{d}=1\right) \right\vert V$ for any $d \in \left\lbrace 0, 1 \right\rbrace$, while Subsection \ref{Mixtures} shows that the distribution of the potential outcomes can be seen as a mixture of latent groups, an important feature of the model. In the following subsections, we sharply bound the $MTE^{OO}$ under increasingly restrictive assumptions. First, we bound the $MTE^{OO}$ without imposing any assumption on the selection mechanism (Subsection \ref{noassumption}). We then tighten those bounds by additionally imposing monotone sample selection (Subsection \ref{monotonicity}) and stochastic dominance (Subsection \ref{stodom}). For completeness, we also show that the $MTE^{OO}$ is point-identified under the ``no selection effect'' assumption (Remark \ref{unitmass}) at the end of Subsection \ref{stodom}. Finally, in Subsection \ref{emprel}, we discuss how to sharply bound treatment effect parameters that can be written as weighted averages of the $MTE^{OO}(p)$.

	\subsection{Identifying the Joint Distribution of Potential Outcome and Selection}\label{joint}

	Before we discuss the identification of the $MTE^{OO}$, we point-identify the conditional joint distribution of each potential outcome and sample selection for different levels of individual heterogeneity, $\left. \left(Y_{d}^{*}, S_{d}=1\right) \right\vert V$ for $d \in \left\lbrace 0, 1 \right\rbrace$.

	Under Assumptions \ref{RA}-\ref{Dist}, for any $p \in \text{int }\mathcal{P}$ and any Borel set $A \subseteq \mathcal Y$, we have that
	\begin{align*}
		\mathbb P\left[Y\in A, S=1, D=1|P(Z)=p\right] & = \mathbb P\left[Y^{*}_{1}\in A,S_1=1,V\leq p|P(Z)=p\right]\\
		&= \mathbb P\left[Y^{*}_{1}\in A,S_1=1,V\leq p\right]\\
		&= \mathbb P\left[Y^{*}_{1}\in A,S_1=1|V\leq p\right]\mathbb P\left[V\leq p\right]\\
		&= \left(\int^{p}_0\mathbb P\left[Y^{*}_{1}\in A,S_1=1|V=v\right]\frac{f_{V}(v)}{\mathbb P\left[V\leq p\right]}dv\right) \cdot \mathbb P\left[V\leq p\right]\\
		&= \int^{p}_0\mathbb P\left[Y^{*}_{1}\in A, S_1=1|V=v\right]dv,
	\end{align*}
	where the second equality follows from Assumption \ref{RA}, the third and fourth equalities follow from the Law of Iterated Expectations, and the last equality follows from Assumption \ref{Dist}. By differentiating each side with respect to $p$, we point-identify the conditional distribution of $\left(Y_{1}^{*}, S_{1}=1\right)$ given $V=p$ as follows:
	\begin{eqnarray}\label{eq1}
		\mathbb P\left[Y^{*}_{1}\in A, S_1=1|V=p\right]=\frac{\partial \mathbb P\left[Y\in A, S=1, D=1|P(Z)=p\right]}{\partial p}.
	\end{eqnarray}

	Similarly, we can show that
	\begin{eqnarray}\label{eq2}
		\mathbb P\left[Y^{*}_{0}\in A, S_0=1|V=p\right]=-\frac{\partial \mathbb P\left[Y\in A, S=1,D=0|P(Z)=p\right]}{\partial p}.
	\end{eqnarray}

	Note that since equations \eqref{eq1} and \eqref{eq2} reflect probabilities, they generate two testable implications for Assumptions \ref{RA} and \ref{Dist}:
	\begin{eqnarray}
		&& 0 \leq \frac{\partial \mathbb E[\mathbbm{1}\left\{Y\in A\right\}SD|P(Z)=p]}{\partial p} \leq 1,\label{test1}\\
		&& 0 \leq -\frac{\partial \mathbb E[\mathbbm{1}\left\{Y\in A\right\}S(1-D)|P(Z)=p]}{\partial p} \leq 1,\label{test2}
	\end{eqnarray}
	for all Borel sets $A \subset \mathbb R$ and $p\in (0,1).$ Intuitively, for people with observable characteristics ($Z$) that indicate a higher likelihood of being treated, the share of treated (untreated) individuals that self-select into the sample increases (decreases) for any range of the outcome.\footnote{A formal characterization of the sharp testable implications implied by our model is given in Appendix \ref{testable}.}

	Similarly to the local IV approach proposed by \cite{heckman2005structural}, equations \eqref{eq1} and \eqref{eq2} can be used to point-identify the MTE on the probability of being observed $\left(\mathbb E[S_{1}-S_{0}|V=p]\right)$, capturing the extensive margin of the treatment. Note that, for $A=\mathcal Y$,
	\begin{eqnarray}
		\mathbb P\left[S_{1}=1|V=p\right]&=&\frac{\partial \mathbb P\left[S=1,D=1|P(Z)=p\right]}{\partial p},\label{eq1p}\\
		\mathbb P\left[S_{0}=1|V=p\right]&=&-\frac{\partial \mathbb P\left[S=1,D=0|P(Z)=p\right]}{\partial p}, \label{eq2p}
	\end{eqnarray}
	implying that $\mathbb E[S_{1}-S_{0}|V=p]=\frac{\partial \mathbb E[S|P(Z)=p]}{\partial p}$. This effect could be of interest in itself: for example, the researcher may want to evaluate whether a training program increases employment levels. We can also identify the MTE on the observed outcome,
	\begin{align*}
		\mathbb E[Y^{*}_{1}S_{1}-Y^{*}_{0}S_{0} | V=p] = \frac{\partial \mathbb E[YS|P(Z)=p]}{\partial p}.
	\end{align*}

	We would like to disentangle the marginal treatment on the observed outcome into the extensive margin and the intensive margin. While the extensive margin is point-identified, we show that the intensive margin ($MTE^{OO}$) is partially identified by considering that the distribution of potential outcomes is a mixture of latent groups.

	\subsection{Potential Outcomes as Mixtures of Latent Groups}\label{Mixtures}

	Fundamental to our identification strategy is recognizing that the observed treated (untreated) group is composed only by $OO$ and $NO$ ($ON$) types, as described in Table \ref{Table:subpop2}. Hence, the conditional distribution $\left. Y_{1}^{*} \right\vert S_1 = 1, V = p$ can be written as the mixture of these latent distributions. For notational simplicity, let $\alpha(p) \equiv \frac{\prob\left[OO|V=p\right]}{\mathbb P\left[S_{1}=1|V=p\right]}$ be the share of always-observed individuals among those for which $S_{1}=1$ conditional on $V=p$. Naturally, the remainder, $\frac{\mathbb P\left[NO|V=p\right]}{\mathbb P\left[S_{1}=1|V=p\right]}$, can be described as $1-\alpha(p)$. By the Law of Total Probability, we have that:
	\begin{align}
		\label{eqmix} \prob\left[\left. Y^{*}_{1}\in A \right\vert S_{1}=1, V=p \right] & = \alpha(p) \cdot \prob\left[\left. Y^{*}_{1}\in A \right\vert S_{0} = 1, S_{1} = 1, V=p \right] \\
		& \hspace{20pt} + \left(1 - \alpha\left(p\right)\right) \cdot \prob\left[\left. Y^{*}_{1}\in A \right\vert S_{0} = 0, S_{1} = 1, V=p \right]. \nonumber
	\end{align}

	As a consequence, $\mathbb E[Y^{*}_{1}|S_{1}=1, V=p]$ is also a mixture of the expectation of $Y^{*}_{1}$ for the always-observed and for observed-only-when-treated given $V=p$,
	\begin{align*}
		\mathbb{E}\left[\left. Y^{*}_{1}\right\vert S_1=1, V=p \right] & = \alpha\left(p\right) \cdot \mathbb{E}\left[\left. Y^{*}_{1} \right\vert S_{0} = 1, S_{1} = 1, V=p \right] \\
		& \hspace{20pt} + \left(1 - \alpha\left(p\right)\right) \cdot \mathbb{E}\left[\left. Y^{*}_{1} \right\vert S_{0} = 0, S_{1} = 1, V=p \right]. \nonumber
	\end{align*}

	Similarly, the conditional distribution of $\left. Y_{0}^{*} \right\vert S_{0} = 1, V = p$ is the mixture of $\left. Y_{d}^{*} \right\vert V = p$ for two latent groups, the always-observed and the observed-only-when-untreated group:
	\begin{align*}
		\mathbb{E}\left[\left. Y^{*}_{0}\right\vert S_0=1, V=p \right] & = \beta\left(p\right) \cdot \mathbb{E}\left[\left. Y^{*}_{0} \right\vert S_{0} = 1, S_{1} = 1, V=p \right] \\
		& \hspace{20pt} + \left(1 - \beta\left(p\right)\right) \cdot \mathbb{E}\left[\left. Y^{*}_{0} \right\vert S_{0} = 1, S_{1} = 0, V=p \right], \nonumber
	\end{align*}
	where $\beta(p) \equiv \frac{\prob\left[OO|V=p\right]}{\mathbb P\left[S_{0}=1|V=p\right]}$.

	We exploit these mixture representations to bound the marginal treatment response of the censored treated outcome within the always-observed subpopulation $\left(\mathbb{E}\left[\left. Y^{*}_{1} \right\vert S_{0} = 1, S_{1} = 1, V=p \right]\right)$ by considering the tails of the observed outcomes' distribution for treated individuals. The smallest attainable value of $\mathbb E[Y^{*}_{1}|S_{0} = 1, S_{1} = 1, V=p]$ is obtained when we consider the scenario in which the always-observed individuals are contained entirely in the left tail of mass $\alpha(p)$ of the outcome distribution, i.e., the lowest values of $Y^{*}_{1}$ among the subpopulation $\{S_{1}=1\}$ conditional on $V$ being equal to $p$. Respectively, the largest attainable value of $\mathbb E[Y^{*}_{1}|S_{0} = 1, S_{1} = 1, V=p]$ is obtained in the case that the always-observed individuals would be the right tail of the same distribution, getting the highest values of $Y^{*}_{1}$ on that subpopulation. This is the same intuition behind the trimming procedure suggested by \cite{lee2009training} and \cite{chen2015bounds}, but, differently from them, the trimmed distribution is conditional on a specific value for the latent heterogeneity variable. This type of trimming approach, used in the current and above papers, relies on results derived in a more general mixture model by \cite{horowitz1995}.

	Hence, $\mathbb{E}\left[\left. Y^{*}_{1} \right\vert S_0 = 1, S_1 = 1, V=p \right]$ lies within the interval $[LB_{1}(p),UB_{1}(p)],$ where
	\begin{align}
		\label{eqboundlb1} LB_{1}(p) & = \mathbb E\left[Y^{*}_{1}|S_1=1,V=p, Y^{*}_{1}\leq F^{-1}_{Y^{*}_{1}|S_{1}=1,V=p}\left(\alpha(p)\right)\right], \\
		\label{eqboundub1} UB_{1}(p) & = \mathbb E\left[Y^{*}_{1}|S_1=1,V=p, Y^{*}_{1} > F^{-1}_{Y^{*}_{1}|S_{1}=1,V=p}\left(1-\alpha(p)\right)\right]
	\end{align}
	and $F^{-1}_{Y^{*}_{d}|S_{d}=1,V=p}(\cdot)$ is the quantile function of the distribution of $Y^{*}_{d}$ given $S_{d}=1 \text{ and } V=p$.

	Similarly, the conditional distribution of $\left. Y_{0}^{*} \right\vert S_0 = 1, V = p$ can be written as the mixture of $\left. Y_{d}^{*} \right\vert V = p$ for two latent groups, the always-observed and the observed-only-when-untreated group. Analogously to the treated outcome, the marginal treatment response of the untreated outcome within the always-observed subpopulation $\left(\mathbb{E}\left[\left. Y^{*}_{0} \right\vert S_0 = 1, S_1 = 1, V=p \right]\right)$ lies within the interval $[LB_{0}(p),UB_{0}(p)],$ where
	\begin{align}
		\label{eqboundlb0} LB_{0}(p) & = \mathbb E\left[Y^{*}_{0}|S_0=1,V=p, Y^{*}_{0}\leq F^{-1}_{Y^{*}_{0}|S_{0}=1,V=p}\left(\beta(p)\right)\right], \\
		\label{eqboundub0} UB_{0}(p) & = \mathbb E\left[Y^{*}_{0}|S_0=1,V=p, Y^{*}_{0} > F^{-1}_{Y^{*}_{0}|S_{0}=1,V=p}\left(1-\beta(p)\right)\right].
	\end{align}

	Combining the bounds around $\mathbb{E}\left[\left. Y^{*}_{1} \right\vert S_{0} = 1, S_{1} = 1, V=p \right]$ and $\mathbb{E}\left[\left. Y^{*}_{0} \right\vert S_{0} = 1, S_{1} = 1, V=p \right]$, we find that $MTE^{OO}\left(p\right)$ lies within the interval $$[LB_{1}(p) - UB_{0}(p), UB_{1}(p) - LB_{0}(p)].$$

	\begin{remark}\label{remarkalpha}
		The issue central to identification of the target parameter is what can be learned about the mixture weights $\left(\alpha(p), \beta(p)\right)$, and $\mathbb{E}\left[\left. Y^{*}_{d}\right\vert S_d=1, V=p \right]$. Note that the bounds on the MTE of interest will be tighter for higher values of $\alpha(p)$ and $\beta(p)$  because we learn about $E[Y^{*}_{1}-Y^{*}_{0}|S_{0} = 1, S_{1} = 1, V=p]$ by considering that the worst- and best-case outcomes of observed treated and untreated individuals are fully attributed to the always-observed. So, as $\alpha(p)$ increases, the share of the observed sample of treated individuals that are from our group of interest increases, providing more information about their conditional expectation of the outcomes. In the extreme case in which $\alpha(p)\rightarrow 1$, the $\mathbb{E}\left[\left. Y^{*}_{1} \right\vert S_0 = 1, S_1 = 1, V=p \right]$ will be point identified. Similarly, if $\alpha(p)\rightarrow 0$, the observed sample is uninformative about the always-observed group. A similar intuition holds regarding $\beta(p)$.
	\end{remark}

	In the next subsections, we investigate the bounds that are generated under the alternative sets of assumptions described in Section \ref{anaF}. Intuitively, those assumptions impose different restrictions on the possible values of the mixture weights $\left(\alpha(p), \beta(p)\right)$, providing different sets of information about $E[Y^{*}_{1}-Y^{*}_{0}|S_{0} = 1, S_{1} = 1, V=p]$.

	\subsection{Identification with No Assumption on the Sample Selection Mechanism}\label{noassumption}

	Initially, consider the case in which the researcher is only willing to consider Assumptions \ref{RA}-\ref{Dist}, leaving the sample selection mechanism unrestricted. To learn about $MTE^{OO}$, we need information about the share of always-observed individuals in the total population, $\prob \left[S_{0} = 1, S_{1} = 1\right]$ and, hence, the conditional joint distribution of $\left. \left(S_{0}, S_{1}\right) \right\vert V = p$. However, we only have information about the conditional marginal distributions $\left. S_{0} \right\vert V = p$ and $\left. S_{1} \right\vert V = p$ based on equations \eqref{eq1p} and \eqref{eq2p}. According to \cite{Imai2008} and \cite{Mullahy2018}, the following Boole-Fréchet bounds are sharp around the share of always-observed individuals:
	\begin{align}
		\prob \left[S_0 = 1, S_1 = 1 \vert V = p\right] & \in \left[\max \left\lbrace \prob \left[S_0 = 1 \vert V = p \right] + \prob \left[S_1 = 1 \vert V = p \right] - 1, 0  \right\rbrace, \right. \nonumber \\
		& \hspace{20pt} \left. \min \left\lbrace \prob \left[S_0 = 1 \vert V = p \right] , \prob \left[S_1 = 1 \vert V = p \right] \right\rbrace  \right].
	\end{align}
	Combining this information with Equations \eqref{eq1p} and \eqref{eq2p}, leads to Lemma \ref{BFbounds}.
	\begin{lemma}\label{BFbounds}
		Under Assumptions \ref{RA}-\ref{Dist}, the share of always-observed individuals is partially identified:
		\begin{align*}
			& \prob \left[S_0 = 1, S_1 = 1 \vert V = p\right] \\
			& \hspace{20pt} \in \left[\max \left\lbrace -\frac{\partial \mathbb P\left[S=1,D=0|P(Z)=p\right]}{\partial p} + \frac{\partial P\left[S=1,D=1|P(Z)=p\right]}{\partial p} - 1, 0  \right\rbrace, \right. \\
			& \hspace{40pt} \left. \min \left\lbrace -\frac{\partial \mathbb P\left[S=1,D=0|P(Z)=p\right]}{\partial p} , \frac{\partial \prob \left[S=1,D=1|P(Z)=p\right]}{\partial p} \right\rbrace  \right] \\
			& \hspace{20pt} \texteqcolon \Upsilon\left(p\right).
		\end{align*}
		These bounds are sharp.
	\end{lemma}

	Note that, since $\Upsilon\left(p\right)$ provides the identified set of possible values for the share of always-observed individuals, we can obtain the equivalent range of possible values for $\alpha(p)$ and $\beta(p)$, the mixture weights described in Subsection \ref{Mixtures}. For brevity, let $\prob \left[S_{0} = 1, S_{1} = 1 \vert V = p\right]$ take any particular value, $\upsilon \in \Upsilon\left(p\right)$. Define,
	\begin{align*}
		\alpha\left(p, \upsilon\right) \coloneqq \prob \left[S_{0} = 1 \vert S_{1} = 1, V = p\right] = \dfrac{\upsilon}{\prob \left[S_1 = 1 \vert V = p \right]} = \dfrac{\upsilon}{\frac{\partial \prob [S=1,D=1|P(Z)=p]}{\partial p}},\\
		\beta\left(p, \upsilon\right) \coloneqq \prob \left[S_{1} = 1 \vert S_{0} = 1, V = p\right] = \dfrac{\upsilon}{\prob \left[S_0 = 1 \vert V = p \right]} = - \dfrac{\upsilon}{\frac{\partial \prob [S=1,D=0|P(Z)=p]}{\partial p}}.
	\end{align*}
	Let the bounds in Equations \eqref{eqboundlb1}-\eqref{eqboundub0}, for specific values of $\alpha(p, \upsilon)$ and $\beta(p, \upsilon)$ in the identified set be written as:
	\begin{align*}
		LB_{1}(p, \upsilon) & = \mathbb E\left[Y^{*}_{1}|S_1=1,V=p, Y^{*}_{1}\leq F^{-1}_{Y^{*}_{1}|S_{1}=1,V=p}\left(\alpha(p, \upsilon)\right)\right], \\
		UB_{1}(p, \upsilon) & = \mathbb E\left[Y^{*}_{1}|S_1=1,V=p, Y^{*}_{1} > F^{-1}_{Y^{*}_{1}|S_{1}=1,V=p}\left(1-\alpha(p, \upsilon)\right)\right],\\
		LB_{0}(p, \upsilon) & = \mathbb E\left[Y^{*}_{0}|S_0=1,V=p, Y^{*}_{0}\leq F^{-1}_{Y^{*}_{0}|S_{0}=1,V=p}\left(\beta(p, \upsilon)\right)\right], \\
		UB_{0}(p, \upsilon) & = \mathbb E\left[Y^{*}_{0}|S_0=1,V=p, Y^{*}_{0} > F^{-1}_{Y^{*}_{0}|S_{0}=1,V=p}\left(1-\beta(p, \upsilon)\right)\right].
	\end{align*}
	Combining the bounds around $\mathbb{E}\left[\left. Y^{*}_{1} \right\vert S_0 = 1, S_1 = 1, V=p \right]$ and $\mathbb{E}\left[\left. Y^{*}_{0} \right\vert S_0 = 1, S_1 = 1, V=p \right]$, we find that  $MTE^{OO}\left(p\right)$ lies within the interval $[LB_{1}(p, \upsilon) - UB_{0}(p, \upsilon), UB_{1}(p, \upsilon) - LB_{0}(p, \upsilon)]$
	for a particular $\prob \left[S_0 = 1, S_1 = 1 \vert V = p\right] = \upsilon$.

	To bound the target parameter, we find worst- and best-case scenarios by varying the value $\upsilon$. Explicitly, $MTE^{OO}\left(p\right)$ is partially identified and lies within the interval
	\begin{align*}
		\left[\min_{\upsilon \in \Upsilon\left(p\right)} \left\lbrace LB_{1}(p, \upsilon) - UB_{0}(p, \upsilon) \right\rbrace, \max_{\upsilon \in \Upsilon\left(p\right)} \left\lbrace UB_{1}(p, \upsilon) - LB_{0}(p, \upsilon) \right\rbrace\right].
	\end{align*}

	Note that $\upsilon$ has a monotone relationship to the mixture weights, which define the trimming points in the bounds. As previously discussed, higher values for $\alpha(p)$ ($\beta(p)$) indicate that a bigger share of the observed treated (untreated) population belongs to the always-observed latent group, thus providing more information and tighter bounds for the parameter of interest. Hence, we only need to focus on the scenario that generates the wider bounds, that is, the smallest admissible $\alpha(p)$ and $\beta(p)$. Let $\upsilon^\ell$ be the lower bound of $\Upsilon(p)$. We have:
	\begin{align*}
		\min_{\upsilon \in \Upsilon\left(p\right)} LB_{1}(p, \upsilon) - \max_{\upsilon \in \Upsilon\left(p\right)}  UB_{0}(p, \upsilon) \leq \min_{\upsilon \in \Upsilon\left(p\right)} \left\lbrace LB_{1}(p, \upsilon) - UB_{0}(p, \upsilon) \right\rbrace,\\
		\min_{\upsilon \in \Upsilon\left(p\right)} LB_{1}(p, \upsilon)=LB_{1}(p,\upsilon^\ell), \text{ and }\max_{\upsilon \in \Upsilon\left(p\right)}  UB_{0}(p, \upsilon)=UB_{0}(p, \upsilon^\ell).
	\end{align*}

	Making the same argument to the upper bound, we can rewrite them as,
	\begin{align*}
		\min_{\upsilon \in \Upsilon\left(p\right)} \left\lbrace LB_{1}(p, \upsilon) - UB_{0}(p, \upsilon) \right\rbrace=LB_1(p,\upsilon^\ell)-UB_0(p,\upsilon^\ell),\\
		\max_{\upsilon \in \Upsilon\left(p\right)} \left\lbrace UB_{1}(p, \upsilon) - LB_{0}(p, \upsilon) \right\rbrace=UB_1(p,\upsilon^\ell)-LB_0(p,\upsilon^\ell),
	\end{align*}
	greatly simplifying our bounds, which need only to be evaluated at the end point of $\Upsilon(p)$.

	We can combine these facts with equations \eqref{eq1}, \eqref{eq2}, \eqref{eq1p} and \eqref{eq2p} to propose the first identification result for $MTE^{OO}$, which does not impose meaningful restrictions on the sample selection mechanism.
	\begin{proposition}\label{THMnoassumption}
		Under Assumptions \ref{RA}-\ref{Dist}, the MTE is partially identified for the always-observed, i.e.,
		$$\underline{\Delta}_{1}\left(p\right) \leq MTE^{OO}\left(p\right) \leq \overline{\Delta}_{1}\left(p\right)$$ for any $p \in \mathcal{P}$, where $\underline{\Delta}_{1} \colon \mathcal{P} \rightarrow \mathbb{R}$ and $\overline{\Delta}_{1} \colon \mathcal{P} \rightarrow \mathbb{R}$ are given by
		\begin{align*}
			\underline{\Delta}_{1}\left(p\right) & \coloneqq \mathbb E\left[\tilde{Y}_1|S=1,D=1,P(Z)=p,\tilde{Y}_1\leq F^{-1}_{\tilde{Y}_1|S=1,D=1,P(Z)=p}\left(\alpha(p, \upsilon^\ell)\right)\right] \\
			& \hspace{20pt}- \mathbb E\left[\tilde{Y}_0|S=1,D=0,P(Z)=p,\tilde{Y}_0> F^{-1}_{\tilde{Y}_0|S=1,D=0,P(Z)=p}\left(1-\beta(p, \upsilon^\ell)\right)\right],\\
			\overline{\Delta}_{1}\left(p\right) & \coloneqq \mathbb E\left[\tilde{Y}_1|S=1,D=1,P(Z)=p,\tilde{Y}_1> F^{-1}_{\tilde{Y}_1|S=1,D=1,P(Z)=p}\left(1-\alpha(p, \upsilon^\ell)\right)\right] \\
			& \hspace{20pt} - \mathbb E\left[\tilde{Y}_0|S=1,D=0,P(Z)=p,\tilde{Y}_0\leq F^{-1}_{\tilde{Y}_0|S=1,D=0,P(Z)=p}\left(\beta(p, \upsilon^\ell)\right)\right]
		\end{align*}
		for any $p \in \mathcal{P}$, the conditional distribution of $\tilde{Y}_d$ is given by
		\begin{equation*}
			\tilde{Y}_d| S=1,D=d,P(Z)=p\ \sim F_{\tilde{Y}_d|S=1,D=d,P(Z)=p}(y) = \frac{\frac{\partial \mathbb P\left[Y\leq y, S=1,D=d|P(Z)=p\right]}{\partial p}}{\frac{\partial \mathbb P\left[S=1,D=d|P(Z)=p\right]}{\partial p}}
		\end{equation*}
		for any $d \in \left\lbrace 0, 1 \right\rbrace$, and
		\begin{align*}
			\alpha\left(p, \upsilon^\ell\right) = \dfrac{\max \left\lbrace -\frac{\partial \mathbb P\left[S=1,D=0|P(Z)=p\right]}{\partial p} + \frac{P\left[S=1,D=1|P(Z)=p\right]}{\partial p} - 1, 0  \right\rbrace}{\frac{\partial \prob \left[S=1,D=1|P(Z)=p\right]}{\partial p}},\\
			\beta\left(p, \upsilon^\ell\right) = - \dfrac{\max \left\lbrace -\frac{\partial \mathbb P\left[S=1,D=0|P(Z)=p\right]}{\partial p} + \frac{P\left[S=1,D=1|P(Z)=p\right]}{\partial p} - 1, 0  \right\rbrace}{\frac{\partial \prob \left[S=1,D=0|P(Z)=p\right]}{\partial p}}.
		\end{align*}
		Moreover, these bounds are uniformly sharp.
	\end{proposition}

	\begin{remark}
		The definition of uniform sharpness \citep{Firpo2019} used in this article states that, for any function $\delta \colon \mathcal{P} \rightarrow \mathbb{R}$ such that $\delta\left(p\right) \in \left[\underline{\Delta}_{1}\left(p\right), \overline{\Delta}_{1}\left(p\right)\right]$ for any $p \in \mathcal{P}$, it is possible to construct random variables $(\tilde{Y}^{*}_{0}, \tilde{Y}^*_1,\tilde{S}_0, \tilde{S}_1, \tilde{V}, Z)$ that satisfy all the restrictions imposed on the data by the underlying assumptions, induce the joint distribution on the data $(Y,S,D,Z)$ and achieve the candidate target parameter $\delta\left(p\right) = \mathbb{E}\left[\left. \tilde{Y}^*_1 - \tilde{Y}^*_0 \right\vert \tilde{S}_0 = 1, \tilde{S}_1 = 1, \tilde{V}=p \right]$ for every $p \in \mathcal{P}$.
	\end{remark}

	\begin{remark} In the special case in which the treatment $D$ is independent of the potential outcomes for $Y^{*}$ and $S$, and that $\Upsilon(p)=\Upsilon$ for any $p \in \mathcal{P} = \left[0, 1\right]$, the MTE on the censored outcome will be constant and equal to the ATE for the always-observed. Then, the bounds derived in Proposition \ref{THMnoassumption} simplify to those derived by \citet[Proposition~1]{Imai2008}.
	\end{remark}

	\subsection{Bounds under the Monotonicity Assumption}\label{monotonicity}

	In this subsection, we introduce monotonicity of sample selection in the treatment (Assumption \ref{MONS}), which can considerably shrink the identified set for $MTE^{OO}$. As discussed in Section \ref{anaF}, under the monotonicity assumption, individuals who self-select into the sample when untreated $\left(S_{0} = 1\right)$ would also be observed if they had been treated, ruling out the subgroup $ON$. In other words, any untreated individuals observed on the sample are members of the always-observed latent subpopulation $\left(S_{0} = 1, S_{1} = 1\right)$.
	Formally, the following two events are identical: $\left\{S_{0}=1\right\}=\left\{S_{0}=1,S_{1}=1\right\}$, and the mixture weight for the untreated group, $\beta(p)$, equals one.

	Consequently, $\mathbb P\left[S_{0}=1,S_{1} = 1|V=p\right]$ is point-identified by equation (\ref{eq2p}), and we no longer need to rely on the partial identification results in Lemma \ref{BFbounds}. Specifically, we have that
	\begin{align}
		\mathbb P\left[S_{0}=1,S_{1}=1|V=p\right]& = \label{eq3}-\frac{\partial \mathbb P\left[S=1,D=0|P(Z)=p\right]}{\partial p}\\
		& = \frac{\partial \mathbb P\left[S=1,D=1|P(Z)=p\right]}{\partial p}-\frac{\partial \mathbb P\left[S=1|P(Z)=p\right]}{\partial p}.\nonumber
	\end{align}
	This result connects the conditional share of always observed individuals to changes on the conditional mass of observed untreated individuals when the propensity score increases. Looking at the second equality, we find that the conditional probability of being always observed is the difference between the increase in the share of observed treated individuals and the increase in the share of observed individuals when the propensity score is equal to $p$.

	Since $\left\{S_{0}=1\right\}=\left\{S_{0}=1,S_{1}=1\right\}$ ($\beta(p)=1$), the distribution of $\left. \left( Y_0^{*}, S_{0}=1,S_{1}=1 \right)\right\vert V$ is equal to the distribution of $\left. \left( Y_0^{*}, S_{0}=1 \right)\right\vert V$, implying that
	\begin{equation}\label{Y0MONS}
		\mathbb P\left[Y^{*}_{0}\in A| S_{0}=1,S_{1}=1, V=p\right]=\frac{\mathbb P\left[Y^{*}_{0}\in A, S_0=1|V=p\right]}{\mathbb P\left[S_{0}=1,S_{1}=1|V=p\right]}.
	\end{equation}
	Note that the right-hand side of equation \eqref{Y0MONS} is point-identified according to equations \eqref{eq2} and \eqref{eq3}. Consequently, the expectation $\mathbb E[Y^{*}_{0}|S_0 = 1, S_1 = 1,V=p]$ is also point-identified. Monotonicity also leads to point identification of the mixture weight, $\alpha(p)=\dfrac{\mathbb P\left[S_{0}=1,S_{1}=1|V=p\right]}{\prob \left[S_1 = 1 \vert V = p \right]}$ by Equations \eqref{eq1p} and \eqref{eq3}.

	Then, under monotonicity, the researcher has to obtain bounds only for the expected potential outcomes under treatment, which still can be written as a mixture of the always-observed and observed-only-when-treated latent subpopulations.

	As discussed in Section \ref{Mixtures}, the expectation $\mathbb E[Y^{*}_{1}|S_{0}=1,S_{1}=1, V=p]$ lies in the interval $[LB_{1}(p),UB_{1}(p)]$, given in Equations \eqref{eqboundlb1}-\eqref{eqboundub1}. Combining the bounds and identification results in equations \eqref{eq1}, \eqref{eq2}, \eqref{eq1p}, \eqref{eq3} and \eqref{Y0MONS}, the following proposition holds:
	\begin{proposition}\label{thm1}
		Under Assumptions \ref{RA}-\ref{MONS}, the MTE is partially identified for the always-observed, i.e.,
		$$\underline{\Delta}_{2}\left(p\right) \leq MTE^{OO}\left(p\right) \leq \overline{\Delta}_{2}\left(p\right)$$ for any $p \in \mathcal{P}$, where $\underline{\Delta}_{2}\colon \mathcal{P}\rightarrow\mathbb{R}$ and $\overline{\Delta}_{2}\colon \mathcal{P}\rightarrow\mathbb{R}$ are given by
		\begin{align*}
			\underline{\Delta}_{2}\left(p\right) & \coloneqq \mathbb E\left[\tilde{Y}_1|S=1,D=1,P(Z)=p,\tilde{Y}_1\leq F^{-1}_{\tilde{Y}_1|S=1,D=1,P(Z)=p}\left(\alpha(p)\right)\right] \\
			& \hspace{20pt} - \mathbb E\left[\tilde{Y}_0|S=1,D=0,P(Z)=p\right],\\
			\overline{\Delta}_{2}\left(p\right) & \coloneqq \mathbb E\left[\tilde{Y}_1|S=1,D=1,P(Z)=p,\tilde{Y}_1> F^{-1}_{\tilde{Y}_1|S=1,D=1,P(Z)=p}\left(1-\alpha(p)\right)\right] \\
			& \hspace{20pt} - \mathbb E\left[\tilde{Y}_0|S=1,D=0,P(Z)=p\right]
		\end{align*}
		for any $p \in \mathcal{P}$, the conditional distribution of $\tilde{Y}_d$ is given by
		\begin{equation*}
			\tilde{Y}_d| S=1,D=d,P(Z)=p\ \sim F_{\tilde{Y}_d|S=1,D=d,P(Z)=p}(y) = \frac{\frac{\partial \mathbb P\left[Y\leq y, S=1,D=d|P(Z)=p\right]}{\partial p}}{\frac{\partial \mathbb P\left[S=1,D=d|P(Z)=p\right]}{\partial p}}
		\end{equation*}
		for any $d \in \left\lbrace 0, 1 \right\rbrace$, and
		\begin{equation*}
			\alpha\left(p\right) = - \dfrac{\frac{\partial \mathbb P\left[S=1,D=0|P(Z)=p\right]}{\partial p}}{\frac{\partial \mathbb P\left[S=1,D=1|P(Z)=p\right]}{\partial p}}.
		\end{equation*}
		Moreover, these bounds are uniformly sharp.
	\end{proposition}
	\begin{remark}
		By adding the monotonicity assumption, we increase the lower bound and decrease the upper bound stated in Proposition \ref{THMnoassumption}. The length of the identified set here is strictly shorter than the length of the identified set in Proposition \ref{THMnoassumption} when the mixture weights are not point identified.\footnote{That is, the set $\Upsilon\left(p\right)$ in Lemma \ref{BFbounds} is not a singleton and the distributions of $\left. Y_{0}^{*} \right\vert S_1 = 1, V$ and $\left. Y_{1}^{*} \right\vert S_1 = 1, V$ are not degenerate.} This improvement shows the identifying power of Assumption \ref{MONS}.
	\end{remark}
	\begin{remark} In the special case in which the treatment $D$ is independent of the potential outcomes for $Y^{*}$ and $S$, and $\alpha(p)=\alpha$ for any $p \in \mathcal{P} = \left[0,1\right]$, the MTE will be constant and equal to the average treatment effect for the always-observed and the bounds in Proposition \ref{thm1} simplify to the ones proposed by \citet[Proposition 1a]{lee2009training}.
	\end{remark}

	\subsection{\textbf{Bounds under the Monotonicity and Dominance Assumptions}}\label{stodom}

	In this section, we add the stochastic mean dominance assumption to tighten the identified set for $MTE^{OO}$ under Assumptions \ref{RA}-\ref{DOM}. Stochastic dominance and equation \eqref{eqmix} imply that
	\begin{align*}
		\prob\left[\left. Y^{*}_{1} \leq y \right\vert S_1=1, V=p \right] \geq \prob\left[\left. Y^{*}_{1} \leq y \right\vert S_0 = 1, S_1 = 1, V=p \right]
	\end{align*}
	for any $y \in \mathcal{Y}$. As a consequence, the following inequality holds
	\begin{equation*}
		\mathbb{E}\left[\left. Y^{*}_{1}\right\vert S_1=1, V=p \right] \leq \mathbb{E}\left[\left. Y^{*}_{1} \right\vert S_0 = 1, S_1 = 1, V=p \right].
	\end{equation*}

	This tightens the lower bound for $\mathbb E[Y^{*}_{1}|S_{0}=1,S_{1}=1, V=p]$ as we no longer need to focus on the lowest $\alpha(p)$ mass of outcomes as the lower bound, since the stochastic dominance assumption guarantees that the expectation of outcomes for the always observed subpopulation will be larger than the one of the observed treated individuals which mixes $OO$ and $NO$ types. Hence, $\mathbb E[Y^{*}_{1}|S_{0}=1,S_{1}=1, V=p]$ lies within the interval $[LB_{3}(p),UB_{3}(p)],$ where
	\begin{eqnarray*}
		LB_3(p)&=&\mathbb{E}\left[\left. Y^{*}_{1} \right\vert S_1=1, V=p \right],\\
		UB_3(p)&=&\mathbb E\left[Y^{*}_{1}|S_1=1,V=p, Y^{*}_{1} > F^{-1}_{Y^{*}_{1}|S_{1}=1,V=p}\left(1-\alpha(p)\right)\right].
	\end{eqnarray*}
	The upper bound remains unchanged.
	Naturally, that leads to tighter identified sets relative to the ones in Proposition \ref{thm1}, which are presented in the following proposition.
	\begin{proposition}\label{thmDOM}
		Under Assumptions \ref{RA}-\ref{DOM}, the MTE is partially identified for the always-observed, i.e.,
		 \begin{align*}
		\underline{\Delta}_{3}\left(p\right) \leq MTE^{OO}\left(p\right) \leq \overline{\Delta}_{3}\left(p\right)
		\end{align*}
	 for any $p \in \mathcal{P}$, where $\underline{\Delta}_{3}\left(p\right) \colon \mathcal{P} \rightarrow \mathbb{R}$ and $\overline{\Delta}_{3}\left(p\right) \colon \mathcal{P} \rightarrow \mathbb{R}$ are given by
	\begin{align*}
			\underline{\Delta}_{3}\left(p\right) & \coloneqq \mathbb E\left[\tilde{Y}_1|S=1,D=1,P(Z)=p\right] - \mathbb E\left[\tilde{Y}_0|S=1,D=0,P(Z)=p\right],\\
			\overline{\Delta}_{3}\left(p\right) & \coloneqq \mathbb E\left[\tilde{Y}_1|S=1,D=1,P(Z)=p,\tilde{Y}_1> F^{-1}_{\tilde{Y}_1|S=1,D=1,P(Z)=p}\left(1-\alpha(p)\right)\right] \\
			& \hspace{20pt} - \mathbb E\left[\tilde{Y}_0|S=1,D=0,P(Z)=p\right]
		\end{align*}
		for any $p \in \mathcal{P}$, where the conditional distribution of $\tilde{Y}_d$ is given by
		\begin{equation*}
			\tilde{Y}_d| S=1,D=d,P(Z)=p\ \sim F_{\tilde{Y}_d|S=1,D=d,P(Z)=p}(y) = \frac{\frac{\partial \mathbb P\left[Y\leq y, S=1,D=d|P(Z)=p\right]}{\partial p}}{\frac{\partial \mathbb P\left[S=1,D=d|P(Z)=p\right]}{\partial p}}
		\end{equation*}
		for any $d \in \left\lbrace 0, 1 \right\rbrace$.
		and
		\begin{equation*}
			\alpha\left(p\right) = - \dfrac{\frac{\partial \mathbb P\left[S=1,D=0|P(Z)=p\right]}{\partial p}}{\frac{\partial \mathbb P\left[S=1,D=1|P(Z)=p\right]}{\partial p}}.
		\end{equation*}
		Moreover, these bounds are uniformly sharp.
	\end{proposition}
	\begin{remark}
		The lower bound proposed here is strictly greater than the one in Proposition \ref{thm1} when $\alpha\left(p\right) \in \left(0, 1\right)$ and the distribution of $\left. Y_{1}^{*} \right\vert S_1 = 1, V$ is not degenerate.  This improvement shows the identifying power of Assumption \ref{DOM}.
	\end{remark}
	\begin{remark}
		In the special case in which the treatment $D$ is independent of the potential outcomes for $Y^{*}$ and $S$, and $\alpha(p)=\alpha$ for any $p \in \mathcal{P} = \left[0,1\right]$, the MTE will be constant and equal to the average treatment effect for the always-observed and the bounds in Proposition \ref{thmDOM} simplify to those proposed in \citet[Equation (8)]{Imai2008}.
	\end{remark}

	For completeness in the identification discussion, note that point identification of $MTE^{OO}$ is achieved if, in addition to assumptions \ref{RA}-\ref{Dist}, we assume that the always-observed and never-observed subpopulations are the only existing groups, i.e., $S_{0}=S_{1}$. This ``no selection effect'' assumption imposes that the treatment has no impact on sample selection. In the job training program context, this implies that workers' employment  would not be affected by the program.

Under Assumptions \ref{RA}-\ref{Dist} and ``no selection effect'', the distributions of $\left. Y_{d}^{*} \right\vert S_{1} = 1, V$ for $d \in \left\lbrace 0, 1 \right\rbrace$ are exclusively composed of always-observed individuals ($\alpha(p)=\beta(p)=1$). Then,
	\begin{align*}
		\mathbb P\left[Y^{*}_{d}\in A| S_{0}=1,S_{1}=1, V=p\right]=\frac{\mathbb P\left[Y^{*}_{d}\in A, S_{d}=1|V=p\right]}{\mathbb P\left[S_{0}=1|V=p\right]} \text{ for } d=\{0,1\},
	\end{align*}
	where point-identification follows from equations \eqref{eq1}, \eqref{eq2}, \eqref{eq1p} and \eqref{eq2p}. The expectations $\mathbb{E}[Y^{*}_{d}|S_0 = 1, S_1 = 1,V=p]$ are also point-identified. Then, $MTE^{OO}\left(p\right) = \frac{\frac{\partial \mathbb E[YS|P(Z)=p]}{\partial p}}{\frac{\partial \mathbb E[SD|P(Z)=p]}{\partial p}}$.

	Alternatively, point-identification of the unconditional MTE is achieved if potential sample selection status and outcomes are independent given the unobserved characteristics,$(S_{0},S_{1})\ \indep\ (Y^{*}_{0},Y^{*}_{1})|V$. In this case, the distributions $\mathbb P\left[Y^{*}_{d}\leq y|V=p\right]$ are point-identified from Equations \eqref{eq1} and \eqref{eq2}, and the unconditional MTE is point-identified as:
		\begin{align*}
			\mathbb E[Y_{1}^{*}-Y_{0}^{*}|V=p] =\frac{\frac{\partial \mathbb E[YSD|P(Z)=p]}{\partial p}}{\frac{\partial \mathbb E[SD|P(Z)=p]}{\partial p}}-\frac{\frac{\partial \mathbb E[YS(1-D)|P(Z)=p]}{\partial p}}{\frac{\partial \mathbb E[S(1-D)|P(Z)=p]}{\partial p}}.
		\end{align*}

	\subsection{Empirical Relevance of Bounds for the \emph{MTE}\textsuperscript{\emph{OO}}}\label{emprel}

	The partial identification results for the $MTE^{OO}$ are relevant for a vast array of empirical objectives. First, bounds for the $MTE^{OO}$ can illuminate the treatment effect's heterogeneity, allowing researchers to better understand who would benefit from a specific treatment. This feature is important because common parameters (e.g., $ATE$, $ATT$, $ATU$, and $LATE$ within the always-observed subpopulation) can be positive even when most people are adversely affected by a policy. Moreover, knowing, even partially, the $MTE^{OO}$ function can be useful to design policies that provide incentives to agents to take some treatment.

	Second, the $MTE^{OO}$ bounds can be used to partially identify alternative treatment effect parameters that are described as a weighted integral of $MTE^{OO}$ because
	\begin{equation*}
		\int_{\mathcal{P}} \left(\underline{\Delta}_{t}\left(p\right) \cdot \omega\left(p\right)\right) \, \text{d} p \leq \int_{\mathcal{P}} \left(MTE^{OO}\left(p\right) \cdot \omega\left(p\right)\right) \, \text{d} p \leq \int_{\mathcal{P}} \left(\overline{\Delta}_{t}\left(p\right) \cdot \omega\left(p\right) \right) \, \text{d} p,
	\end{equation*}
	where $t \in \left\lbrace 1, 2, 3 \right\rbrace$, $\underline{\Delta}_{t}$ and $\overline{\Delta}_{t}$ are described in Propositions \ref{THMnoassumption}-\ref{thmDOM}, and $\omega(\cdot)$ is a known or identifiable weighting function. Those bounds are sharp, as summarized in Proposition \ref{thmTE}.\footnote{We focus on the case under the monotonicity restriction (Assumptions \ref{RA}-\ref{MONS}) for brevity. Similar results hold under our other identifying sets of assumptions.}
	\begin{proposition}\label{thmTE}
		Let $\omega_{TE}\colon \mathcal{P} \rightarrow \mathbb{R}$ be a known or identifiable weighting function and define the treatment effect parameter $TE \coloneqq \int_{\mathcal{P}} MTE^{OO}\left(p\right) \cdot \omega_{TE}\left(p\right) \, \text{d} p$. Under Assumptions \ref{RA}-\ref{MONS}, the treatment effect parameter TE is partially identified:
		\begin{equation*}
			\int_{\mathcal{P}} \left(\underline{\Delta}_{2}\left(p\right) \cdot \omega_{TE}\left(p\right)\right) \, \text{d} p \leq TE \leq \int_{\mathcal{P}} \left(\overline{\Delta}_{2}\left(p\right) \cdot \omega_{TE}\left(p\right)\right) \, \text{d} p.
		\end{equation*}
		Moreover, these bounds are sharp.
	\end{proposition}

	Under Assumptions \ref{RA}-\ref{MONS}, Table \ref{integral} shows the most relevant treatment effect parameters that are partially identified using Proposition \ref{thmTE}. The weights in Table \ref{weights} are derived in Appendix \ref{PROOFweights} and are identified if the support of the propensity score is the full unit interval, i.e., $\mathcal{P} = \left[0, 1\right]$.

	\begin{table}[!htbp]
		\centering
		\caption{{Treatment Effects as Weighted Integrals of the $MTE^{OO}$}} \label{integral}
		\begin{tabular}{l}
			\hline
			\hline

			\\

			$ATE^{OO} = \mathbb{E}\left[Y_{1}^{*} - Y_{0}^{*} \left\vert S_{0} = 1, S_{1} = 1 \right.\right] = \int_{0}^{1} MTE^{OO}\left(p\right) \cdot \omega_{ATE}\left(p\right) \, \text{d} p$ \\

			\\

			\\

			$ATT^{OO} = \mathbb{E}\left[Y_{1}^{*} - Y_{0}^{*} \left\vert D = 1, S_{0} = 1, S_{1} = 1 \right.\right] = \int_{0}^{1} MTE^{OO}\left(p\right) \cdot \omega_{ATT}\left(p\right) \, \text{d} p$ \\

			\\

			\\

			$ATU^{OO} = \mathbb{E}\left[Y_{1}^{*} - Y_{0}^{*} \left\vert D = 0, S_{0} = 1, S_{1} = 1 \right.\right] = \int_{0}^{1} MTE^{OO}\left(p\right) \cdot \omega_{ATU}\left(p\right) \, \text{d} p$ \\

			\\

			\\

			$LATE^{OO}(\underline{p}, \overline{p}) = \mathbb{E}\left[Y_{1}^{*} - Y_{0}^{*} \left\vert V \in \left[\underline{p}, \overline{p}\right], S_{0} = 1, S_{1} = 1 \right.\right] = \int_{\underline{p}}^{\overline{p}} MTE^{OO}\left(p\right) \cdot \omega_{LATE}\left(p\right) \, \text{d} p$ \\

			\\

			\\

			$PRTE^{OO} = \dfrac{\mathbb{E}\left[Y_{a}^{*} - Y_{a^{\prime}}^{*} \left\vert S_{0} = 1, S_{1} = 1 \right.\right]}{\int_{0}^{1} \left(F_{P_{a^{\prime}}}\left(p\right) - F_{P_{a}}\left(p\right)\right) \cdot f_{\left. V \right\vert S_{0} = 1, S_{1} = 1}\left(p\right) \, \text{d} p } = \int_{0}^{1} MTE^{OO}\left(p\right) \cdot \omega_{PRTE}\left(p, a, a^{\prime}\right) \, \text{d} p$\\

			\\
			for two policies $a$ and $a^{\prime}$ that affect only $Z$ (See notational details in Appendix \ref{PROOFweights}.)\\

			\\

			\hline
		\end{tabular}
	\end{table}

	\begin{table}[!htb]
		\centering
		\caption{{Weights}} \label{weights}
		\begin{tabular}{l}
			\hline
			\hline

			\\

			$\omega_{ATE}\left(p\right) = \dfrac{\dfrac{\partial \mathbb{P}\left[\left. S = 1, D = 0 \right\vert P\left(Z\right) = p \right]}{\partial p}}{\bigint_{\hspace{3pt} 0}^{1} \dfrac{\partial\mathbb{P}\left[\left. S = 1, D = 0 \right\vert P\left(Z\right) = p \right]}{\partial p} \, \text{d} p}$ \\

			\\

			\\

			\\

			\\

			$\omega_{ATT}\left(p\right) = \dfrac{\left(\int_{p}^{1} f_{P\left(Z\right)}\left(u\right) \, \text{d} u \right) \cdot \dfrac{\partial\mathbb{P}\left[\left. S = 1, D = 0 \right\vert P\left(Z\right) = p \right]}{\partial p}}{\bigint_{\hspace{3pt} 0}^{1} \left(\int_{p}^{1} f_{P\left(Z\right)}\left(u\right) \, \text{d} u \right) \cdot \dfrac{\partial\mathbb{P}\left[\left. S = 1, D = 0 \right\vert P\left(Z\right) = p \right]}{\partial p} \, \text{d} p}$ \\

			\\

			\\

			\\

			\\

			$\omega_{ATU}\left(p\right) = \dfrac{\left(\int_{0}^{p} f_{P\left(Z\right)}\left(u\right) \, \text{d} u \right) \cdot \dfrac{\partial\mathbb{P}\left[\left. S = 1, D = 0 \right\vert P\left(Z\right) = p \right]}{\partial p}}{\bigint_{\hspace{3pt} 0}^{1} \left(\int_{0}^{p} f_{P\left(Z\right)}\left(u\right) \, \text{d} u \right) \cdot \dfrac{\partial\mathbb{P}\left[\left. S = 1, D = 0 \right\vert P\left(Z\right) = p \right]}{\partial p} \, \text{d} p}$ \\

			\\

			\\

			\\

			\\

			$\omega_{LATE}\left(p\right) = \dfrac{ \dfrac{\partial\mathbb{P}\left[\left. S = 1, D = 0 \right\vert P\left(Z\right) = p \right]}{\partial p}}{\bigint_{\hspace{3pt} \underline{p}}^{\overline{p}}  \dfrac{\partial\mathbb{P}\left[\left. S = 1, D = 0 \right\vert P\left(Z\right) = p \right]}{\partial p} \, \text{d} p}$ \\

			\\

			\\

			\\

			\\

			$\omega_{PRTE}\left(p, a, a^{\prime}\right) = \dfrac{\left( F_{P_{a^{\prime}}}\left(p\right) - F_{P_{a}}\left(p\right) \right) \cdot \dfrac{\partial\mathbb{P}\left[\left. S = 1, D = 0 \right\vert P\left(Z\right) = p \right]}{\partial p}}{\bigint_{\hspace{3pt} 0}^{1} \left( F_{P_{a^{\prime}}}\left(p\right) - F_{P_{a}}\left(p\right) \right) \cdot \dfrac{\partial\mathbb{P}\left[\left. S = 1, D = 0 \right\vert P\left(Z\right) = p \right]}{\partial p} \, \text{d} p}$ \\

			\\

			\hline
		\end{tabular}
	\end{table}

	Importantly, note that, differently from the weights for the case without sample selection (\citealp{heckman2005structural}, \citealp{carneiro2009estimating}, and \citealp{carneiro2011estimating}), these weights must be integrated over the distribution of the latent heterogeneity for the always-observed subpopulation instead of its unconditional distribution. 

	\section{Estimation}\label{Inf}

		This section describes the general estimation steps for the bounds proposed in Proposition~\ref{thm1}, while details on two proposed methods are provided in Appendix \ref{App_est} and \ref{empirical_app}. For brevity, we focus on the bounds identified under monotonicity of sample selection in the treatment (Assumptions \ref{RA}-\ref{MONS}), as it is the most relevant (and feasible) case empirically. Estimators for the bounds in Proposition \ref{THMnoassumption} and Proposition \ref{thmDOM} are natural extensions. Appendix \ref{mcsimulation} presents a Monte Carlo Simulation that evaluates the small sample properties of the estimator.

	To estimate the bounds in Proposition \ref{thm1}, it is useful to focus on the building blocks that are the foundation for the identification results. In particular, we need the CDFs:
	\begin{equation*}
		\tilde{Y}_d| S=1,D=d,P(Z)=p\ \sim F_{\tilde{Y}_d|S=1,D=d,P(Z)=p}(y) = \frac{\frac{\partial \mathbb P\left[Y\leq y, S=1,D=d|P(Z)=p\right]}{\partial p}}{\frac{\partial \mathbb P\left[S=1,D=d|P(Z)=p\right]}{\partial p}}
	\end{equation*}
	for any $d \in \left\lbrace 0, 1 \right\rbrace$ and
	\begin{equation*}
		\alpha\left(p\right) = - \dfrac{\frac{\partial \mathbb P\left[S=1,D=0|P(Z)=p\right]}{\partial p}}{\frac{\partial \mathbb P\left[S=1,D=1|P(Z)=p\right]}{\partial p}}.
	\end{equation*}
	Consequently, we need to estimate:
	\begin{align*}
		\Gamma_{1}\left(p, y\right) & \coloneqq \frac{\partial \mathbb P\left[\left. Y \leq y, S=1,D=1 \right\vert P(Z)=p\right]}{\partial p}, &  \pi_{1}\left(p\right) & \coloneqq \frac{\partial \mathbb P\left[\left. S=1,D=1 \right\vert P(Z)=p\right]}{\partial p}, \\
		\Gamma_{0}\left(p, y\right) & \coloneqq - \frac{\partial \mathbb P\left[\left. Y \leq y, S=1,D=0 \right\vert P(Z)=p\right]}{\partial p}, &  \pi_{0}\left(p\right) & \coloneqq - \frac{\partial \mathbb P\left[\left. S=1,D=0 \right\vert P(Z)=p\right]}{\partial p}.
	\end{align*}
Furthermore, the estimation of the propensity score, $P(Z)$, is necessary to obtain the moments of the conditional distribution of the observed outcome.

Each of these components can be estimated by standard approaches, and multiple procedures might be valid depending on the assumptions and model structure the researcher is willing to impose. For example, one could resort to nonparametric methods to estimate $P(Z)$, $ \pi_{0}(p)$, $\pi_{1}(p)$, $\Gamma_{0}(p,y)$, and $\Gamma_{1}(p,y)$, avoiding functional form choices as discussed in Appendix \ref{App_est}. While appealing, nonparametric estimation can be quite challenging in practice especially when covariates are added, as illustrated in the empirical application. Alternatively, we could estimate the parameters based on parametric functions for $P(Z)$,  $\mathbb{P}\left[\left. S = 1, D = d \right\vert P\left(Z\right) = p \right]$, and a partition of the outcome's support  $\gamma_{d}(p,k)=\mathbb{P}\left[\left. y_{k-1} \leq Y < y_{k}, S = 1, D = d \right\vert P\left(Z\right) = p \right]$ for $d=\{0,1\}$, $p \in \left[0, 1 \right]$, $k \in \left\lbrace 1,\ldots,K \right\rbrace$.  The choice of estimator should be guided by the nature of the problem being studied and the data available to the researcher.

With estimates $\hat{P}(Z)$, $ \hat{\pi}_{0}(p)$, $\hat{\pi}_{1}(p)$, $\hat{\Gamma}_{d}(p,y)$, and $\hat{\gamma}_{d}(p,y)$ at hand, we can estimate $\alpha\left(p\right)$, by its sample analog $\hat{\alpha}\left(p\right) \coloneqq \dfrac{\hat{\pi}_{0}\left(p\right)}{\hat{\pi}_{1}\left(p\right)}.$ Finally, the estimated bounds $LB_{2}(p)$, $UB_{2}(p)$, can be obtained as
	\begin{align}
	\widehat{LB}_{2}(p) & \coloneqq \label{estLB} \sum_{k = 2}^{K_{N}} \overline{y}_{k} \cdot \mathbbm{1}\left\lbrace \hat{F}_{1}\left(p, y_{k}\right) \leq \hat{\alpha}\left(p\right) \right\rbrace \cdot \dfrac{\hat{f}_{1}(p, k)}{\hat{\alpha}\left(p\right)} \\
	\widehat{UB}_{2}(p) & \coloneqq \label{estUB} \sum_{k = 2}^{K_{N}} \overline{y}_{k} \cdot \mathbbm{1}\left\lbrace 1 - \hat{F}_{1}\left(p, y_{k}\right) < \hat{\alpha}\left(p\right) \right\rbrace \cdot \dfrac{\hat{f}_{1}(p, k)}{\hat{\alpha}\left(p\right)},
	\end{align}
	where $\hat{f}_{1}(p,k) \coloneqq \frac{\hat{\gamma}_{1}(p,k)}{\hat{\pi}_{1}(p)}$, $\hat{F}_{1}\left(p, y_{k}\right) \coloneqq \sum_{j = 2}^{k} \hat{f}_{1}(p, j)$ and $\overline{y}_{k}$ is the center point of each bin $[y_{k-1}, y_{k}]$  for any $k \in \left\lbrace 2,\ldots,K_{N} \right\rbrace$. Moreover, we can estimate $\mathbb E\left[\tilde{Y}_0|S=1,D=0,P(Z)=p\right]$ in Proposition \ref{thm1} using $
	\hat{\Xi}_{OO, 0}(p) \coloneqq \sum_{k = 2}^{K_{N}} \overline{y}_{k} \cdot \hat{f}_{0}(p, k)$, where $\hat{f}_{0}(p,k) \coloneqq \frac{\hat{\gamma}_{0}(p,k)}{\hat{\pi}_{0}(p)}$. Naturally, the estimated $MTE^{OO}$ bounds can then be obtained by  $\hat{\underline{\Delta}}_{2}\left(p\right) \coloneqq  \widehat{LB}_{2}(p)-\hat{\Xi}_{OO, 0}(p)$ and $\hat{\overline{\Delta}}_{2}\left(p\right) \coloneqq \widehat{UB}_{2}(p)-\hat{\Xi}_{OO, 0}(p)$.

	We summarize the estimation procedure in the following steps:
	\setcounter{bean}{0}
	\begin{center}
		\begin{list}
			{\textsc{Step} \arabic{bean}.}{\usecounter{bean}}
			\item Estimate  $\mathbbm{P}(D=1|Z=z)$ and obtain  $\hat{P}_{i}=\hat{\mathbbm{P}}(Z_{i})$ for all observations.
			\item Estimate $\hat{\gamma}_{0}(p, y_{k})$, $\hat{\gamma}_{1}(p, y_{k})$,  $\hat{\pi}_{0}(p)$ and $\hat{\pi}_{1}(p)$ for the value $p$ of interest.
			\item Estimate $\hat{\alpha}(p)$  for $p$.
			\item Implement $\widehat{LB}_{2}(p)$,  $\widehat{UB}_{2}(p)$ and $\hat{\Xi}_{OO, 0}(p)$.
			\item Calculate the bounds for $MTE^{OO}(p)$ using $\hat{\underline{\Delta}}_{2}\left(p\right)$ and $\hat{\overline{\Delta}}_{2}\left(p\right)$.
		\end{list}
	\end{center}

	\section{Empirical Illustration: Managed Care and Ambulatory Expenditures}\label{empirical}

	To illustrate the empirical usefulness of our partial identification strategy in a concrete application, we analyze the impact of insurance plan choice on ambulatory expenditures using the data set made available by \cite{Deb2006} through the Journal of Applied Econometrics' Data Archive.\footnote{The Journal of Applied Econometrics' Data Archive can be accessed at http://qed.econ.queensu.ca/jae/.} We simplify their analyzes in two important dimensions. To enforce a binary treatment, we follow \cite{Papadoulos2012} and combine Health Maintenance Organizations (HMO) and Preferred Provider Organization (PPO) in one treatment category (managed care) while the control group contains all individuals who choose fee-for-service (FFS) plans. Second, while \cite{Deb2006} analyze ambulatory and hospital expenditures, we focus only on the former since the large share of zero hospital expenditures \citep[Table 1]{Deb2006} imply that bounds around the $MTE^{OO}$ would be very wide as explained in Appendix \ref{Sim}.

	In this application, the treatment variable $D$ is equal to one if the person chooses a managed care plan and equal to zero if the person chooses an FFS plan. $Y_{0}^{*}$ and $Y_{1}^{*}$ represent the potential ambulatorial expenditures in dollars, that is only observed if the person seeks care. $S_{0}$ and $S_{1}$ represent the potential indicators for seeking care, i.e., for spending a positive amount of money on ambulatory services. The latent heterogeneity $V$ in our treatment choice model (Equation \eqref{seq1}) can be interpreted as the individual's relative cost of choosing a managed care plan over FFS plan. Importantly, our treatment effect of interest $\left(MTE^{OO}\left(p\right) \coloneqq \mathbb{E}\left[Y_{1}^{*} - Y_{0}^{*} \left\vert V = p, S_{0} = 1, S_{1} = 1 \right.\right]\right)$ captures the intensive margin of the impact of managed care plans on ambulatory expenditures, that is, the increased intensity in use of services by those individuals that would seek ambulatory care in both insurance scenarios.

	We use data from the 1996-2001 waves of the Medical Expenditure Panel Survey (MEPS) made available by \cite{Deb2006}. The sample is restricted to employed individuals who bought a private insurance plan and whose age is between 21 and 64 years. In this data set, we observe 24 covariates.\footnote{The covariate variables are family size, age, squared age, years of schooling, income, female indicator, interaction term between age and female, African-American indicator, Hispanic indicator, marriage indicator, three geographic region dummies, metropolitan area indicator, three subjective health dummies, physical limitation indicator, number of chronic conditions, injury indicator and four year dummies. The descriptive statistics of all variables can be found in Table 1 in \cite{Deb2006}.} Moreover, our instruments are spouse's age $\left(Z_{1}\right)$ and spouse's insurance plan type in the previous year $\left(Z_{2}\right)$.\footnote{Differently from \cite{Deb2006}, we also combine the spouse’s insurance 	plan choice in a binary variable that is equal to one if the spouse chose a managed care plan and equal to zero if the spouse chose a FFS plan.} The discussion about instruments' validity follows \cite{Deb2006}. Regarding instrument relevance, the argument is that insurance plans cover an entire family, so spouse's age and lagged choice of the spouse's insurance plan should be a determinant of plan choice. The instruments' independence relies on the assertion that \textit{conditional on own age and other individual characteristics}, spouse's age should not affect personal medical expenditures directly. Moreover, since lagged spouse's choice was pre-determined, it should not impact personal medical expenditures directly either. Therefore, conditioning on covariates is crucial for the instrument's credibility and can be more clearly dealt by specifying parametric functions for the probabilities underlying the DGP.

	We calculate bounds for $MTE^{OO}\left(p, x\right)$ based on parametric estimates for the functions $\mathbb{P}\left[\left. S = 1, D = d \right\vert X = x, P\left(Z\right) = p \right]$, and $\mathbb{P}\left[\left. y_{k-1} \leq Y < y_{k}, S = 1, D = d \right\vert X = x, P\left(Z\right) = p \right]$ for $d=\{0,1\}$, $p \in \left[0, 1 \right]$, $k \in \left\lbrace 1,\ldots,K \right\rbrace$ and covariate values $x$. These probabilities are modeled as logit functions that depend on a linear index of the covariates and a quadratic function of the propensity score, using 20 grid points for the outcome variable. The propensity score is also estimated using a logit model that depends linearly on covariate and instrumental variables. To enforce the common support assumption, we trim the top and bottom 1\% of the overlapping estimated propensity score distribution. After estimating those probabilities, we estimate $\alpha\left(p, x\right)$, $\beta\left(p, x\right)$ and the bounds around $MTE^{OO}\left(p, x\right)$ for each covariate value $x$. Then, following \cite{Deb2006}, we assume that the covariates are independent of $(S_{0}, S_{1},V)$ and average the bounds for $MTE^{OO}\left(p, x\right)$ across the sample using observed covariates values. By doing so, we recover the bounds for the unconditional $MTE^{OO}(p)$.\footnote{By averaging with respect to the observed density of the covariates, we compute bounds around a summary measure of the conditional $MTE$ for the always-observed subgroup: $SCMTE^{OO}\left(p\right) \coloneqq \int \mathbb{E}\left[\left. Y_{1}^{*} - Y_{0}^{*} \right\vert V = p, S_{0} = 1, S_{1} = 1, X = x^{\prime} \right] \, \text{d}F_{X}\left(x^{\prime}\right)$. If $X \indep \left(V = p, S_{0} = 1, S_{1} = 1\right)$ holds, then $SCMTE^{OO}\left(p\right) = MTE^{OO}\left(p\right)$, implying that the summary bounds are valid for the unconditional $MTE$ function for the always-observed subgroup. Importantly, \cite{Deb2006} assumed that the covariates are fully exogenous, implying that $X \indep \left(V = p, S_{0} = 1, S_{1} = 1\right)$ holds. \cite{carneiro2009estimating} assume a similar exogeneity assumption. Alternatively, we can analyze the conditional $MTE^{OO}$ function for pre-specified values of the covariates. These results are available upon request.} For details on this parametric approach, see Appendix \ref{empirical_app}.

	We estimate bounds around the $MTE^{OO}$ function under three sets of assumptions: (i) no restrictions on the sample selection mechanism (Assumptions \ref{RA}-\ref{Dist}), (ii) ``monotonicity of sample selection in the treatment'' (Assumptions \ref{RA}-\ref{MONS}), and (iii) ``monotonicity of sample selection in the treatment'' and ``stochastic dominance'' (Assumptions \ref{RA}-\ref{DOM}). We are interested in understanding how the different sets of assumptions impact the identified set for $MTE^{OO}$ and interpreting the heterogeneity captured by the MTE.

\begin{figure}[htbp]
		\begin{center}
			\subfigure[Within Observed-when-Treated Group and under Assumptions \ref{RA}-\ref{Dist}: $\alpha\left(p, \upsilon^\ell\right)$]{
				\includegraphics[width = 0.31 \textwidth]{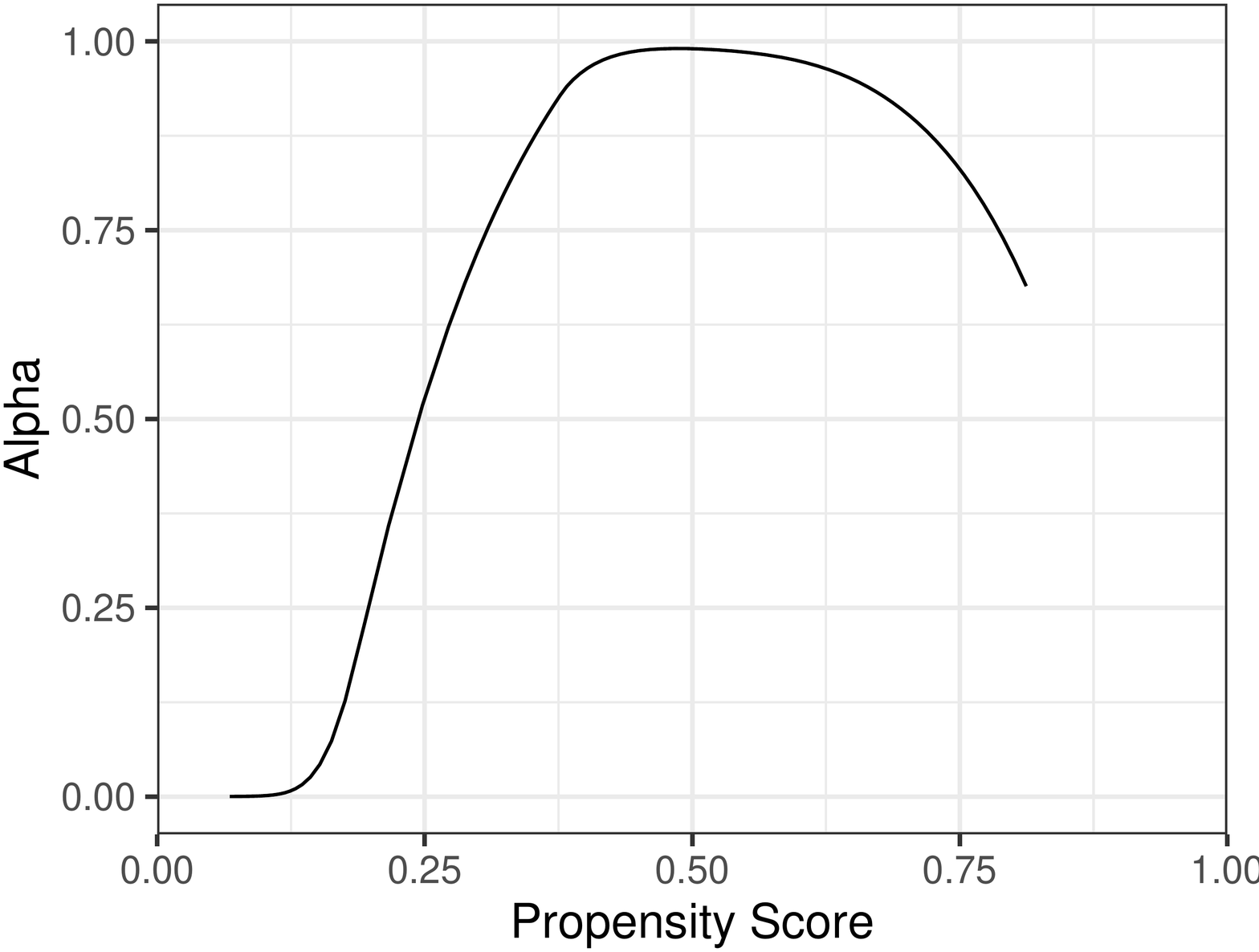}
				\label{fig_alpha_no_assumption_parametric}
			}
			\subfigure[Within Observed-when-Untreated Group and under Assumptions \ref{RA}-\ref{Dist}: $\beta\left(p, \upsilon^\ell\right)$]{
				\includegraphics[width = 0.31 \textwidth]{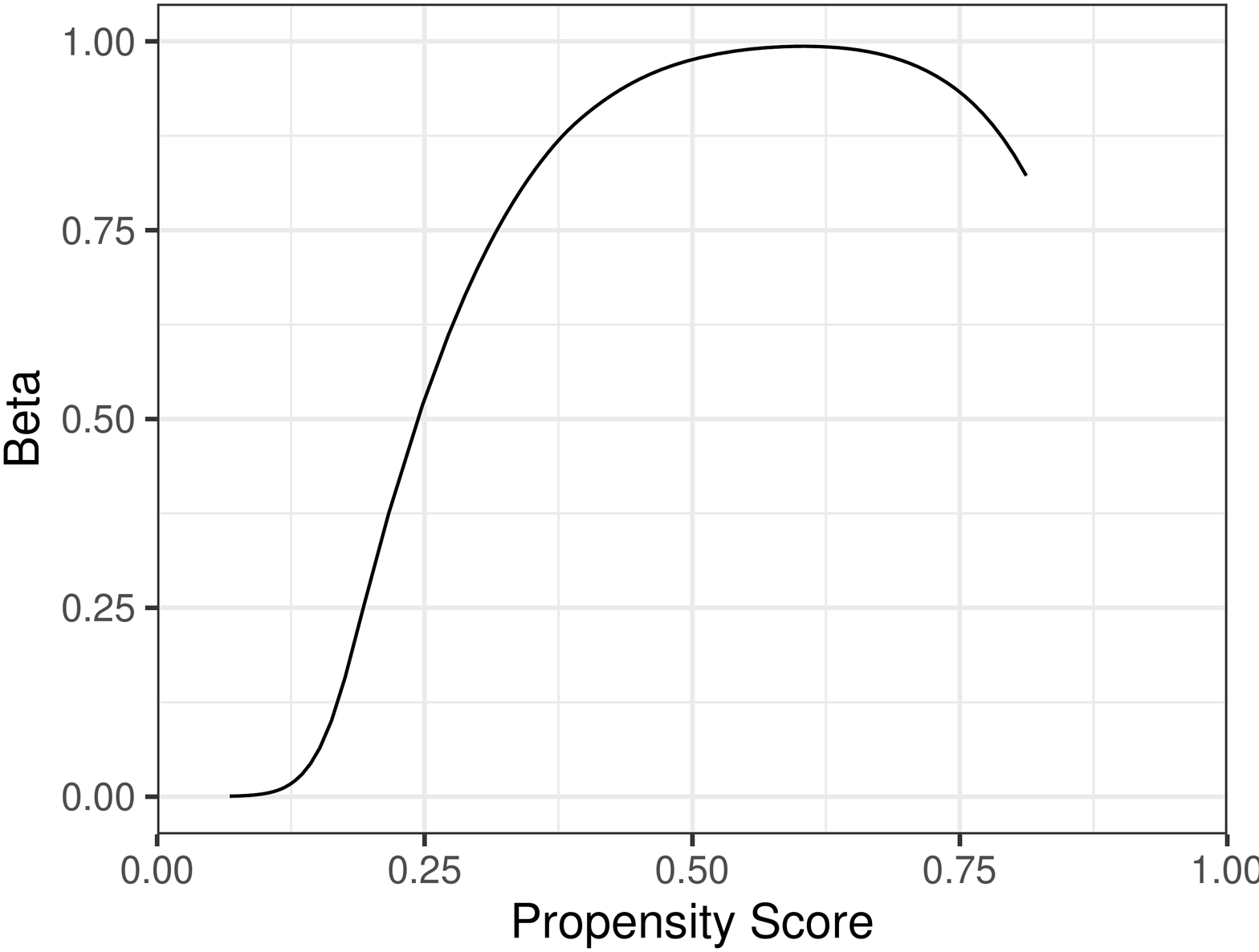}
				\label{fig_beta_no_assumption_parametric}
			}
			\subfigure[Within Observed-when-Treated Group under Assumptions \ref{RA}-\ref{MONS}: $\alpha\left(p\right)$]{
				\includegraphics[width = 0.31 \textwidth]{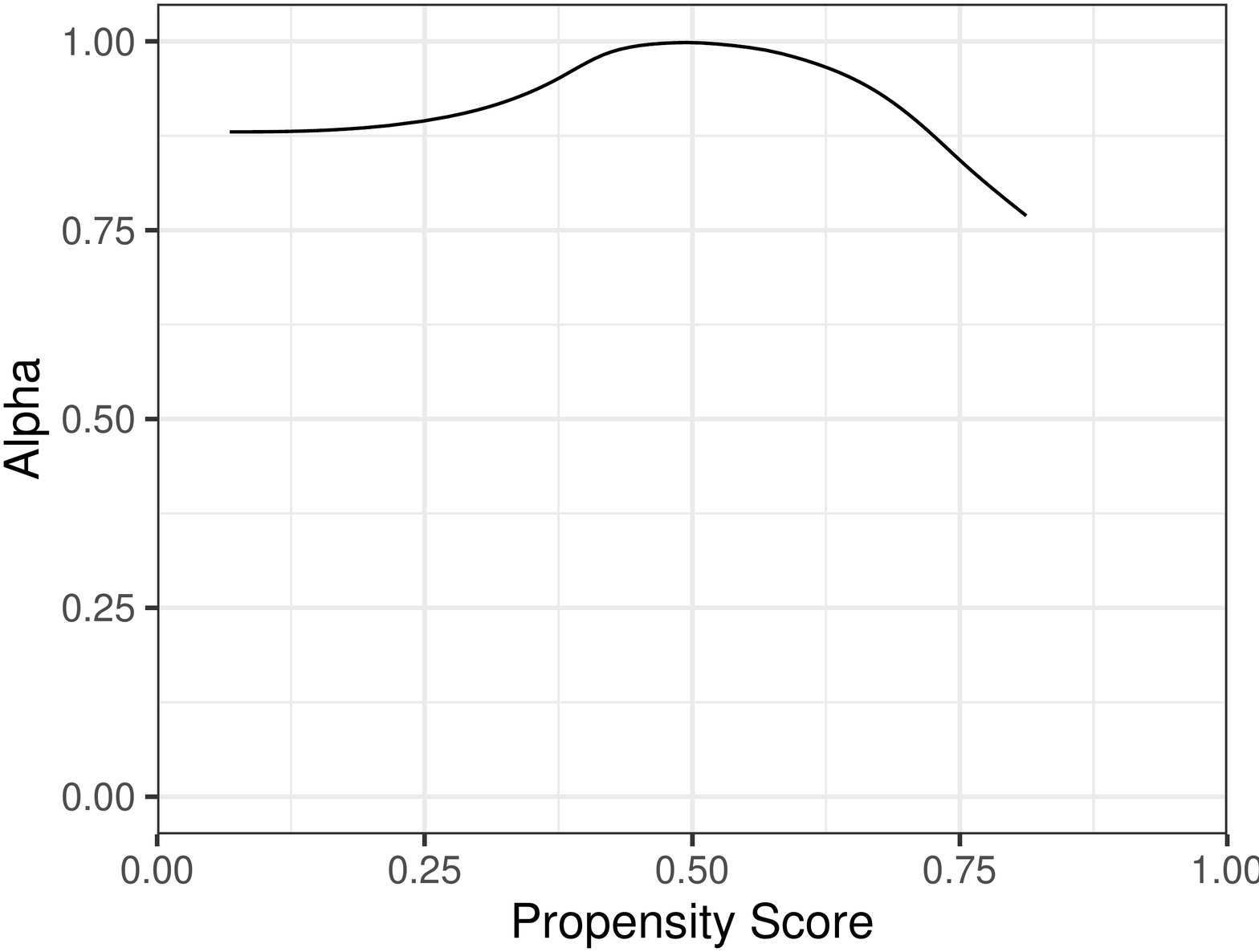}
				\label{fig_alpha_mon_parametric}
			}
		\end{center}
		\footnotesize{Notes: Solid lines are the parametric estimated proportions of the always-observed group. }
		\caption[]{Unconditional Proportion of the Always-Observed Group}
		\label{fig_alpha_parametric}
	\end{figure}

\begin{figure}[htbp]
		\begin{center}
			\subfigure[Under Assumptions \ref{RA}-\ref{Dist}]{
				\includegraphics[width = 0.31 \textwidth]{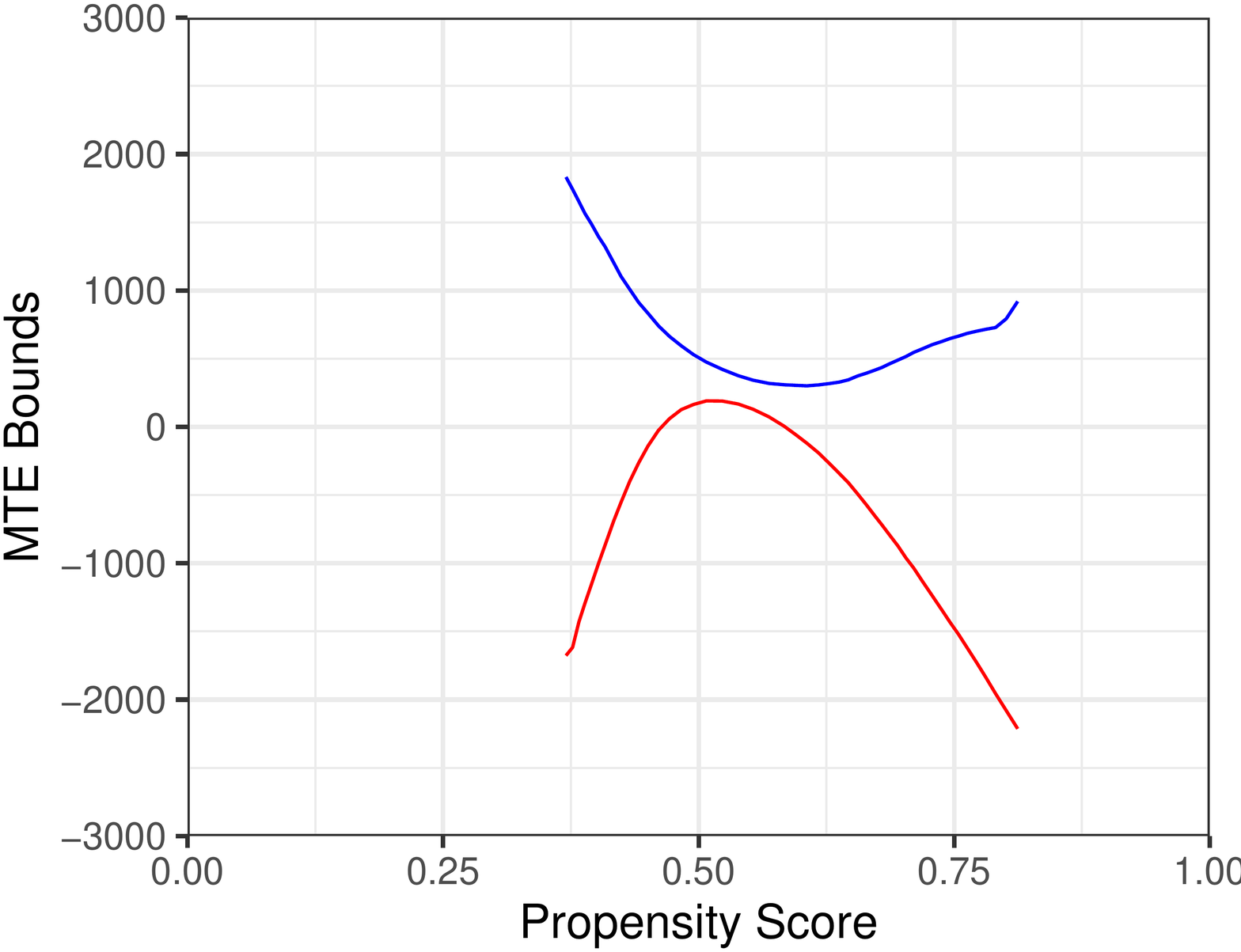}
				\label{fig_MTE_no_assumption_parametric}
			}
			\subfigure[Under Assumptions \ref{RA}-\ref{MONS}]{
				\includegraphics[width = 0.31 \textwidth]{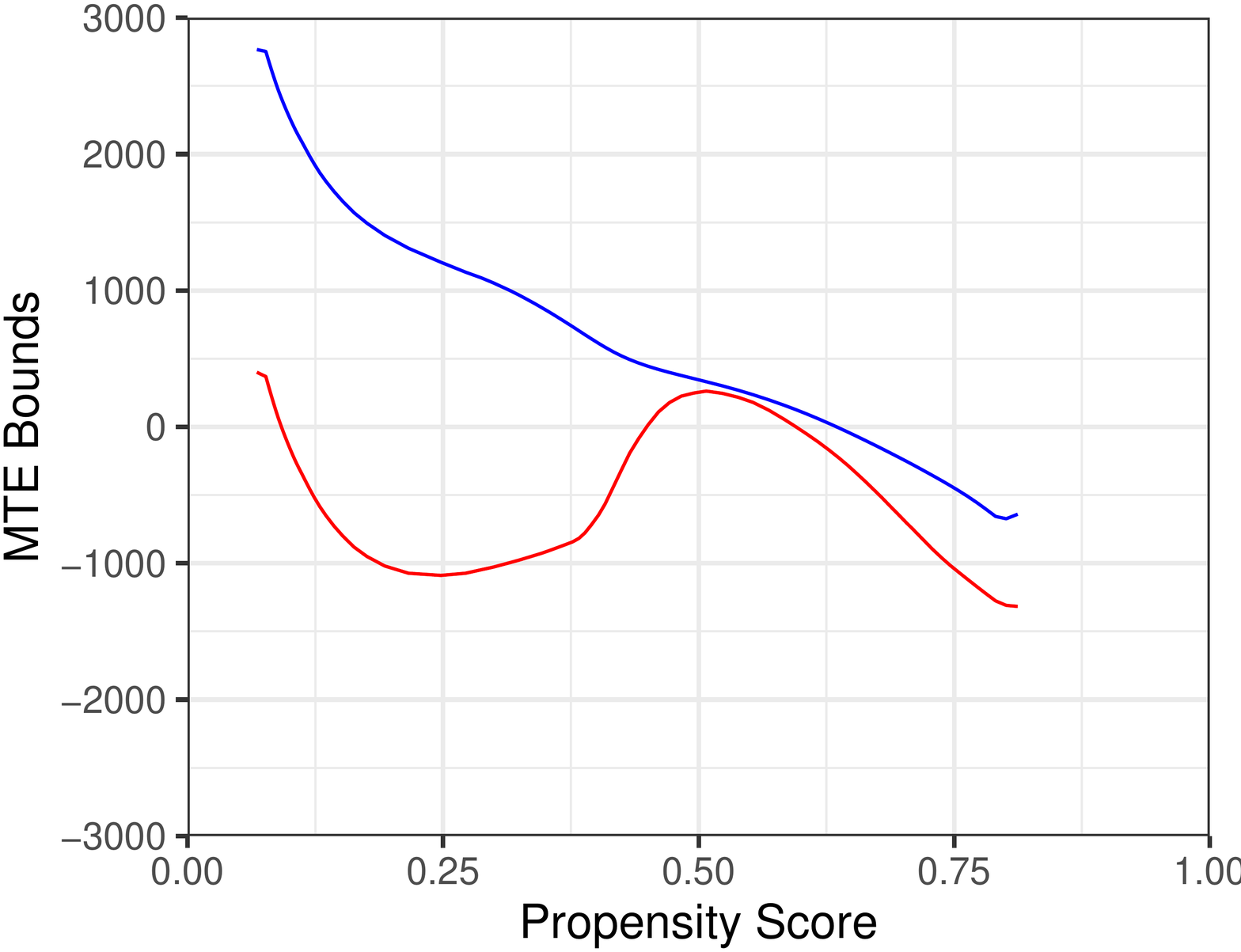}
				\label{fig_MTE_mon_parametric}
			}
			\subfigure[Under Assumptions \ref{RA}-\ref{DOM}]{
				\includegraphics[width = 0.31 \textwidth]{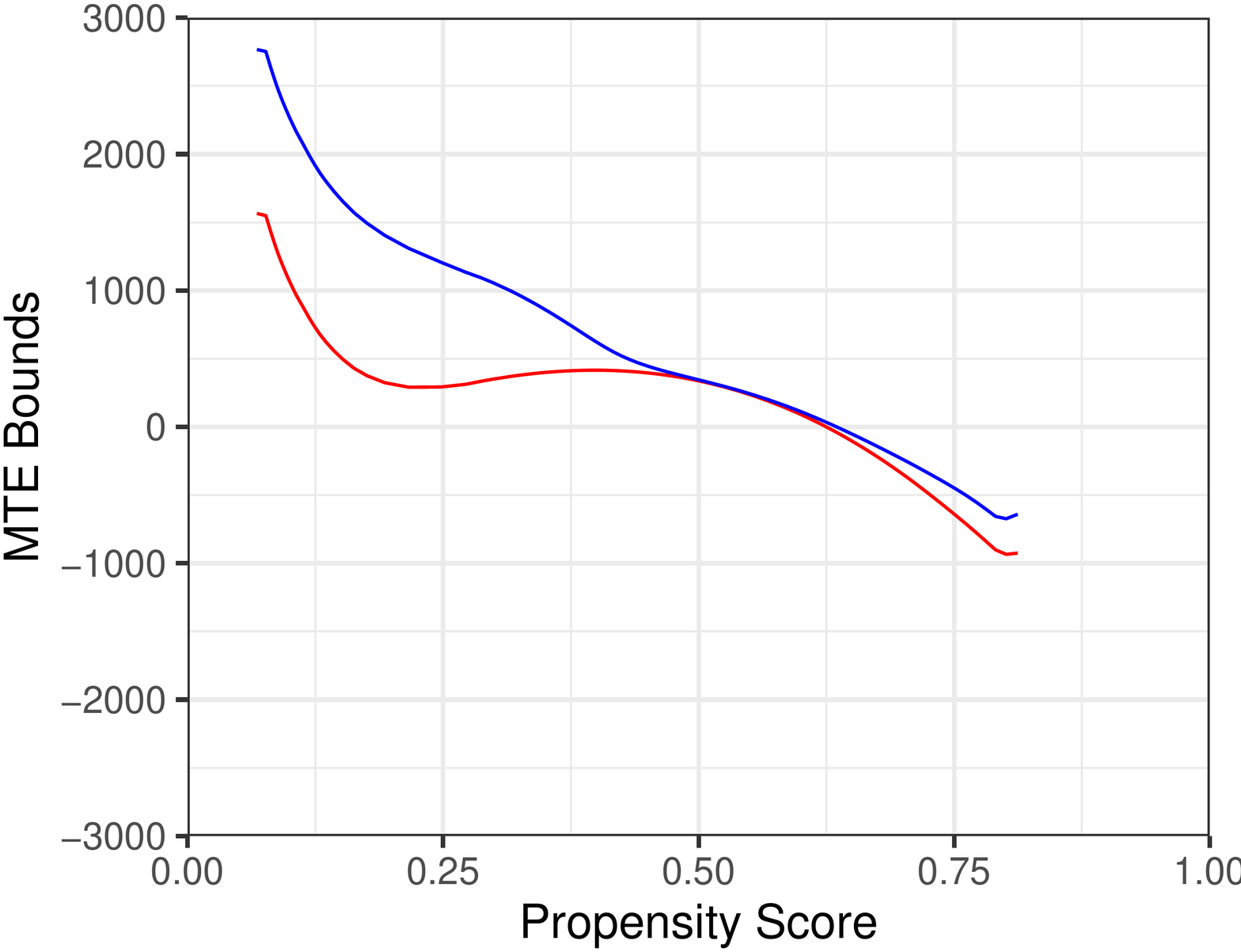}
				\label{fig_MTE_mon_dom_parametric}
			}
		\end{center}
		\footnotesize{Notes: The solid red (blue) line is the estimated lower (upper) bound of the $MTE^{OO}$.}
		\caption[]{Unconditional Bounds around the $MTE^{OO}$}
		\label{fig_MTE_parametric}
	\end{figure}

	First, we analyze the $MTE^{OO}$ bounds based only on Assumptions \ref{RA}-\ref{Dist} (Subsection \ref{noassumption}). Subfigures \ref{fig_alpha_no_assumption_parametric} and \ref{fig_beta_no_assumption_parametric} show the estimated proportion of the always-observed subpopulation within the observed-when-treated and observed-when-untreated groups, that is, $\alpha\left(p, \upsilon^\ell\right)$ and $\beta\left(p, \upsilon^\ell\right)$ in Proposition \ref{THMnoassumption}. Importantly, in some regions of the support those proportions are equal to one or zero, implying that $MTE^{OO}$ is point-identified or not-identified, respectively. Subfigure \ref{fig_MTE_no_assumption_parametric} shows the estimated bounds. For most of the propensity score's support, the bounds without restrictions on the sample selection mechanism are very wide or identification is lost. The estimated bounds do not rule out the possibility of homogeneous treatment effects, i.e., we can still place a constant function inside the bounds in Subfigure \ref{fig_MTE_no_assumption_parametric}.

	In order to tighten those bounds, we consider restrictions on the sample selection mechanism. The ``monotonicity of sample selection in the treatment'' condition (Assumption \ref{MONS}) imposes that agents who would spend a positive amount of money on ambulatory services if allocated to a FFS plan would also have positive ambulatory expenditures if allocated to a managed care plan. This assumption's direction is in accordance with the results described by \cite{Deb2006}, who found that individuals enrolled in HMOs and PPOs are more likely to seek care than FFS enrollees. Note that, under Assumption \ref{MONS}, $\beta(p)$ is always equal to 1. Subfigure \ref{fig_alpha_mon_parametric} shows that, under Assumptions \ref{RA}-\ref{MONS}, the estimated proportion of the always-observed subpopulation within the observed-when-treated group $\left(\alpha\left(p\right)\right)$ is strictly positive everywhere in this example. Consequently, the $MTE^{OO}$ function is at least partially identified for the entire support of the propensity score, as can be seen in the bounds reported Subfigure \ref{fig_MTE_mon_parametric}, illustrating the identifying power of the ``monotonicity of sample selection in the treatment'' assumption. As a result of tighter bounds, we can rule out the possibility of homogeneous treatment effects, i.e., we cannot fit a constant function inside the bounds in Subfigure \ref{fig_MTE_mon_parametric}. Interestingly, the upper bound under monotonicity is decreasing. Hence, if the $MTE^{OO}$ function followed the pattern from the upper bound, our results would suggest that the agents who are more likely to enroll in a managed care plan are the ones who incur larger additional ambulatory expenses due to their choice of insurance coverage. This interpretation is compatible with individuals taking into account their potential expenditures when selecting their insurance plans, at least among those for which those expenditures will be positive regardless of their plan (always-observed), and provides extra support for the selectivity result found by \cite{Deb2006}.

	In order to further tighten the bounds around the $MTE^{OO}$ function, we impose the stochastic dominance assumption (Assumption \ref{DOM}). To interpret this assumption, recall that the ``always-observed'' ($OO$) individuals are those for whom ambulatory expenditures would be positive regardless of their insurance plans, while the ``observed-only-when-treated''($NO$) population encompasses people who would incur expenses only if enrolled in managed care plans. Assumption \ref{DOM} compares the (counterfactual) expenditures that would take place if all employees were enrolled in a managed care plan ($Y^{*}_{1}$) between those two groups. Formally, it says that for any particular level of expenditures, say 1000\$, the share of individuals that spend less than 1000\$ will be larger for the $NO$ group compared to the $OO$ type. Alternatively, it implies that the average expenditures among the lowest 25\% (or any particular quantile) of spenders, will be smaller for the $NO$ than for the $OO$ subgroup. Intuitively, if everyone had a managed care plan, typical patients who would go to ambulatories only if insured by a managed care plan spend less in services than those that would go regardless of their plan. In Appendix~\ref{apx:stoch_dom}, we provide a simple theoretical framework in which this assumption holds.\footnote{As pointed out by a referee, this assumption is hard to interpret and motivate empirically. However, in a layered policy analysis \citep{Manski2011}, we offer a menu of estimates based on different assumptions, that may or may not be plausible according to each researcher's own expertise and beliefs \citep{Tamer2010}.} Importantly, Assumption~\ref{DOM} has no impact on the information about $\alpha(p)$. Consequently, it only impacts the results by increasing the lower bound for the $MTE^{OO}$, which is much tighter as can be seen in Subfigure \ref{fig_MTE_mon_dom_parametric}, illustrating the identifying power of the stochastic dominance assumption. If we consider Assumption \ref{DOM} plausible, we find that the lower bound is positive for most values of the propensity score. This finding reinforces the results in \cite{Deb2006}, who obtained a positive effect of PPO choice on ambulatory expenditures and a zero effect of HMO choice after controlling for selection. Furthermore, our results also suggest that the choice of managed care plan reduces ambulatory expenditures for individuals who face high latent costs.


	\section{Extension: Identification with discrete instruments}\label{extensions}

 	This section extends Proposition \ref{thm1} to the case with multi-valued discrete instruments, sharply bounding many LATE parameters for the always-observed subpopulation. We focus on the case under the monotonicity restriction (Assumption \ref{RA}-\ref{MONS}). Similar results hold under our other identifying sets of assumptions.

	In many applications, the only instruments available are discrete, e.g., treatment eligibility, number of children in the household and quarter of birth. In this section, we provide sharp identification results when the instrument is multi-valued discrete, implying the support of the propensity score is finite. The results here can be seen as an extension of \cite{chen2015bounds}, who provide an outer set for the LATE parameter when the instrument is binary.
	\begin{assumption}\label{Discrete}
		The instrument $Z$ is discrete with support $\{z_1, z_2, \ldots, z_K\}$ and the propensity score $p_\ell \equiv \mathbb P\left[D=1|Z=z_\ell\right]$ satisfies $0 < p_{1} < p_{2} < \ldots < p_{K} < 1.$
	\end{assumption}

	Assumption \ref{Discrete} requires that one can rank the probabilities of receiving treatment for the points of $P(Z)$ that are available, allowing the researcher to partition the $[0,1]$ interval into regions $[p_\ell - p_{\ell-1}]$ for $\ell=2,...,K$.  The researcher can only identify an average of the MTE within each region, i.e., a LATE. Naturally, if the instrument has more points of positive mass (providing finer partitions of the probabilities), one could obtain averages of the MTE for more specific ranges of the unobservable characteristic $V$.\footnote{If for some values of $Z$, the probabilities are the same, $p_\ell=p_{\ell-1}$, we cannot refine the partition of the unit interval describing the probabilities and, hence, cannot improve on the detail level of the MTE identified.}

	The identification argument is similar to the one presented for the continuous instrument case in Subsection \ref{monotonicity}. Under Assumption \ref{RA}, we have $p_\ell=\mathbb P\left[V\leq P(z_\ell)\right]$ and $\mathbb P\left[P(z_{\ell-1}) < V\leq P(z_\ell)\right]=p_{\ell}-p_{\ell-1}$. If Assumptions~\ref{RA} and \ref{Dist} hold, then $P(z_\ell)=p_\ell$. To ease the exposition, we use the shorthand $P \coloneqq P(Z)$.

	We have  $\mathbb P\left[Y\in A,S=1,D=1|P=p_\ell\right]=\mathbb P\left[Y_1^* \in A, S_1=1,V \leq p_\ell\right]$. Therefore,
	\begin{eqnarray*}
		\mathbb P\left[Y_1^* \in A, S_1=1,p_{\ell -1} < V \leq p_\ell\right]&=&\mathbb P\left[Y\in A,S=1,D=1|P=p_\ell\right]\\
		&&\qquad - \mathbb P\left[Y\in A,S=1,D=1|P=p_{\ell-1}\right],
	\end{eqnarray*}
	which implies that
	\begin{eqnarray*}
		&&\mathbb P\left[Y_1^* \in A, S_1=1|p_{\ell -1} < V \leq p_\ell\right]=\nonumber\\
		&&\qquad \qquad \frac{\mathbb P\left[Y\in A,S=1,D=1|P=p_\ell\right] - \mathbb P\left[Y\in A,S=1,D=1|P=p_{\ell-1}\right]}{p_\ell - p_{\ell-1}}.\label{eq:discys1}
	\end{eqnarray*}
	Similarly, we have
	\begin{eqnarray*}
		&&\mathbb P\left[Y_0^* \in A, S_0=1|p_{\ell -1} < V \leq p_\ell\right]=\nonumber\\
		&&\qquad \qquad - \frac{\mathbb P\left[Y\in A,S=1,D=0|P=p_\ell\right] - \mathbb P\left[Y\in A,S=1,D=0|P=p_{\ell-1}\right]}{p_\ell - p_{\ell-1}}.\label{eq:discys0}
	\end{eqnarray*}
	Thus for $A=\mathcal Y$, we can write
	\begin{eqnarray*}
		\mathbb P\left[S_1=1|p_{\ell -1} < V \leq p_\ell\right] &=& \frac{\mathbb P\left[S=1,D=1|P=p_\ell\right] - \mathbb P\left[S=1,D=1|P=p_{\ell-1}\right]}{p_\ell - p_{\ell-1}},\label{eq:discs1}\\
		\mathbb P\left[S_0=1|p_{\ell -1} < V \leq p_\ell\right] &=&- \frac{\mathbb P\left[S=1,D=0|P=p_\ell\right] - \mathbb P\left[S=1,D=0|P=p_{\ell-1}\right]}{p_\ell - p_{\ell-1}}. \label{eq:discs0}
	\end{eqnarray*}
	We know that $\mathbb P\left[Y_d^* \in A|S_d=1,p_{\ell -1} < V \leq p_\ell\right]=\frac{\mathbb P\left[Y_d^* \in A, S_d=1|p_{\ell -1} < V \leq p_\ell\right]}{\mathbb P\left[S_d=1|p_{\ell -1} < V \leq p_\ell\right]}$ for $d\in \{0,1\}$. Under Assumption \ref{MONS}, we identify $\mathbb P\left[S_{0}=1,S_{1}=1|p_{\ell -1} < V \leq p_\ell\right]=\mathbb P\left[S_0=1|p_{\ell -1} < V \leq p_\ell\right]$.

	To implement the trimming in this setting, we define the discrete case analog of $\alpha(p)$, denoted by $\tilde{\alpha}(p_{\ell-1},p_\ell)$,
	$$\tilde{\alpha}(p_{\ell-1},p_{\ell}) \coloneqq \frac{\mathbb P\left[S_{0}=1,S_{1}=1|p_{\ell -1} < V \leq p_{\ell}\right]}{\mathbb P\left[S_{1}=1|p_{\ell -1} < V \leq p_{\ell}\right]}=\frac{\mathbb E[S(1-D)|P=p_{\ell -1}] - \mathbb E[S(1-D)|P=p_{\ell}]}{\mathbb E[SD|P=p_{\ell}] - \mathbb E[SD|P=p_{\ell-1}]}.$$
	To find bounds around $\mathbb E[Y_{1}^{*}-Y_{0}^{*}|S_{0}=1,S_{1}=1,p_{\ell-1}<V\leq p_{\ell}]$, we follow the steps in Subsection \ref{monotonicity} to derive the following proposition:
	\begin{proposition}\label{thm2}
		Under Assumptions \ref{RA}, \ref{positive}-\ref{MONS} and \ref{Discrete}, the $LATE$ parameters are partially identified for the always-observed, i.e., $$\underline{\Delta}_{LATE}\left(\ell\right) \leq LATE^{OO}\left(\ell\right) \leq \overline{\Delta}_{LATE}\left(\ell\right)$$ for any $\ell \in \left\lbrace 2, \ldots, K \right\rbrace$, where $LATE^{OO}\colon \left\lbrace 2, \ldots, K \right\rbrace \rightarrow \mathbb{R}$, $\underline{\Delta}_{LATE} \colon \left\lbrace 2, \ldots, K \right\rbrace \rightarrow \mathbb{R}$ and $\overline{\Delta}_{LATE} \colon \left\lbrace 2, \ldots, K \right\rbrace \rightarrow \mathbb{R}$ are given by
		\begin{align*}
			LATE^{OO}\left(\ell\right) & \coloneqq E[Y^{*}_{1}-Y^{*}_{0}|S_{0}=1,S_{1}=1, p_{\ell-1}<V\leq p_\ell], \\
			\underline{\Delta}_{LATE}\left(\ell\right) & \coloneqq \mathbb E\left[Y^{*}_{1}|S_1=1,p_{\ell-1}<V\leq p_\ell, Y^{*}_{1}\leq F^{-1}_{Y^{*}_{1}|S_1=1,p_{\ell-1}<V\leq p_\ell}\left(\tilde{\alpha}(p_{\ell-1},p_\ell)\right)\right] \\
			& \hspace{20pt} - \mathbb E[Y^{*}_{0}|S_{0}=1,S_{1}=1,p_{\ell-1}<V\leq p_\ell], \text{ and } \\
			\overline{\Delta}_{LATE}\left(\ell\right) & \coloneqq \mathbb E\left[Y^{*}_{1}|S_1=1,p_{\ell-1}<V\leq p_\ell, Y^{*}_{1} > F^{-1}_{Y^{*}_{1}|S_1=1,p_{\ell-1}<V\leq p_\ell}\left(1-\tilde{\alpha}(p_{\ell-1},p_\ell)\right)\right] \\
			& \hspace{20pt} - \mathbb E[Y^{*}_{0}|S_{0}=1,S_{1}=1,p_{\ell-1}<V\leq p_\ell].
		\end{align*}
		Moreover, these bounds are uniformly sharp.
	\end{proposition}
	As mentioned above, the quantity for which we derive bounds in Proposition \ref{thm2} is an average of the $MTE(p)$ evaluated at levels of $p$ in the interval $(p_{\ell-1}, p_{\ell}]$, i.e., we partially identify a LATE. More can be said about the MTE if additional assumptions are made. For example, if we assume that the $MTE$ is flat within each interval, then Proposition \ref{thm2} provides sharp bounds on the $MTE$ for the always-observed. This result is related to previous work in which discrete instruments are used to identify $MTE$ in the absence of sample selection. For example, \cite{Brinch2017} leverages additional functional structure for identification, while \cite{Mogstad2017} provided partial identification results for the $MTE$. An extension of their results to the current framework is an interesting topic for future research.

	\section{Conclusion}\label{Con}
	This paper derives sharp bounds for the marginal treatment effect for the always-observed individuals when there is sample selection. We achieve partial identification results under three increasingly restrictive sets of assumptions. First, we impose standard MTE assumptions without any restrictions to the sample selection mechanism. The second case, which is our main result, imposes monotonicity of the sample selection variable with respect to the treatment, considerably shrinking the identified set. Finally, we consider a strong stochastic dominance assumption which tightens the lower bound for the MTE.

	All the results rely on the insight that observed individuals in the sample are a mixture of two possible groups, the ones that would always be observed regardless of treatment status and the ones that would self-select into the sample only when (un)treated. The mixture weights can be identified, leading to a trimming procedure that partially identifies the target parameter, extending  \cite{Imai2008}, \cite{lee2009training} and \cite{chen2015bounds} results to the context of MTE. Moreover, we derive testable implications of our identifying assumptions, and provide extensions to bound LATE parameters with multi-valued discrete instruments. A feasible estimator is proposed and implemented in an empirical illustration analyzing the impacts of managed health care options on health related expenditures, following \cite{Deb2006} and highlighting the practical relevance of the results.

	\section*{Acknowledgment}

	We thank the Editor Elie Tamer, an Associate Editor, and two anonymous referees for constructive feedback that help improve the quality of the paper. We also thank Joseph Altonji, Nathan Barker, Michael Bates, Ivan Canay, Xiaohong Chen, Xuan Chen, Michael Darden, Nino Doghonadze, John Finlay, Carlos A. Flores, Thomas Fujiwara, Dalia Ghanem, John Eric Humphries, Yuichi Kitamura, Marianne Köhli, Helena Laneuville, Jaewon Lee, Giovanni Mellace, Ismael Mourifi\'e, Yusuke Narita, Pedro Sant'anna, Masayuki Sawada, Azeem Shaikh, Edward Vytlacil, Stephanie Weber, Siuyuat Wong and seminar participants at Iowa State University, University of Iowa, Yale University, UC Davis, UC-Riverside, UNC-Chapel Hill, CEME Conference for Young Econometricians 2019, IZA/CREST Conference on Labor Market Policy Evaluation, Southern Denmark University, Statistics Norway, CMStatistics 2019, the Bristol Econometrics Study Group 2019 and the 42\textsuperscript{nd} Meeting of the Brazilian Econometric Society for helpful discussions, and Seung Jin Cho for excellent research assistance.

	\singlespace

	\bibliographystyle{jpe}
	\bibliography{MTEBounds}

	\newpage


	\appendix

	\begin{center}
		\huge
		Supporting Information

		(Online Appendix)

	\end{center}

	\doublespacing

	\section{Proofs}
\subsection{Proof of Lemma \ref{BFbounds}}\label{PROOFBFbounds}
The validity of the bounds is proven in the main text. It remains to show that the bounds are uniformly sharp. Given the restrictions that Assumptions \ref{RA}-\ref{Dist} impose on the data, we need to find a joint distribution on $(\tilde{S}_0, \tilde{S}_1, \tilde{V}, Z)$ that satisfies these restrictions and achieves any value $\upsilon\colon\left[0,1\right] \rightarrow \left[0,1\right]$ such that $\upsilon\left(p\right) \in \Upsilon\left(p\right)$ for any $p \in \mathcal{P}=[\underline{p}, \overline{p}]$.

Define
\begin{eqnarray*}
                 \upsilon(p)=\left\{ \begin{array}{lcl}
		\overline{\upsilon}\left(p\right) \in \Upsilon\left(p\right)\ \text{ if }\ p\in [\underline{p}, \overline{p}]\\ \\
		\epsilon_0 \in [0,1]\ \text{ if }\ p < \underline{p} \\ \\
		\epsilon_1 \in [0,1]\ \text{ if }\ p > \overline{p}
	\end{array} \right.
\end{eqnarray*}
We need to define the joint density (mass) function of $(\tilde{S}_0, \tilde{S}_1, \tilde{V}, Z)$. To do so, we will define the density function $f_{\tilde{V}}$, define the mass function $\tilde{\pi}_{\left(\tilde{S}_0, \tilde{S}_1\right)|\tilde{V}}$ and use the density function of $Z$ --- $f_{Z}$ --- to define $f_{(\tilde{S}_0,\tilde{S}_1),\tilde{V},Z} = \tilde{\pi}_{\left(\tilde{S}_0, \tilde{S}_1\right)|\tilde{V}} \cdot f_{\tilde{V}} \cdot f_{Z}$. Note that, by construction, Assumption \ref{RA} holds. With this goal in mind, fix $\left(s_{0}, s_{1}, p, z\right) \in \left\lbrace0,1\right\rbrace^{2} \times \mathbb{R}^{2}$ arbitrarily. Define $f_{\tilde{V}}\left(p\right) = \mathbbm{1}\left\lbrace p \in \left[0,1\right] \right\rbrace$, ensuring that Assumption \ref{Dist} holds by construction. For brevity, denote the strata by \emph{OO} = always observed, \emph{NO} = observed only when treated, \emph{ON} = observed only when untreated and \emph{NN} = never observed, and the probability of the stratum $k$ conditional on $\tilde{V}=p$ by $\tilde{\pi}_{k|p}$. We now propose the following DGP:
\begin{eqnarray*}\mathbb P\left(\tilde{S}_1=s_1\vert \tilde{V}=p\right) =
	\left\{ \begin{array}{lcl}
		\frac{\partial \mathbb P\left(S=s_1,D=1\vert P(Z)=p\right)}{\partial p}\ \text{ if }\ p\in [\underline{p}, \overline{p}]\\ \\
		\frac{\mathbb P\left(S=s_1,D=1\vert P(Z)=\underline{p}\right)}{ \underline{p}}\ \text{ if }\ p < \underline{p} \\ \\
		\upsilon\left(p\right) \cdot\mathbbm{1}\{s_1=1\}+(1-\upsilon\left(p\right)) \cdot \mathbbm{1}\{s_1=0\}\ \text{ if }\ p > \overline{p}
	\end{array} \right.
\end{eqnarray*}
and
\begin{eqnarray*}\mathbb P\left(\tilde{S}_0=s_0\vert \tilde{V}=p\right) =
	\left\{ \begin{array}{lcl}
		-\frac{\partial \mathbb P\left(S=s_0,D=0\vert P(Z)=p\right)}{\partial p}\ \text{ if }\ p\in [\underline{p}, \overline{p}]\\ \\
		\upsilon\left(p\right) \cdot\mathbbm{1}\{s_0=1\}+(1-\upsilon\left(p\right)) \cdot \mathbbm{1}\{s_0=0\}\ \text{ if }\ p < \underline{p} \\ \\
		\frac{\mathbb P\left(S=s_0,D=0\vert P(Z)=\overline{p}\right)}{1-\overline{p}}\ \text{ if }\ p > \overline{p}
	\end{array} \right..
\end{eqnarray*}
Define
\begin{align*}
\tilde{\pi}_{OO|p} & = \upsilon(p) \\
\tilde{\pi}_{NO|p} & =\mathbb P\left(\tilde{S}_1=1\vert \tilde{V}=p\right) -\upsilon(p) \\
\tilde{\pi}_{ON|p} & = \mathbb P\left(\tilde{S}_0=1\vert \tilde{V}=p\right) - \upsilon(p) \\
\tilde{\pi}_{NN|p} & = 1 - \tilde{\pi}_{OO|p} - \tilde{\pi}_{NO|p} - \tilde{\pi}_{ON|p}.
\end{align*}
We are going to show that the proposed joint distribution satisfies all restrictions on the unconditional joint distribution of $(S_0,S_1)$ and the marginals of $S_0$ and $S_1$.
We define the correspondence $G$ between the unobservables $(S_0,S_1)$ and the observables $(S,D)$:
\begin{eqnarray*}
	&& G\{(0,0)\}=\{(0,0),(0,1)\},\ G\{(0,1)\}=\{(0,0),(1,1)\},\\
	&&G\{(1,0)\}=\{(1,0),(0,1)\},\ G\{(1,1)\}=\{(1,0),(1,1)\}.
\end{eqnarray*}
By \citet[Theorem 1]{henry2011}, we have that all restrictions on the unconditional joint distribution of $(S_0,S_1)$ and the marginals of $S_0$ and $S_1$ are given by: for all $A\subset \{(0,0),(0,1),(1,0),(1,1)\}$,
\begin{eqnarray*}
	\mathbb P((S,D) \in A|Z=z) &\leq& \mathbb P(G(S_0,S_1)\cap A \neq \emptyset \vert Z=z)\\
	&=& \mathbb P(G(S_0,S_1)\cap A \neq \emptyset ), \;  \forall z \in \mathcal Z,
\end{eqnarray*}
that is:\\
for singletons,
\begin{eqnarray}
\mathbb P(S=1,D=1\vert Z=z) &\leq& \pi_{NO}+\pi_{OO} \label{hg1},\\
\mathbb P(S=0,D=1\vert Z=z) &\leq& \pi_{NN}+\pi_{ON} \label{hg2},\\
\mathbb P(S=1,D=0\vert Z=z) &\leq& \pi_{OO}+\pi_{ON} \label{hg3},\\
\mathbb P(S=0,D=0\vert Z=z) &\leq& \pi_{NN}+\pi_{NO} \label{hg4};
\end{eqnarray}
for pairs,
\begin{eqnarray}
\mathbb P(S=1,D=1\vert Z=z)+\mathbb P(S=1,D=0\vert Z=z) &\leq& \pi_{OO}+\pi_{NO}+\pi_{ON} \label{hg5},\\
\mathbb P(S=1,D=1\vert Z=z)+\mathbb P(S=0,D=1\vert Z=z) &\leq& 1 \label{hg6}, \\
\mathbb P(S=1,D=1\vert Z=z)+\mathbb P(S=0,D=0\vert Z=z) &\leq& \pi_{OO}+\pi_{NN}+\pi_{NO} \label{hg7},\\
\mathbb P(S=1,D=0\vert Z=z)+\mathbb P(S=0,D=1\vert Z=z) &\leq& \pi_{NN}+\pi_{OO}+\pi_{ON} \label{hg8},\\
\mathbb P(S=1,D=0\vert Z=z)+\mathbb P(S=0,D=0\vert Z=z) &\leq& 1\label{hg9},\\
\mathbb P(S=0,D=1\vert Z=z)+\mathbb P(S=0,D=0\vert Z=z) &\leq& \pi_{NN}+\pi_{NO}+\pi_{ON} \label{hg10};
\end{eqnarray}
for triplets,
\begin{eqnarray}
\mathbb P(S=1,D=1\vert Z=z)+\mathbb P(S=1,D=0\vert Z=z)+\mathbb P(S=0,D=1\vert Z=z) &\leq& 1\label{hg11},\\
\mathbb P(S=1,D=1\vert Z=z)+\mathbb P(S=1,D=0\vert Z=z)+\mathbb P(S=0,D=0\vert Z=z) &\leq& 1\label{hg12},\\
\mathbb P(S=1,D=1\vert Z=z)+\mathbb P(S=0,D=1\vert Z=z)+\mathbb P(S=0,D=0\vert Z=z) &\leq& 1\label{hg13},\\
\mathbb P(S=1,D=0\vert Z=z)+\mathbb P(S=0,D=1\vert Z=z)+\mathbb P(S=0,D=0\vert Z=z) &\leq& 1\label{hg14}.
\end{eqnarray}
Restrictions (\ref{hg6}), (\ref{hg9}), (\ref{hg11})-(\ref{hg14}) are redundant. We only need to check that $\tilde{\pi}$ satisfies (\ref{hg1})-(\ref{hg4}), (\ref{hg5}), (\ref{hg7}), (\ref{hg8}), and (\ref{hg10}).
We have
\begin{eqnarray*}
	\tilde{\pi}_{OO}+\tilde{\pi}_{NO}&=&\int_0^1 \tilde{\pi}_{OO\vert p}+\tilde{\pi}_{NO\vert p}dp=\int_0^1 \mathbb P\left(\tilde{S}_1=1\vert \tilde{V}=p\right) dp,\\
	&=& \mathbb P(S=1,D=1\vert P(Z)=\overline{p})+\epsilon_1(1-\overline{p}),\\
	&\geq& \mathbb P(S=1,D=1\vert P(Z)=\overline{p}) = \sup_{p\in \mathcal P} \left\{\mathbb P(S=1,D=1\vert P(Z)=p)\right\},\\
	&=& \sup_{P(z) \in \mathcal P} \left\{\mathbb P(S=1,D=1\vert P(Z)=P(z))\right\}= \sup_{z \in \mathcal Z} \left\{\mathbb P(S=1,D=1\vert Z=z)\right\}
\end{eqnarray*}
because $\mathbb P(S=1,D=1\vert P(Z)=p)$ is increasing in $p$. Hence, condition (\ref{hg1}) is satisfied. Similarly, we can show that
\begin{eqnarray*}
	\tilde{\pi}_{NN}+\tilde{\pi}_{ON}&=&\int_0^1 \mathbb P\left(\tilde{S}_1=0\vert \tilde{V}=p\right) dp,\\
	&=& \mathbb P(S=0,D=1\vert P(Z)=\overline{p})+(1-\epsilon_1)(1-\overline{p}),\\
	&\geq& \sup_{z \in \mathcal Z} \left\{\mathbb P(S=0,D=1\vert Z=z)\right\}.
\end{eqnarray*}
Therefore, $\tilde{\pi}$ satisfies condition (\ref{hg2}). Moreover, we have:
\begin{eqnarray*}
	\tilde{\pi}_{OO}+\tilde{\pi}_{ON}&=&\int_0^1 \tilde{\pi}_{OO\vert p}+\tilde{\pi}_{ON\vert p}dp=\int_0^1 \mathbb P\left(\tilde{S}_0=1\vert \tilde{V}=p\right) dp,\\
	&=& \mathbb P(S=1,D=0\vert P(Z)=\underline{p})+\epsilon_0\underline{p},\\
	&\geq& \mathbb P(S=1,D=0\vert P(Z)=\underline{p}) = \sup_{p\in \mathcal P} \left\{\mathbb P(S=1,D=0\vert P(Z)=p)\right\},\\
	&=& \sup_{P(z) \in \mathcal P} \left\{\mathbb P(S=1,D=0\vert P(Z)=P(z))\right\}= \sup_{z \in \mathcal Z} \left\{\mathbb P(S=1,D=0\vert Z=z)\right\}
\end{eqnarray*}
because $\mathbb P(S=1,D=0\vert P(Z)=p)$ is decreasing in $p$. Thus, condition (\ref{hg3}) is satisfied. Similarly, we have
\begin{eqnarray*}
	\tilde{\pi}_{NN}+\tilde{\pi}_{NO}&=&\int_0^1 \mathbb P\left(\tilde{S}_0=0\vert \tilde{V}=p\right) dp,\\
	&=& \mathbb P(S=0,D=0\vert P(Z)=\underline{p})+(1-\epsilon_0)\underline{p},\\
	&\geq& \sup_{z \in \mathcal Z} \left\{\mathbb P(S=0,D=0\vert Z=z)\right\}.
\end{eqnarray*}
Hence, condition (\ref{hg4}) is verified.\\
Condition (\ref{hg5}) is equivalent to $\pi_{NN} \leq \mathbb P(S=0,D=1\vert Z=z)+\mathbb P(S=0,D=0\vert Z=z).$ By definition,
\begin{eqnarray*}
	\tilde{\pi}_{NN\vert p} &=& 1-\mathbb P(\tilde{S}_1=1\vert \tilde{V}=p)-\mathbb P(\tilde{S}_0=1\vert \tilde{V}=p)+\upsilon(p),\\
	&=& \left\{ \begin{array}{lcl}
		\mathbb P(\tilde{S}_1=0\vert \tilde{V}=p)-\mathbb P(\tilde{S}_0=1\vert \tilde{V}=p)+\upsilon(p)	\\ \\
		\mathbb P(\tilde{S}_0=0\vert \tilde{V}=p)-\mathbb P(\tilde{S}_1=1\vert \tilde{V}=p)+\upsilon(p)
	\end{array} \right.\\
	&\leq&
	\left\{ \begin{array}{lcl}
		\mathbb P(\tilde{S}_1=0\vert \tilde{V}=p)	\\ \\
		\mathbb P(\tilde{S}_0=0\vert \tilde{V}=p)
	\end{array} \right.
\end{eqnarray*}
since $\upsilon(p) \leq \min\left\{\mathbb P(\tilde{S}_1=1\vert \tilde{V}=p),\mathbb P(\tilde{S}_0=1\vert \tilde{V}=p)\right\}$. Thus,
\begin{eqnarray*}
	\tilde{\pi}_{NN\vert p} &\leq& \min\left\{\mathbb P(\tilde{S}_1=0\vert \tilde{V}=p),\mathbb P(\tilde{S}_0=0\vert \tilde{V}=p)\right\},\\
	&\leq& \mathbb P(\tilde{S}_1=0\vert \tilde{V}=p) \lambda(z) + \mathbb P(\tilde{S}_0=0\vert \tilde{V}=p) (1-\lambda(z))\ \text{ for all }\ \lambda(z)\in[0,1],\\
	&\leq& \mathbb P(\tilde{S}_1=0\vert \tilde{V}=p) \mathbbm{1}\{p\leq P(z)\} + \mathbb P(\tilde{S}_0=0\vert \tilde{V}=p) \mathbbm{1}\{p> P(z)\}\ \text{ for }\ \lambda(z)=\mathbbm{1}\{p\leq P(z)\},
\end{eqnarray*}
which implies
\begin{eqnarray*}
	\int^1_0 \tilde{\pi}_{NN\vert p} dp &\leq& \int^1_0\mathbb P(\tilde{S}_1=0\vert \tilde{V}=p) \mathbbm{1}\{p\leq P(z)\} dp + \int^1_0 \mathbb P(\tilde{S}_0=0\vert \tilde{V}=p) \mathbbm{1}\{p> P(z)\} dp,\\
	&=& \mathbb P(S=0,D=1\vert P(Z)=P(z))+\mathbb P(S=0,D=0\vert P(Z)=P(z)),\\
	&=& \mathbb P(S=0,D=1\vert Z=z)+\mathbb P(S=0,D=0\vert Z=z).
\end{eqnarray*}
Hence, $\tilde{\pi}_{NN} \leq \mathbb P(S=0,D=1\vert Z=z)+\mathbb P(S=0,D=0\vert Z=z)$ for all $z\in \mathcal Z$ and Condition \eqref{hg5} holds.

Condition (\ref{hg10}) is equivalent to $\pi_{OO} \leq \mathbb P(S=1,D=1\vert Z=z)+\mathbb P(S=1,D=0\vert Z=z).$ We have $\tilde{\pi}_{OO\vert p}=\upsilon(p)\leq \min\left\{\mathbb P(\tilde{S}_1=1\vert \tilde{V}=p),\mathbb P(\tilde{S}_0=1\vert \tilde{V}=p)\right\}$. Using a reasoning similar to the previous derivation, we obtain $\tilde{\pi}_{OO} \leq \mathbb P(S=1,D=1\vert Z=z)+\mathbb P(S=1,D=0\vert Z=z).$\\
Condition (\ref{hg7}) is equivalent to $\pi_{ON} \leq \mathbb P(S=1,D=0\vert Z=z)+\mathbb P(S=0,D=1\vert Z=z),$ while condition (\ref{hg8}) is equivalent to $\pi_{NO} \leq \mathbb P(S=0,D=0\vert Z=z)+\mathbb P(S=1,D=1\vert Z=z).$ We are going to show that (\ref{hg8}) is satisfied, and similar derivation holds for (\ref{hg7}). By definition,
\begin{eqnarray*}
	\tilde{\pi}_{NO|p} &=& \mathbb P(\tilde{S}_1=1\vert \tilde{V}=p) -\upsilon(p),\\
	&\leq&
	\left\{ \begin{array}{lcl}
		\mathbb P(\tilde{S}_1=1\vert \tilde{V}=p)\ \text{ since }\ \upsilon(p) \geq 0\\ \\
		\mathbb P(\tilde{S}_0=0\vert \tilde{V}=p)\ \text{ since }\ \upsilon(p) \geq \mathbb P(\tilde{S}_1=1\vert \tilde{V}=p)+\mathbb P(\tilde{S}_0=1\vert \tilde{V}=p)-1
	\end{array} \right. \\
	&\leq& \min\left\{\mathbb P(\tilde{S}_1=1\vert \tilde{V}=p),\mathbb P(\tilde{S}_0=0\vert \tilde{V}=p)\right\},\\
	&\leq& \mathbb P(\tilde{S}_1=1\vert \tilde{V}=p) \mathbbm{1}\{p\leq P(z)\} + \mathbb P(\tilde{S}_0=0\vert \tilde{V}=p) \mathbbm{1}\{p> P(z)\}.
\end{eqnarray*}
Therefore, by taking the integral of each side over [0,1], the result follows as for $\tilde{\pi}_{NO}$:
$$\tilde{\pi}_{NO} \leq \mathbb P(S=1,D=1\vert Z=z)+\mathbb P(S=0,D=0\vert Z=z).$$

	\subsection{Proof of Proposition \ref{THMnoassumption}}\label{PROOFnoassumption}

	The validity of the bounds is proven in the main text. It remains to show that the bounds are uniformly sharp. Given the restrictions that Assumptions \ref{RA}-\ref{Dist} impose on the data (i.e., equations \eqref{eq1} and \eqref{eq2}), we need to find a joint distribution on $(\tilde{Y}^*_0, \tilde{Y}^*_1,\tilde{S}_0, \tilde{S}_1, \tilde{V}, Z)$ that satisfies these restrictions, induces the joint distribution on the data $(Y,S,D,Z)$, and achieves any value $\delta \in \left[\underline{\Delta}_1, \overline{\Delta}_1\right]$. To do so, assume that $Y^{*}$ is absolutely continuous and has a strictly positive density.

	First, we show that the lower bound $\underline{\Delta}_1 \colon \mathcal{P} \rightarrow \mathbb{R}$ is attainable. We need to define the joint density (mass) function of $(\tilde{Y}^*_0, \tilde{Y}^*_1,\tilde{S}_0, \tilde{S}_1, \tilde{V}, Z)$. To do so, we will define the density functions $f_{\left. \tilde{Y}^*_0 \right\vert \tilde{S}_0, \tilde{S}_1, \tilde{V}}$, $f_{\left. \tilde{Y}^*_1  \right\vert \tilde{S}_0, \tilde{S}_1, \tilde{V}}$ and $f_{\tilde{V}}$, define the mass function $\tilde{\pi}_{\left(\tilde{S}_0, \tilde{S}_1\right)|\tilde{V}}$ and use the density function of $Z$ --- $f_{Z}$ --- to define $f_{\tilde{Y}^*_0,\tilde{Y}^*_1,(\tilde{S}_0,\tilde{S}_1),\tilde{V},Z} = f_{\left. \tilde{Y}^*_0 \right\vert \tilde{S}_0, \tilde{S}_1, \tilde{V}} \cdot f_{\left. \tilde{Y}^*_1 \right\vert \tilde{S}_0, \tilde{S}_1, \tilde{V}} \cdot \tilde{\pi}_{\left(\tilde{S}_0, \tilde{S}_1\right)|\tilde{V}} \cdot f_{\tilde{V}} \cdot f_{Z}$. Note that, by construction, Assumption \ref{RA} holds. With this goal in mind, fix $\left(s_{0}, s_{1}, y_{0}, y_{1}, p, z\right) \in \left\lbrace0,1\right\rbrace^{2} \times \mathbb{R}^{4}$ arbitrarily. Define $f_{\tilde{V}}\left(p\right) = \mathbbm{1}\left\lbrace p \in \left[0,1\right] \right\rbrace$, ensuring that Assumption \ref{Dist} holds by construction. Define
\begin{eqnarray*}
                 \upsilon(p)=\left\{ \begin{array}{lcl}
	 \overline{\upsilon}(p) = \argmin_{\upsilon \in \Upsilon\left(p\right)} \left\lbrace LB_{1}(p, \upsilon) - UB_{0}(p, \upsilon) \right\rbrace\ \text{ if }\ p\in [\underline{p}, \overline{p}]\\ \\
		\epsilon_0 \in [0,1]\ \text{ if }\ p < \underline{p} \\ \\
		\epsilon_1 \in [0,1]\ \text{ if }\ p > \overline{p}
	\end{array} \right.
\end{eqnarray*}
where $LB_{1}(p, \upsilon)$ and $UB_{0}(p, \upsilon)$ are defined in Subsection \ref{noassumption}. For brevity, denote the strata by \emph{OO} = always observed, \emph{NO} = observed only when treated, \emph{ON} = observed only when untreated and \emph{NN} = never observed, and the probability of the stratum $k$ conditional on $\tilde{V}=p$ by $\tilde{\pi}_{k|p}$.
Define
\begin{eqnarray*}\mathbb P\left(\tilde{S}_1=s_1\vert \tilde{V}=p\right) =
	\left\{ \begin{array}{lcl}
		\frac{\partial \mathbb P\left(S=s_1,D=1\vert P(Z)=p\right)}{\partial p}\ \text{ if }\ p\in [\underline{p}, \overline{p}]\\ \\
		\frac{\mathbb P\left(S=s_1,D=1\vert P(Z)=\underline{p}\right)}{ \underline{p}}\ \text{ if }\ p < \underline{p} \\ \\
		\upsilon(p) \cdot \mathbbm{1}\{s_1=1\}+(1-\upsilon(p)) \cdot \mathbbm{1}\{s_1=0\}\ \text{ if }\ p > \overline{p}
	\end{array} \right.
\end{eqnarray*}
and
\begin{eqnarray*}\mathbb P\left(\tilde{S}_0=s_0\vert \tilde{V}=p\right) =
	\left\{ \begin{array}{lcl}
		-\frac{\partial \mathbb P\left(S=s_0,D=0\vert P(Z)=p\right)}{\partial p}\ \text{ if }\ p\in [\underline{p}, \overline{p}]\\ \\
		\upsilon(p) \cdot \mathbbm{1}\{s_0=1\}+(1-\upsilon(p)) \cdot \mathbbm{1}\{s_0=0\}\ \text{ if }\ p < \underline{p} \\ \\
		\frac{\mathbb P\left(S=s_0,D=0\vert P(Z)=\overline{p}\right)}{1-\overline{p}}\ \text{ if }\ p > \overline{p}
	\end{array} \right.
\end{eqnarray*}
The probabilities $\tilde{\pi}_{k|p}$ are given by:
\begin{align*}
\tilde{\pi}_{OO|p} & = \upsilon(p) \\
\tilde{\pi}_{NO|p} & =\mathbb P\left(\tilde{S}_1=1\vert \tilde{V}=p\right) -\upsilon(p) \\
\tilde{\pi}_{ON|p} & = \mathbb P\left(\tilde{S}_0=1\vert \tilde{V}=p\right) - \upsilon(p) \\
\tilde{\pi}_{NN|p} & = 1 - \tilde{\pi}_{OO|p} - \tilde{\pi}_{NO|p} - \tilde{\pi}_{ON|p}.
\end{align*}
	Note that the above quantities are positive according to Lemma \ref{BFbounds} and add up to 1 by construction.

Define
\begin{eqnarray*}\mathbb P\left(\tilde{Y}^*_1 \leq y_1, \tilde{S}_1=s_1\vert \tilde{V}=p\right) =
	\left\{ \begin{array}{lcl}
		\frac{\partial \mathbb P\left(Y\leq y_1, S=s_1,D=1\vert P(Z)=p\right)}{\partial p}\ \text{ if }\ p\in [\underline{p}, \overline{p}]\\ \\
		\frac{\mathbb P\left(Y \leq y_1, S=s_1,D=1\vert P(Z)=\underline{p}\right)}{ \underline{p}}\ \text{ if }\ p < \underline{p} \\ \\
		\left(\upsilon(p) \cdot \mathbbm{1}\{s_1=1\}+(1-\upsilon(p)) \cdot \mathbbm{1}\{s_1=0\}\right)\mathbb P(Y\leq y_1)\ \text{ if }\ p > \overline{p}
	\end{array} \right.
\end{eqnarray*}
and
\begin{eqnarray*}\mathbb P\left(\tilde{Y}^*_0 \leq y_0, \tilde{S}_0=s_0\vert \tilde{V}=p\right) =
	\left\{ \begin{array}{lcl}
		-\frac{\partial \mathbb P\left(Y \leq y_0, S=s_0,D=0\vert P(Z)=p\right)}{\partial p}\ \text{ if }\ p\in [\underline{p}, \overline{p}]\\ \\
		\left(\upsilon(p) \cdot \mathbbm{1}\{s_0=1\}+(1-\upsilon(p)) \cdot \mathbbm{1}\{s_0=0\}\right)\mathbb P(Y\leq y_0)\ \text{ if }\ p < \underline{p} \\ \\
		\frac{\mathbb P\left(Y\leq y_0, S=s_0,D=0\vert P(Z)=\overline{p}\right)}{1-\overline{p}}\ \text{ if }\ p > \overline{p}
	\end{array} \right.
\end{eqnarray*}

Now, we define $f_{\left. \tilde{Y}^*_0 \right\vert \tilde{S}_0, \tilde{S}_1, \tilde{V}}\left(\left. y_{0} \right\vert k, p \right) = \frac{\partial \mathbb P(\tilde{Y}^*_0\leq y_0|k,\tilde{V}=p)}{\partial y_0}$ and $f_{\left. \tilde{Y}^*_1 \right\vert \tilde{S}_0, \tilde{S}_1, \tilde{V}}\left(\left. y_{1} \right\vert k, p \right) = \frac{\partial \mathbb P(\tilde{Y}^*_1\leq y_1|k,\tilde{V}=p)}{\partial y_1}$ for any $k \in \left\lbrace OO, NO, ON, NN \right\rbrace$, where we only need to define $\frac{\partial \mathbb P(\tilde{Y}^*_0\leq y_0|k,\tilde{V}=p)}{\partial y_0}$ and $\frac{\partial \mathbb P(\tilde{Y}^*_1\leq y_1|k,\tilde{V}=p)}{\partial y_1}$. Note that the data restriction imposed by equation \eqref{eq1} is satisfied by $(\tilde{Y}^*_1, \tilde{S}_1, \tilde{V}, Z)$, implying that $F_{\left. \tilde{Y}_{1}, \tilde{S}_{1} \right\vert \tilde{V}}$ is a proper C.D.F.. Note also that the data restriction imposed by equation \eqref{eq2} is satisfied by $(\tilde{Y}^*_0, \tilde{S}_0, \tilde{V}, Z)$ , implying that $F_{\left. \tilde{Y}_{0}, \tilde{S}_{0} \right\vert \tilde{V}}$ is a proper C.D.F.. Suppose that $\tilde{Y}_0\sim F_{\tilde{Y}^*_0|\tilde{S}_0=1,\tilde{V}=p}$ and $\tilde{Y}_1\sim F_{\tilde{Y}^*_1|\tilde{S}_1=1,\tilde{V}=p}$. Define
	\begin{align*}
		\mathbb P(\tilde{Y}^*_1\leq y_1|OO, \tilde{V}=p)& = \mathbb P\left(\tilde{Y}_1\leq y_1|\tilde{Y}_1\leq F^{-1}_{\tilde{Y}_1}\left(\frac{\tilde{\pi}_{OO|p}}{\tilde{\pi}_{OO|p}+\tilde{\pi}_{NO|p}}\right)\right),\\
		\mathbb P(\tilde{Y}^*_1\leq y_1|NO,\tilde{V}=p)&= \mathbb P\left(\tilde{Y}_1\leq y_1|\tilde{Y}_1 > F^{-1}_{\tilde{Y}_1}\left(\frac{\tilde{\pi}_{OO|p}}{\tilde{\pi}_{OO|p}+\tilde{\pi}_{NO|p}}\right)\right), \\
		\mathbb P(\tilde{Y}^*_1\leq y_1|k,\tilde{V}=p)&= \frac{\mathbb P(\tilde{Y}^*_1 \leq y_1, \tilde{S}_1=1\vert \tilde{V}=p)}{\mathbb P(\tilde{S}_1=1\vert \tilde{V}=p)}, \ \ k \in \left\lbrace ON, NN \right\rbrace \\
		\mathbb P(\tilde{Y}^*_0\leq y_0|OO, \tilde{V}=p)& = \mathbb P\left(\tilde{Y}_0\leq y_0|\tilde{Y}_0 > F^{-1}_{\tilde{Y}_0}\left(\frac{\tilde{\pi}_{ON|p}}{\tilde{\pi}_{OO|p}+\tilde{\pi}_{ON|p}}\right)\right), \\
		\mathbb P(\tilde{Y}^*_0 \leq y_0|ON,\tilde{V}=p)&= \mathbb P\left(\tilde{Y}_0\leq y_0|\tilde{Y}_0 \leq F^{-1}_{\tilde{Y}_0}\left(\frac{\tilde{\pi}_{ON|p}}{\tilde{\pi}_{OO|p}+\tilde{\pi}_{ON|p}}\right)\right), \\
		\mathbb P(\tilde{Y}^*_0\leq y_0|k,\tilde{V}=p)&= \frac{\mathbb P(\tilde{Y}^*_0 \leq y_0, \tilde{S}_0=1\vert \tilde{V}=p)}{\mathbb P(\tilde{S}_0=1\vert \tilde{V}=p)}, \ \ k\in \left\{NO,NN\right\}.
	\end{align*}

	Notice that the lower bound in Proposition \ref{THMnoassumption} is attained by the distributions of $\left. \tilde{Y}_{0}^{*} \right\vert OO, \tilde{V}$ and $\left. \tilde{Y}_{1}^{*} \right\vert OO, \tilde{V}$, i.e.,
	\begin{equation*}
		\underline{\Delta}_{1}\left(p\right) = \mathbb{E}\left[\tilde{Y}^*_1 -\tilde{Y}^*_0 |OO, \tilde{V}=p\right]
	\end{equation*}
	for any $p \in \mathcal{P}$.
	Finally, we show that the joint distribution of $(\tilde{Y}^*_0, \tilde{Y}^*_1,\tilde{S}_0, \tilde{S}_1, \tilde{V}, Z)$ induces the joint distribution on the data $(Y,S,D,Z)$. Define $\tilde{D} = \mathbbm{1}\left\lbrace \tilde{V} \leq P\left(Z\right) \right\rbrace$. For any $\left(y, z\right) \in \mathbb{R} \times \mathcal{Z}$,
	\begin{align*}
		& \mathbb{P} \left(\left. \tilde{Y} \leq y, \tilde{S} = 1, \tilde{D} = 1 \right\vert Z = z \right) \\
		& \hspace{20pt} = \mathbb{P} \left(\left. \tilde{Y}_{1}^{*} \leq y, \tilde{S}_{1} = 1, \tilde{V} \leq P\left(z\right) \right\vert  Z = z \right) \\
		& \hspace{20pt} = \mathbb{P} \left(\tilde{Y}_{1}^{*} \leq y, \tilde{S}_{1} = 1, \tilde{V} \leq P\left(z\right)\right) \\
		& \hspace{20pt} = \int_{0}^{P\left(z\right)} \mathbb{P} \left(\left. \tilde{Y}_{1}^{*} \leq y, \tilde{S}_{1} = 1 \right\vert  \tilde{V} = v \right) \, \text{d} v \\
		& \hspace{20pt} = \int_{0}^{\underline{p}} \frac{\mathbb P(Y\leq y, S=1,D=1|P(Z)=\underline{p})}{\underline{p}} \, \text{d} v + \int_{\underline{p}}^{P\left(z\right)} \frac{\partial \mathbb P(Y\leq y, S=1,D=1|P(Z)=v)}{\partial p} \, \text{d} v \\
		& \hspace{20pt} = \mathbb{P}\left(\left. Y\leq y, S = 1, D = 1 \right\vert P(Z) = P\left(z\right)\right) \\
		& \hspace{20pt} = \mathbb{P}\left(\left. Y\leq y, S = 1, D = 1 \right\vert Z = z\right)
	\end{align*}
	where the last equality is a testable implication of the model (see Appendix \ref{testable}). Analogously,
	\begin{align*}
		\mathbb{P} \left(\left. \tilde{Y} \leq y, \tilde{S} = 1, \tilde{D} = 0 \right\vert Z = z \right) & = \mathbb{P}\left(\left. Y\leq y, S = 1, D = 0 \right\vert Z = z\right), \\
		\mathbb{P} \left(\left. \tilde{S} = 0, \tilde{D} = 1 \right\vert Z = z \right) & = \mathbb{P}\left(\left. S = 0, D = 1 \right\vert Z = z\right), \\
		\mathbb{P} \left(\left. \tilde{S} = 0, \tilde{D} = 0 \right\vert Z = z \right) & = \mathbb{P}\left(\left. S = 0, D = 0 \right\vert Z = z\right).
	\end{align*}

	Similar reasoning holds for the upper bound, $\overline{\Delta}_{1}$. To attain any function $\delta \in \left(\underline{\Delta}_{1}, \overline{\Delta}_{1}\right)$, we can use convex combinations of the joint distributions that achieve the lower and upper bounds, where the weights of the convex combination may depend on the value $p \in \mathcal{P}$.

	\subsection{Proof of Proposition \ref{thm1}}\label{PROOF1}

	The validity of the bounds is proven in the main text. It remains to show that the bounds are uniformly sharp. Given the restrictions that Assumptions \ref{RA}-\ref{MONS} impose on the data (i.e., equations \eqref{eq1} and \eqref{eq2}), we need to find a joint distribution on $(\tilde{Y}^*_0, \tilde{Y}^*_1,\tilde{S}_0, \tilde{S}_1, \tilde{V}, Z)$ that satisfies these restrictions, induces the joint distribution on the data $(Y,S,D,Z)$, and achieves any value $\delta \in \left[\underline{\Delta}_{2}, \overline{\Delta}_{2}\right]$. Assume that $Y^{*}$ is absolutely continuous with strictly positive density.

	First, we show that the lower bound $\underline{\Delta}_2 \colon \mathcal{P} \rightarrow \mathbb{R}$ is attainable. We need to define the joint density (mass) function of $(\tilde{Y}^*_0, \tilde{Y}^*_1,\tilde{S}_0, \tilde{S}_1, \tilde{V}, Z)$. To do so, define the density functions $f_{\left. \tilde{Y}^*_0 \right\vert \tilde{S}_0, \tilde{S}_1, \tilde{V}}$, $f_{\left. \tilde{Y}^*_1  \right\vert \tilde{S}_0, \tilde{S}_1, \tilde{V}}$ and $f_{\tilde{V}}$, define the mass function $\tilde{\pi}_{\left(\tilde{S}_0, \tilde{S}_1\right)|\tilde{V}}$ and use the density function of $Z$ --- $f_{Z}$ --- to obtain $f_{\tilde{Y}^*_0,\tilde{Y}^*_1,(\tilde{S}_0,\tilde{S}_1),\tilde{V},Z} = f_{\left. \tilde{Y}^*_0 \right\vert \tilde{S}_0, \tilde{S}_1, \tilde{V}} \cdot f_{\left. \tilde{Y}^*_1 \right\vert \tilde{S}_0, \tilde{S}_1, \tilde{V}} \cdot \tilde{\pi}_{\left(\tilde{S}_0, \tilde{S}_1\right)|\tilde{V}} \cdot f_{\tilde{V}} \cdot f_{Z}$. Then, by construction, Assumption \ref{RA} holds. Fix $\left(s_{0}, s_{1}, y_{0}, y_{1}, p, z\right) \in \left\lbrace0,1\right\rbrace^{2} \times \mathbb{R}^{4}$ arbitrarily. Define $f_{\tilde{V}}\left(p\right) = \mathbbm{1}\left\lbrace p \in \left[0,1\right] \right\rbrace$, ensuring Assumption \ref{Dist} holds by construction. For brevity, denote the strata by \emph{OO} = always observed, \emph{NO} = observed only when treated and \emph{NN} = never observed, and the probability of the stratum $k$ conditional on $\tilde{V}=p$ by $\tilde{\pi}_{k|p}$. Define
\begin{eqnarray*}\mathbb P\left(\tilde{Y}^*_1 \leq y_1, \tilde{S}_1=1\vert \tilde{V}=p\right) =
	\left\{ \begin{array}{lcl}
		\frac{\partial \mathbb P\left(Y\leq y_1, S=1,D=1\vert P(Z)=p\right)}{\partial p}\ \text{ if }\ p\in [\underline{p}, \overline{p}]\\ \\
		\frac{\mathbb P\left(Y \leq y_1, S=1,D=1\vert P(Z)=\underline{p}\right)}{ \underline{p}}\ \text{ if }\ p < \underline{p} \\ \\
		\frac{\mathbb P\left(Y\leq y_1, S=1,D=0\vert P(Z)=\overline{p}\right)}{1-\overline{p}}\ \text{ if }\ p > \overline{p}
	\end{array} \right.
\end{eqnarray*}
\begin{eqnarray*}\mathbb P\left(\tilde{S}_1=0\vert \tilde{V}=p\right) =
	\left\{ \begin{array}{lcl}
		\frac{\partial \mathbb P\left(S=0,D=1\vert P(Z)=p\right)}{\partial p}\ \text{ if }\ p\in [\underline{p}, \overline{p}]\\ \\
		\frac{\mathbb P\left(S=0,D=1\vert P(Z)=\underline{p}\right)}{ \underline{p}}\ \text{ if }\ p < \underline{p} \\ \\
		\frac{\mathbb P\left(S=0,D=0\vert P(Z)=\overline{p}\right)}{1-\overline{p}}\ \text{ if }\ p > \overline{p}
	\end{array} \right.
\end{eqnarray*}
\begin{eqnarray*}\mathbb P\left(\tilde{Y}^*_0 \leq y_0, \tilde{S}_0=1\vert \tilde{V}=p\right) =
	\left\{ \begin{array}{lcl}
		-\frac{\partial \mathbb P\left(Y \leq y_0, S=1,D=0\vert P(Z)=p\right)}{\partial p}\ \text{ if }\ p\in [\underline{p}, \overline{p}]\\ \\
		\epsilon \frac{\mathbb P\left(Y \leq y_0, S=1,D=1\vert P(Z)=\underline{p}\right)}{ \underline{p}}\ \text{ if }\ p < \underline{p} \\ \\
		\frac{\mathbb P\left(Y\leq y_0, S=1,D=0\vert P(Z)=\overline{p}\right)}{1-\overline{p}}\ \text{ if }\ p > \overline{p}
	\end{array} \right.
\end{eqnarray*}
\begin{eqnarray*}\mathbb P\left(\tilde{S}_0=0\vert \tilde{V}=p\right) =
	\left\{ \begin{array}{lcl}
		-\frac{\partial \mathbb P\left(S=0,D=0\vert P(Z)=p\right)}{\partial p}\ \text{ if }\ p\in [\underline{p}, \overline{p}]\\ \\
		1-\epsilon \frac{\mathbb P\left(S=1,D=1\vert P(Z)=\underline{p}\right)}{ \underline{p}}\ \text{ if }\ p < \underline{p} \\ \\
		\frac{\mathbb P\left(S=0,D=0\vert P(Z)=\overline{p}\right)}{1-\overline{p}}\ \text{ if }\ p > \overline{p}
	\end{array} \right.
\end{eqnarray*}
where $\epsilon \in (0,1)$.\\
	The probabilities $\tilde{\pi}_{k|p}$ are given by:
	\begin{align*}
		\tilde{\pi}_{OO|p} & = \mathbb P(\tilde{S}_0=1\vert \tilde{V}=p) \\
		\tilde{\pi}_{NO|p} & =\mathbb P(\tilde{S}_1=1\vert \tilde{V}=p) -\mathbb P(\tilde{S}_0=1\vert \tilde{V}=p)  \\
		\tilde{\pi}_{NN|p} & = \mathbb P(\tilde{S}_1=0\vert \tilde{V}=p).
	\end{align*}

	Under Assumptions \ref{RA}-\ref{MONS}, the above quantities are positive, and add to one. For $p \in \mathcal{P}$,
	\begin{eqnarray*}
		\tilde{\pi}_{OO|p}+\tilde{\pi}_{NO|p} &=&\mathbb P(\tilde{S}_1=1\vert \tilde{V}=p)\\
		&=& \frac{\partial \mathbb P(S=1,D=1|P(Z)=p)}{\partial p},
	\end{eqnarray*}
	implying that
	\begin{eqnarray*}
		\tilde{\pi}_{OO|p}+\tilde{\pi}_{NO|p} + \tilde{\pi}_{NN|p} &=& \frac{\partial \mathbb P(S=0,D=1|P(Z)=p)}{\partial p} + \frac{\partial \mathbb P(S=1,D=1|P(Z)=p)}{\partial p},\\
		&=& \frac{\partial \mathbb P(D=1|P(Z)=p)}{\partial p}.
	\end{eqnarray*}
	We also have
	\begin{eqnarray*}
		\mathbb P(D=1|P(Z)=p) &=& \mathbb P(V\leq p|P(Z)=p) \ \ \ \text{by definition},\\
		&=& \mathbb P(V\leq p)\ \ \ \ \text{under Assumption \ref{RA}},\\
		&=& p\ \ \  \text{under Assumption \ref{Dist}}.
	\end{eqnarray*}
	Therefore, $\tilde{\pi}_{OO|p,z}+\tilde{\pi}_{NO|p,z} + \tilde{\pi}_{NN|p,z}=1.$ The same result holds for $p<\underline{p}$ and $p> \overline{p}$ from the fact that $\mathbb P(D=1\vert P(Z)=\underline{p})=\underline{p}$ and $\mathbb P(D=0\vert P(Z)=\overline{p})=1-\overline{p}$.

	Now, define $f_{\left. \tilde{Y}^*_0 \right\vert \tilde{S}_0, \tilde{S}_1, \tilde{V}}\left(\left. y_{0} \right\vert k, p \right) = \frac{\partial \mathbb P(\tilde{Y}^*_0\leq y_0|k,\tilde{V}=p)}{\partial y_0}$ and $f_{\left. \tilde{Y}^*_1 \right\vert \tilde{S}_0, \tilde{S}_1, \tilde{V}}\left(\left. y_{1} \right\vert k, p \right) = \frac{\partial \mathbb P(\tilde{Y}^*_1\leq y_1|k,\tilde{V}=p)}{\partial y_1}$ for any $k \in \left\lbrace OO, NO, NN \right\rbrace$, where we only need to characterize $\frac{\partial \mathbb P(\tilde{Y}^*_0\leq y_0|k,\tilde{V}=p)}{\partial y_0}$ and $\frac{\partial \mathbb P(\tilde{Y}^*_1\leq y_1|k,\tilde{V}=p)}{\partial y_1}$. The data restriction imposed by equation \eqref{eq1} is satisfied by $(\tilde{Y}^*_1, \tilde{S}_1, \tilde{V}, Z)$, implying that $F_{\left. \tilde{Y}_{1}, \tilde{S}_{1} \right\vert \tilde{V}}$ is a proper CDF. Similarly, the data restriction imposed by equation \eqref{eq2} is satisfied by $(\tilde{Y}^*_0, \tilde{S}_0, \tilde{V}, Z)$, and $F_{\left. \tilde{Y}_{0}, \tilde{S}_{0} \right\vert \tilde{V}}$ is a proper CDF. Suppose that $\tilde{Y}_1\sim F_{\tilde{Y}^*_1|\tilde{S}_1=1,\tilde{V}=p}$. Define
	\begin{align}
		\mathbb P(\tilde{Y}^*_1\leq y_1|OO, \tilde{V}=p)& = \mathbb P\left(\tilde{Y}_1\leq y_1|\tilde{Y}_1\leq F^{-1}_{\tilde{Y}_1}\left(\frac{\tilde{\pi}_{OO|p}}{\tilde{\pi}_{OO|p}+\tilde{\pi}_{NO|p}}\right)\right), \label{proofOO}\\
		\mathbb P(\tilde{Y}^*_1\leq y_1|NO,\tilde{V}=p)&= \mathbb P\left(\tilde{Y}_1\leq y_1|\tilde{Y}_1 > F^{-1}_{\tilde{Y}_1}\left(\frac{\tilde{\pi}_{OO|p}}{\tilde{\pi}_{OO|p}+\tilde{\pi}_{NO|p}}\right)\right), \label{proofNO} \\
		\mathbb P(\tilde{Y}^*_1\leq y_1|NN,\tilde{V}=p)&= \frac{\mathbb P(\tilde{Y}^*_1 \leq y_1, \tilde{S}_1=1\vert \tilde{V}=p)}{\mathbb P(\tilde{S}_1=1\vert \tilde{V}=p)}, \nonumber \\
		\mathbb P(\tilde{Y}^*_0\leq y_0|k,\tilde{V}=p)&= \frac{\mathbb P(\tilde{Y}^*_0 \leq y_0, \tilde{S}_0=1\vert \tilde{V}=p)}{\mathbb P(\tilde{S}_0=1\vert \tilde{V}=p)}, \ \ k\in \left\{OO,NO,NN\right\}. \nonumber
	\end{align}

	Notice that the lower bound in Proposition \ref{thm1} is attained by the distributions of $\left. \tilde{Y}_{0}^{*} \right\vert OO, \tilde{V}$ and $\left. \tilde{Y}_{1}^{*} \right\vert OO, \tilde{V}$ because
	\begin{equation*}
		\underline{\Delta}_{2}\left(p\right) = \mathbb{E}\left[\tilde{Y}^*_1 -\tilde{Y}^*_0 |OO, \tilde{V}=p\right] \text{ for any $p \in \mathcal{P}$.}
	\end{equation*}
	Finally, the joint distribution of $(\tilde{Y}^*_0, \tilde{Y}^*_1,\tilde{S}_0, \tilde{S}_1, \tilde{V}, Z)$ induces the joint distribution on the data $(Y,S,D,Z)$ by construction. The proof is identical to the proof in Appendix \ref{PROOFnoassumption}.

	Similar reasoning holds for the upper bound, $\overline{\Delta}_{2}$. To attain any function $\delta \in \left(\underline{\Delta}_{2}, \overline{\Delta}_{2}\right)$, we can use convex combinations of the joint distributions that achieve the lower and upper bounds, where the weights of the convex combination may depend on the value $p \in \mathcal{P}$.

	\subsection{Proof of Proposition \ref{thmDOM}}\label{PROOFDOM}

	The validity of the bounds is proven in the main text. It remains to show that the bounds are sharp. Given the restrictions that Assumptions \ref{RA}-\ref{DOM} impose on the data (i.e., equations \eqref{eq1} and \eqref{eq2}), we need to find joint distributions on $(\tilde{Y}^*_0, \tilde{Y}^*_1,\tilde{S}_0, \tilde{S}_1, \tilde{V}, Z)$ that satisfy these restrictions, satisfy the stochastic dominance assumption, induce the joint distribution on the data $(Y,S,D,Z)$, and achieve any value $\delta \in \left[\underline{\Delta}_3, \overline{\Delta}_3\right]$.

	The proof of Proposition \ref{thmDOM} is very similar to the one in Appendix \ref{PROOF1}. We only have to modify equations \eqref{proofOO} and \eqref{proofNO} to:
	\begin{eqnarray*}
		\mathbb P(\tilde{Y}^*_1\leq y_1|OO, \tilde{V}=p)&=& \mathbb P\left(\tilde{Y}_1\leq y_1\right),\\
		\mathbb P(\tilde{Y}^*_1\leq y_1|NO,\tilde{V}=p)&=& \mathbb P\left(\tilde{Y}_1\leq y_1\right).
	\end{eqnarray*}
	These changes ensure that the joint distribution of  $(\tilde{Y}^*_0, \tilde{Y}^*_1,\tilde{S}_0, \tilde{S}_1, \tilde{V}, Z)$ satisfy the Stochastic Dominance Assumption by construction.

	\subsection{Proof of Proposition \ref{thmTE}}\label{PROOFTE}

	The validity of the bounds is proven in the main text. We first prove that the bounds around $TE$ under Assumptions \eqref{RA}-\eqref{MONS} are sharp. Pick any treatment effect parameter $TE \coloneqq \int_{\mathcal{P}} MTE^{OO}\left(p\right) \cdot \omega\left(p\right) \, \text{d} p$, where the weighting function $\omega\colon\mathcal{P}\rightarrow\mathbb{R}$ is known or identified. Given the restrictions that Assumptions \eqref{RA}-\eqref{MONS} impose on the data (i.e., equations \eqref{eq1} and \eqref{eq2}), we need to find a joint distribution on $(\tilde{Y}^*_0, \tilde{Y}^*_1,\tilde{S}_0, \tilde{S}_1, \tilde{V}, Z)$ that satisfies these restrictions, induces the joint distribution on the data $(Y,S,D,Z)$, and achieves any value $\delta \in \left[\int_{\mathcal{P}} \left(\underline{\Delta}_{2}\left(p\right) \cdot \omega\left(p\right)\right) \, \text{d} p, \int_{\mathcal{P}} \left(\underline{\Delta}_{2}\left(p\right) \cdot \omega\left(p\right)\right) \, \text{d} p\right]$. To do so, assume that $Y^{*}$ is absolutely continuous and has a strictly positive density.

	First, we show that the lower bound $\underline{\Delta}_{2} \colon \mathcal{P} \rightarrow \mathbb{R}$ is attainable. The joint density (mass) function of $(\tilde{Y}^*_0, \tilde{Y}^*_1,\tilde{S}_0, \tilde{S}_1, \tilde{V}, Z)$ is defined as the same distribution used for Proposition \ref{thm1}.

	Now, observe that the lower bound $\int_{\mathcal{P}} \left(\underline{\Delta}_{2}\left(p\right) \cdot \omega\left(p\right)\right) \, \text{d} p$ in Proposition \ref{thmTE} is attained by the distributions of $\left. \tilde{Y}_{0}^{*} \right\vert OO, \tilde{V}$ and $\left. \tilde{Y}_{1}^{*} \right\vert OO, \tilde{V}$ because
	\begin{equation*}
		\underline{\Delta}_{2}\left(p\right) = \mathbb{E}\left[\tilde{Y}^*_1 -\tilde{Y}^*_0 |OO, \tilde{V}=p\right] \text{ for any $p \in \mathcal{P}$, and}
	\end{equation*}
	\begin{equation*}
		\int_{\mathcal{P}} \left(\underline{\Delta}_{2}\left(p\right) \cdot \omega\left(p\right)\right) \, \text{d} p = \int_{\mathcal{P}} \left(\mathbb{E}\left[\tilde{Y}^*_1 -\tilde{Y}^*_0 |OO, \tilde{V}=p\right] \cdot \omega\left(p\right)\right) \, \text{d} p.
	\end{equation*}

	Finally, the joint distribution of $(\tilde{Y}^*_0, \tilde{Y}^*_1,\tilde{S}_0, \tilde{S}_1, \tilde{V}, Z)$ induces the joint distribution on the data $(Y,S,D,Z)$ by construction. The proof is identical to the proof in Appendix \ref{PROOF1}.

	Similar reasoning holds for the upper bound, $\int_{\mathcal{P}} \left(\overline{\Delta}_{2}\left(p\right) \cdot \omega\left(p\right)\right) \, \text{d} p$. To attain any value $\delta \in \left(\int_{\mathcal{P}} \left(\underline{\Delta}_{2}\left(p\right) \cdot \omega\left(p\right)\right) \, \text{d} p, \int_{\mathcal{P}} \left(\overline{\Delta}_{2}\left(p\right) \cdot \omega\left(p\right)\right) \, \text{d} p\right)$, we can use convex combinations of the joint distributions that achieve the lower and upper bounds, where the weights of the convex combination does not depend on the value $p \in \mathcal{P}$.

	The proofs that the bounds around $TE$ under Assumptions \ref{RA}-\ref{Dist} and \ref{RA}-\ref{DOM} are similar to the proof above and they use the distributions described in Appendices \ref{PROOFnoassumption} and \ref{PROOFDOM}.

	\subsection{Deriving Identifiable Weights for $ATE^{OO}$, $ATT^{OO}$, $ATU^{OO}$, $LATE^{OO}$ and $PRTE^{OO}$ (Tables \ref{integral} and \ref{weights})}\label{PROOFweights}

	In this section, we refer to the Law of Iterated Expectations using its acronym, LIE.

	\subsubsection{ATE within the always-observed population}\label{PROOFATE}
	Observe that
	\begin{align*}
		ATE^{OO} & \coloneqq \mathbb{E}\left[\left. Y_{1}^{*} - Y_{0}^{*} \right\vert S_{0} = 1, S_{1} = 1 \right] \\
		& = \mathbb{E}\left[\left. \mathbb{E}\left[\left. Y_{1}^{*} - Y_{0}^{*} \right\vert V, S_{0} = 1, S_{1} = 1 \right] \right\vert S_{0} = 1, S_{1} = 1 \right] \text{ (LIE)} \\
		& = \int_{0}^{1} MTE^{OO}\left(p\right) \cdot f_{\left. V \right\vert S_{0} = 1, S_{1} = 1}\left(p\right) \, \text{d} p \text{ (Equation \eqref{target} and the expectation operator)} \\
		& = \bigintssss_{0}^{1} MTE^{OO}\left(p\right) \cdot \dfrac{\mathbb{P}\left[\left. S_{0} = 1, S_{1} = 1 \right\vert V = p \right]}{\mathbb{P}\left[S_{0} = 1, S_{1} = 1\right]} \cdot f_{V }\left(p\right) \, \text{d} p \text{ (Bayes' rule)} \\
		& = \bigintssss_{0}^{1} MTE^{OO}\left(p\right) \cdot \dfrac{\mathbb{P}\left[\left. S_{0} = 1, S_{1} = 1 \right\vert V = p \right]}{\mathbb{P}\left[S_{0} = 1, S_{1} = 1\right]}\, \text{d} p \text{ (Assumption  \ref{Dist})} \\
		& = \bigintss_{0}^{1} MTE^{OO}\left(p\right) \cdot \dfrac{\mathbb{P}\left[\left. S_{0} = 1, S_{1} = 1 \right\vert V = p \right]}{\int_{0}^{1} \mathbb{P}\left[\left. S_{0} = 1, S_{1} = 1 \right\vert V = p \right] \, \text{d} p}\, \text{d} p \text{ (Assumption \ref{Dist})} \\
		& = \bigint_{0}^{1} MTE^{OO}\left(p\right) \cdot \left[\dfrac{\dfrac{\partial\mathbb{P}\left[\left. S = 1, D = 0 \right\vert P\left(Z\right) = p \right]}{\partial p}}{\bigint_{\hspace{3pt} 0}^{1} \dfrac{\partial\mathbb{P}\left[\left. S = 1, D = 0 \right\vert P\left(Z\right) = p \right]}{\partial p} \, \text{d} p}\right]\, \text{d} p \text{ (Equation \eqref{eq3})},
	\end{align*}
	implying that $\omega_{ATE}\left(p\right) = \dfrac{\dfrac{\partial\mathbb{P}\left[\left. S = 1, D = 0 \right\vert P\left(Z\right) = p \right]}{\partial p}}{\bigint_{\hspace{3pt} 0}^{1} \dfrac{\partial\mathbb{P}\left[\left. S = 1, D = 0 \right\vert P\left(Z\right) = p \right]}{\partial p} \, \text{d} p}$.


	\subsubsection{ATT within the always-observed population}\label{PROOFATT}
	Observe that
	\begin{align*}
		ATT^{OO} & \coloneqq \mathbb{E}\left[\left. Y_{1}^{*} - Y_{0}^{*} \right\vert D = 1, S_{0} = 1, S_{1} = 1 \right] \\
		& = \mathbb{E}\left[\left. Y_{1}^{*} - Y_{0}^{*} \right\vert V \leq P\left(Z\right), S_{0} = 1, S_{1} = 1 \right] \text{ (Equation \eqref{seq1})} \\
		& = \dfrac{\mathbb{E}\left[\left. \mathbbm{1}\left\lbrace V \leq P\left(Z\right) \right\rbrace \cdot \left( Y_{1}^{*} - Y_{0}^{*} \right) \right\vert S_{0} = 1, S_{1} = 1 \right]}{\mathbb{P}\left[\left. V \leq P\left(Z\right) \right\vert S_{0} = 1, S_{1} = 1 \right]} \\
		& = \dfrac{\mathbb{E}\left[\left. \mathbbm{1}\left\lbrace V \leq P\left(Z\right) \right\rbrace \cdot \left( Y_{1}^{*} - Y_{0}^{*} \right) \right\vert S_{0} = 1, S_{1} = 1 \right]}{ \dfrac{\mathbb{P}\left[ V \leq P\left(Z\right), S_{0} = 1, S_{1} = 1 \right]}{\mathbb{P}\left[S_{0} = 1, S_{1} = 1 \right]}} \text{ (by the definition of a conditional expectation)} \\
		& = \dfrac{\mathbb{E}\left[\left. \mathbbm{1}\left\lbrace V \leq P\left(Z\right) \right\rbrace \cdot \mathbb{E}\left[\left. Y_{1}^{*} - Y_{0}^{*} \right\vert V, P\left(Z\right), S_{0} = 1, S_{1} = 1 \right] \right\vert S_{0} = 1, S_{1} = 1 \right]}{ \dfrac{\mathbb{E}\left[ \mathbbm{1}\left\lbrace V \leq P\left(Z\right) \right\rbrace \cdot \mathbb{P}\left[\left. S_{0} = 1, S_{1} = 1 \right\vert V, P\left(Z\right) \right] \right]}{\mathbb{P}\left[S_{0} = 1, S_{1} = 1 \right]}} \text{ (LIE)} \\
		& = \dfrac{\int_{0}^{1} \int_{p}^{1} \mathbb{E}\left[\left. Y_{1}^{*} - Y_{0}^{*} \right\vert V = p, P\left(Z\right) = u, S_{0} = 1, S_{1} = 1 \right] \cdot f_{\left. P\left(Z\right), V \right\vert S_{0} = 1, S_{1} = 1}\left(u, p\right) \, \text{d} u \, \text{d}p}{ \dfrac{\int_{0}^{1} \int_{p}^{1} \mathbb{P}\left[\left. S_{0} = 1, S_{1} = 1 \right\vert V = p, P\left(Z\right) = u \right] \cdot f_{P\left(Z\right), V }\left(u, p\right) \, \text{d} u \, \text{d}p}{\mathbb{P}\left[S_{0} = 1, S_{1} = 1 \right]}} \\
		& = \dfrac{\int_{0}^{1} \int_{p}^{1} \mathbb{E}\left[\left. Y_{1}^{*} - Y_{0}^{*} \right\vert V = p, S_{0} = 1, S_{1} = 1 \right] \cdot f_{P\left(Z\right)}\left(u\right) \cdot f_{\left. V \right\vert S_{0} = 1, S_{1} = 1}\left(p\right) \, \text{d} u \, \text{d}p}{ \dfrac{\int_{0}^{1} \int_{p}^{1} \mathbb{P}\left[\left. S_{0} = 1, S_{1} = 1 \right\vert V = p\right] \cdot f_{P\left(Z\right)}\left(u\right) \cdot f_{V }\left(p\right) \, \text{d} u \, \text{d}p}{\mathbb{P}\left[S_{0} = 1, S_{1} = 1 \right]}} \text{ (Assumption \ref{RA})} \\
		& = \dfrac{\int_{0}^{1} MTE^{OO}\left(p\right) \cdot \left( \int_{p}^{1} f_{P\left(Z\right)}\left(u\right) \, \text{d} u \right) \cdot f_{\left. V \right\vert S_{0} = 1, S_{1} = 1}\left(p\right) \, \text{d}p}{ \dfrac{\int_{0}^{1} \left( \int_{p}^{1} f_{P\left(Z\right)}\left(u\right) \, \text{d} u \right) \cdot \mathbb{P}\left[\left. S_{0} = 1, S_{1} = 1 \right\vert V = p\right] \, \text{d}p}{\mathbb{P}\left[S_{0} = 1, S_{1} = 1 \right]}} \\
		& \hspace{20pt} \text{by Assumption \ref{Dist}, definition \eqref{target} and the linearity of the integral operator,} \\
		& = \dfrac{\bigint_{0}^{1} MTE^{OO}\left(p\right) \cdot \left( \int_{p}^{1} f_{P\left(Z\right)}\left(u\right) \, \text{d} u \right) \cdot \dfrac{\mathbb{P}\left[\left. S_{0} = 1, S_{1} = 1 \right\vert V = p \right]}{\mathbb{P}\left[S_{0} = 1, S_{1} = 1\right]} \, \text{d}p}{ \dfrac{\int_{0}^{1} \left( \int_{p}^{1} f_{P\left(Z\right)}\left(u\right) \, \text{d} u \right) \cdot \mathbb{P}\left[\left. S_{0} = 1, S_{1} = 1 \right\vert V = p\right] \, \text{d}p}{\mathbb{P}\left[S_{0} = 1, S_{1} = 1 \right]}} \\
		& \hspace{20pt} \text{by the argument outlined in Appendix \ref{PROOFATE},} \\
		& = \bigints_{0}^{1} MTE^{OO}\left(p\right) \cdot \dfrac{\left( \int_{p}^{1} f_{P\left(Z\right)}\left(u\right) \, \text{d} u \right) \cdot \mathbb{P}\left[\left. S_{0} = 1, S_{1} = 1 \right\vert V = p \right]}{\int_{0}^{1} \left( \int_{p}^{1} f_{P\left(Z\right)}\left(u\right) \, \text{d} u \right) \cdot \mathbb{P}\left[\left. S_{0} = 1, S_{1} = 1 \right\vert V = p\right] \, \text{d}p} \, \text{d} p \\
		\\
		& = \bigints_{0}^{1} MTE^{OO}\left(p\right) \cdot \left[ \dfrac{\left( \int_{p}^{1} f_{P\left(Z\right)}\left(u\right) \, \text{d} u \right) \cdot \dfrac{\partial\mathbb{P}\left[\left. S = 1, D = 0 \right\vert P\left(Z\right) = p \right]}{\partial p}}{\bigint_{0}^{1} \left( \int_{p}^{1} f_{P\left(Z\right)}\left(u\right) \, \text{d} u \right) \cdot \dfrac{\partial\mathbb{P}\left[\left. S = 1, D = 0 \right\vert P\left(Z\right) = p \right]}{\partial p} \, \text{d}p} \right] \, \text{d} p \\
		& \hspace{20pt} \text{by Equation \eqref{eq3}},
	\end{align*}
	implying that $\omega_{ATT}\left(p\right) = \dfrac{\left(\int_{p}^{1} f_{P\left(Z\right)}\left(u\right) \, \text{d} u \right) \cdot \dfrac{\partial\mathbb{P}\left[\left. S = 1, D = 0 \right\vert P\left(Z\right) = p \right]}{\partial p}}{\bigint_{0}^{1} \left(\int_{p}^{1} f_{P\left(Z\right)}\left(u\right) \, \text{d} u \right) \cdot \dfrac{\partial\mathbb{P}\left[\left. S = 1, D = 0 \right\vert P\left(Z\right) = p \right]}{\partial p} \, \text{d} p}$.


	\subsubsection{ATU within the always-observed population}

	Analogously to Appendix \ref{PROOFATT}, we have that
	\begin{align*}
		ATU^{OO} & \coloneqq \mathbb{E}\left[\left. Y_{1}^{*} - Y_{0}^{*} \right\vert D = 0, S_{0} = 1, S_{1} = 1 \right] \\
		& = \bigint_{0}^{1} MTE^{OO}\left(p\right) \cdot \left[ \dfrac{\left( \int_{0}^{p} f_{P\left(Z\right)}\left(u\right) \, \text{d} u \right) \cdot \dfrac{\partial\mathbb{P}\left[\left. S = 1, D = 0 \right\vert P\left(Z\right) = p \right]}{\partial p}}{\bigint_{0}^{1} \left( \int_{0}^{p} f_{P\left(Z\right)}\left(u\right) \, \text{d} u \right) \cdot \dfrac{\partial\mathbb{P}\left[\left. S = 1, D = 0 \right\vert P\left(Z\right) = p \right]}{\partial p} \, \text{d}p} \right] \, \text{d} p
	\end{align*}
	implying that $\omega_{ATU}\left(p\right) = \dfrac{\left(\int_{0}^{p} f_{P\left(Z\right)}\left(u\right) \, \text{d} u \right) \cdot \dfrac{\partial\mathbb{P}\left[\left. S = 1, D = 0 \right\vert P\left(Z\right) = p \right]}{\partial p}}{\bigint_{0}^{1} \left(\int_{0}^{p} f_{P\left(Z\right)}\left(u\right) \, \text{d} u \right) \cdot \dfrac{\partial\mathbb{P}\left[\left. S = 1, D = 0 \right\vert P\left(Z\right) = p \right]}{\partial p} \, \text{d} p}$.


	\subsubsection{LATE within the always-observed population}

	Observe that, for any $\left(\underline{p}, \overline{p}\right) \in \left[0,1\right]^{2},$
	\begin{align*}
		LATE^{OO}(\underline{p}, \overline{p}) & = \mathbb{E}\left[Y_{1}^{*} - Y_{0}^{*} \left\vert V \in \left[\underline{p}, \overline{p}\right], S_{0} = 1, S_{1} = 1 \right.\right], \\
		& = \int_{\underline{p}}^{\overline{p}} \mathbb{E}\left[Y_{1}^{*}-Y_{0}^{*}\left\vert V=p,S_{0} = 1, S_{1} = 1\right. \right]f_{\left. V \right\vert S_{0} = 1, S_{1} = 1}\left(p\right)\text{d}p.
	\end{align*}
	The proof is similar to that of $ATE^{OO},$ except that the integral is over $\left[\underline{p}, \overline{p}\right]$ instead of $\left[0, 1\right]$. Consequently, we have that $\omega_{LATE}\left(p\right) = \dfrac{ \dfrac{\partial\mathbb{P}\left[\left. S = 1, D = 0 \right\vert P\left(Z\right) = p \right]}{\partial p}}{\bigint_{\underline{p}}^{\overline{p}} \dfrac{\partial\mathbb{P}\left[\left. S = 1, D = 0 \right\vert P\left(Z\right) = p \right]}{\partial p} \, \text{d}p}$.


	\subsubsection{PRTE within the always-observed population}

	First, we need to define a policy. A policy $a$ is a reassignment of the treatment based on $D_{a} = \mathbbm{1}\left\lbrace V \leq P_{a}\right\rbrace$, where the random variable $P_{a}$ is distributed according to the density $f_{P_{a}}$ and to the cumulative distribution function $F_{P_{a}}$. The new random variable $P_{a}$ is still assumed to be independent of $\left(V, Y_{0}^{*}, Y_{1}^{*}, S_{0}, S_{1}\right)$. Moreover, we define the possibly censored outcome under policy $a$ as $Y^{*} \coloneqq D_{a}Y_{1}^{*} + \left(1 - D_{a}\right)Y_{0}^{*}$.

	Now, observe that, for a policy $a$,
	\begin{align*}
		& \mathbb{E}\left[\left. Y_{a}^{*} \right\vert S_{0} = 1, S_{1} = 1 \right] \\
		& \hspace{20pt} = \mathbb{E}\left[\left. \mathbb{E}\left[ \left. Y_{a}^{*} \right\vert P_{a}, S_{0} = 1, S_{1} = 1 \right] \right\vert S_{0} = 1, S_{1} = 1 \right] \text{ (LIE)} \\
		& \hspace{20pt} = \int_{0}^{1} \mathbb{E}\left[ \left. Y_{a}^{*} \right\vert P_{a} = u, S_{0} = 1, S_{1} = 1 \right] \cdot f_{\left. P_{a} \right\vert S_{0} = 1, S_{1} = 1}\left(u\right) \, \text{d} u \\
		& \hspace{20pt} = \int_{0}^{1} \mathbb{E}\left[ \left. Y_{a}^{*} \right\vert P_{a} = u, S_{0} = 1, S_{1} = 1 \right] \cdot f_{P_{a}}\left(u\right) \, \text{d} u \\
		& \hspace{20pt} \hspace{20pt} \text{because } P_{a} \independent \left(V, Y_{0}^{*}, Y_{1}^{*}, S_{0}, S_{1}\right), \\
		& \hspace{20pt} = \int_{0}^{1} \mathbb{E}\left[ \left. D_{a} Y_{1}^{*} + \left(1 - D_{a}\right) Y_{0}^{*} \right\vert P_{a} = u, S_{0} = 1, S_{1} = 1 \right] \cdot f_{P_{a}}\left(u\right) \, \text{d} u \text{ (by definition)} \\
		& \hspace{20pt} = \int_{0}^{1} \mathbb{E}\left[ \left. \mathbbm{1}\left\lbrace V \leq u \right\rbrace \cdot   Y_{1}^{*} \right\vert P_{a} = u, S_{0} = 1, S_{1} = 1 \right] \cdot f_{P_{a}}\left(u\right) \, \text{d} u \\
		& \hspace{20pt} \hspace{20pt} + \int_{0}^{1} \mathbb{E}\left[ \left. \mathbbm{1}\left\lbrace u < V \right\rbrace \cdot   Y_{0}^{*} \right\vert P_{a} = u, S_{0} = 1, S_{1} = 1 \right] \cdot f_{P_{a}}\left(u\right) \, \text{d} u \\
		& \hspace{20pt} = \int_{0}^{1} \mathbb{E}\left[ \left. \mathbbm{1}\left\lbrace V \leq u \right\rbrace \cdot   Y_{1}^{*} \right\vert S_{0} = 1, S_{1} = 1 \right] \cdot f_{P_{a}}\left(u\right) \, \text{d} u \\
		& \hspace{20pt} \hspace{20pt} + \int_{0}^{1} \mathbb{E}\left[ \left. \mathbbm{1}\left\lbrace u < V \right\rbrace \cdot   Y_{0}^{*} \right\vert S_{0} = 1, S_{1} = 1 \right] \cdot f_{P_{a}}\left(u\right) \, \text{d} u \\
		& \hspace{20pt} \hspace{20pt} \text{because } P_{a} \independent \left(V, Y_{0}^{*}, Y_{1}^{*}, S_{0}, S_{1}\right), \\
		& \hspace{20pt} = \int_{0}^{1} \mathbb{E}\left[ \left. \mathbbm{1}\left\lbrace V \leq u \right\rbrace \cdot \mathbb{E} \left[\left.   Y_{1}^{*} \right\vert V, S_{0} = 1, S_{1} = 1 \right] \right\vert S_{0} = 1, S_{1} = 1 \right] \cdot f_{P_{a}}\left(u\right) \, \text{d} u \\
		& \hspace{20pt} \hspace{20pt} + \int_{0}^{1} \mathbb{E}\left[ \left. \mathbbm{1}\left\lbrace u < V \right\rbrace \cdot   \mathbb{E} \left[\left.   Y_{0}^{*} \right\vert V, S_{0} = 1, S_{1} = 1 \right] \right\vert S_{0} = 1, S_{1} = 1 \right] \cdot f_{P_{a}}\left(u\right) \, \text{d} u \text{ (LIE)} \\
		& \hspace{20pt} = \int_{0}^{1} \left[ \int_{0}^{u} \mathbb{E} \left[\left. Y_{1}^{*} \right\vert V = p, S_{0} = 1, S_{1} = 1 \right] \cdot f_{\left. V \right\vert S_{0} = 1, S_{1} = 1}\left(p\right) \, \text{d} p \right] \cdot f_{P_{a}}\left(u\right) \, \text{d} u \\
		& \hspace{20pt} \hspace{20pt} + \int_{0}^{1} \left[ \int_{u}^{1} \mathbb{E} \left[\left.Y_{0}^{*} \right\vert V = p, S_{0} = 1, S_{1} = 1 \right] \cdot f_{\left. V \right\vert S_{0} = 1, S_{1} = 1}\left(p\right) \, \text{d} p \right] \cdot f_{P_{a}}\left(u\right) \, \text{d} u \\
		& \hspace{20pt} = \int_{0}^{1} \mathbb{E} \left[\left. Y_{1}^{*} \right\vert V = p, S_{0} = 1, S_{1} = 1 \right] \cdot \left(\int_{p}^{1}  f_{P_{a}}\left(u\right) \, \text{d} u \right) \cdot f_{\left. V \right\vert S_{0} = 1, S_{1} = 1}\left(p\right) \, \text{d} p \\
		& \hspace{20pt} \hspace{20pt} + \int_{0}^{1} \mathbb{E} \left[\left. Y_{0}^{*} \right\vert V = p, S_{0} = 1, S_{1} = 1 \right] \cdot \left(\int_{0}^{p}  f_{P_{a}}\left(u\right) \, \text{d} u \right) \cdot f_{\left. V \right\vert S_{0} = 1, S_{1} = 1}\left(p\right) \, \text{d} p \\
		& \hspace{20pt} \hspace{20pt} \text{by Assumption \ref{finite} and Fubini's Theorem,} \\
		& \hspace{20pt} = \int_{0}^{1} \mathbb{E} \left[\left. Y_{1}^{*} \right\vert V = p, S_{0} = 1, S_{1} = 1 \right] \cdot \left(1 - F_{P_{a}}\left(p\right) \right) \cdot f_{\left. V \right\vert S_{0} = 1, S_{1} = 1}\left(p\right) \, \text{d} p \\
		& \hspace{20pt} \hspace{20pt} + \int_{0}^{1} \mathbb{E} \left[\left. Y_{0}^{*} \right\vert V = p, S_{0} = 1, S_{1} = 1 \right] \cdot F_{P_{a}}\left(p\right) \cdot f_{\left. V \right\vert S_{0} = 1, S_{1} = 1}\left(p\right) \, \text{d} p \text{ (by definition)} \\
		& \hspace{20pt} = \int_{0}^{1} \mathbb{E} \left[\left. Y_{1}^{*} \right\vert V = p, S_{0} = 1, S_{1} = 1 \right] \cdot f_{\left. V \right\vert S_{0} = 1, S_{1} = 1}\left(p\right) \, \text{d} p \\
		& \hspace{20pt} \hspace{20pt} - \int_{0}^{1} MTE^{OO}\left(p\right) \cdot F_{P_{a}}\left(p\right) \cdot f_{\left. V \right\vert S_{0} = 1, S_{1} = 1}\left(p\right) \, \text{d} p \text{ (Equation \eqref{target}).}
	\end{align*}

	Analogously, we have that, for a policy $a^{\prime}$,
	\begin{align*}
		& \mathbb{E}\left[\left. Y_{a^{\prime}}^{*} \right\vert S_{0} = 1, S_{1} = 1 \right] \\
		& \hspace{20pt} = \int_{0}^{1} \mathbb{E} \left[\left. Y_{1}^{*} \right\vert V = p, S_{0} = 1, S_{1} = 1 \right] \cdot f_{\left. V \right\vert S_{0} = 1, S_{1} = 1}\left(p\right) \, \text{d} p \\
		& \hspace{20pt} \hspace{20pt} - \int_{0}^{1} MTE^{OO}\left(p\right) \cdot F_{P_{a^{\prime}}}\left(p\right) \cdot f_{\left. V \right\vert S_{0} = 1, S_{1} = 1}\left(p\right) \, \text{d} p.
	\end{align*}

	Combining the last two results, we have that
	\begin{align*}
		PRTE^{OO} & \coloneqq \dfrac{\mathbb{E}\left[Y_{a}^{*} - Y_{a^{\prime}}^{*} \left\vert S_{0} = 1, S_{1} = 1 \right.\right]}{\int_{0}^{1} \left(F_{P_{a^{\prime}}}\left(p\right) - F_{P_{a}}\left(p\right)\right) \cdot f_{\left. V \right\vert S_{0} = 1, S_{1} = 1}\left(p\right) \, \text{d} p } \\
		& = \dfrac{\int_{0}^{1} MTE^{OO}\left(p\right) \cdot \left(F_{P_{a^{\prime}}}\left(p\right) - F_{P_{a}}\left(p\right)\right) \cdot f_{\left. V \right\vert S_{0} = 1, S_{1} = 1}\left(p\right) \, \text{d} p}{\int_{0}^{1} \left(F_{P_{a^{\prime}}}\left(p\right) - F_{P_{a}}\left(p\right)\right) \cdot f_{\left. V \right\vert S_{0} = 1, S_{1} = 1}\left(p\right) \, \text{d} p } \\
		& = \dfrac{\int_{0}^{1} MTE^{OO}\left(p\right) \cdot \left(F_{P_{a^{\prime}}}\left(p\right) - F_{P_{a}}\left(p\right)\right) \cdot \mathbb{P}\left[\left. S_{0} = 1, S_{1} = 1 \right\vert V = p\right] \, \text{d} p}{\int_{0}^{1} \left(F_{P_{a^{\prime}}}\left(p\right) - F_{P_{a}}\left(p\right)\right) \cdot \mathbb{P}\left[\left. S_{0} = 1, S_{1} = 1 \right\vert V = p\right] \, \text{d} p } \\
		& \hspace{20pt} \text{by the argument outlined in Appendix \ref{PROOFATE},} \\
		& = \bigint_{0}^{1} MTE^{OO}\left(p\right) \cdot \left[\dfrac{\left( F_{P_{a^{\prime}}}\left(p\right) - F_{P_{a}}\left(p\right) \right) \cdot \dfrac{\partial\mathbb{P}\left[\left. S = 1, D = 0 \right\vert P\left(Z\right) = p \right]}{\partial p}}{\bigint_{\hspace{3pt} 0}^{1} \left( F_{P_{a^{\prime}}}\left(p\right) - F_{P_{a}}\left(p\right) \right) \cdot \dfrac{\partial\mathbb{P}\left[\left. S = 1, D = 0 \right\vert P\left(Z\right) = p \right]}{\partial p} \, \text{d} p} \right] \, \text{d} p \\
		& \hspace{20pt} \text{by Equation \eqref{eq3}},
	\end{align*}
	implying that $\omega_{PRTE}\left(p, a, a^{\prime}\right) = \dfrac{\left( F_{P_{a^{\prime}}}\left(p\right) - F_{P_{a}}\left(p\right) \right) \cdot \dfrac{\partial\mathbb{P}\left[\left. S = 1, D = 0 \right\vert P\left(Z\right) = p \right]}{\partial p}}{\bigint_{\hspace{3pt} 0}^{1} \left( F_{P_{a^{\prime}}}\left(p\right) - F_{P_{a}}\left(p\right) \right) \cdot \dfrac{\partial\mathbb{P}\left[\left. S = 1, D = 0 \right\vert P\left(Z\right) = p \right]}{\partial p} \, \text{d} p}$.

	\subsection{Proof of Proposition \ref{thm2}}
	This proof is similar to that of Proposition \ref{thm1}. It is given for completeness. The validity of the bounds is proven in the main text. It remains to show that the bounds are uniformly sharp. Given the restrictions that Assumptions \ref{RA}, \ref{Dist}, \ref{MONS} and \ref{Discrete} impose on the data, we need to find a joint distribution on $(\tilde{Y}^*_0, \tilde{Y}^*_1,\tilde{S}_0, \tilde{S}_1, \tilde{V}, Z)$ that satisfies these assumptions, induces the joint distribution on the data $(Y,S,D,Z)$, and achieves any value $\delta \in \left[\underline{\Delta}_{LATE}, \overline{\Delta}_{LATE}\right]$. To do so, assume that $Y^{*}$ is absolutely continuous and has a strictly positive density.

	First, we show that the lower bound $\underline{\Delta}_{LATE} \colon \left\lbrace 2, \ldots, K\right\rbrace \rightarrow \mathbb{R}$ is attainable. We need to define the joint density (mass) function of $(\tilde{Y}^*_0, \tilde{Y}^*_1,\tilde{S}_0, \tilde{S}_1, \tilde{V}, Z)$. To do so, we will define the density functions $f_{\left. \tilde{Y}^*_0 \right\vert \tilde{S}_0, \tilde{S}_1, \tilde{V}}$, $f_{\left. \tilde{Y}^*_1  \right\vert \tilde{S}_0, \tilde{S}_1, \tilde{V}}$ and $f_{\tilde{V}}$, define the mass function $\tilde{\pi}_{\left(\tilde{S}_0, \tilde{S}_1\right)|\tilde{V}}$ and use the density function of $Z$ --- $f_{Z}$ --- to define $f_{\tilde{Y}^*_0,\tilde{Y}^*_1,(\tilde{S}_0,\tilde{S}_1),\tilde{V},Z} = f_{\left. \tilde{Y}^*_0 \right\vert \tilde{S}_0, \tilde{S}_1, \tilde{V}} \cdot f_{\left. \tilde{Y}^*_1 \right\vert \tilde{S}_0, \tilde{S}_1, \tilde{V}} \cdot \tilde{\pi}_{\left(\tilde{S}_0, \tilde{S}_1\right)|\tilde{V}} \cdot f_{\tilde{V}} \cdot f_{Z}$. Note that, by construction, Assumption \ref{RA} holds. Fix $\left(y_{0}, y_{1}, p, z\right) \in \mathbb{R}^{4}$ arbitrarily. Define $f_{\tilde{V}}\left(p\right) = \mathbbm{1}\left\lbrace p \in \left[0,1\right] \right\rbrace$, ensuring that Assumption \ref{Dist} holds by construction.

	For brevity, denote the strata by \emph{OO} = always observed, \emph{NO} = observed only when treated and \emph{NN} = never observed, and the probability of the stratum $k$ conditional on $\tilde{V} = p$ by $\tilde{\pi}_{k|p}$. The probabilities $\tilde{\pi}_{k|p}$ are given by
	\begin{align*}
		\tilde{\pi}_{OO|p} & = \sum_{\ell = 2}^{K} \mathbbm{1}\left\lbrace p_{\ell - 1} < p \leq p_{\ell} \right\rbrace \cdot \left(- \frac{\mathbb P(S=1,D=0|P=p_\ell) - \mathbb P(S=1,D=0|P=p_{\ell-1})}{p_\ell - p_{\ell-1}}\right) \\
		& \hspace{20pt} + \mathbbm{1}\left\lbrace p \leq p_{1} \right\rbrace \cdot \dfrac{\mathbb P(S=1,D=1|P=p_1)}{p_1} + \mathbbm{1}\left\lbrace p_{K} < p \right\rbrace \cdot \dfrac{\mathbb P(S=1,D=0|P=p_K)}{1 - p_K}, \\
		\tilde{\pi}_{NO|p} & = \sum_{\ell = 2}^{K} \mathbbm{1}\left\lbrace p_{\ell - 1} < p \leq p_{\ell} \right\rbrace \cdot \left(\frac{\mathbb P(S=1|P=p_\ell) - \mathbb P(S=1|P=p_{\ell-1})}{p_\ell - p_{\ell-1}}\right), \\
		\tilde{\pi}_{NN|p} & = \left(\sum_{\ell = 2}^{K} \mathbbm{1}\left\lbrace p_{\ell - 1} < p \leq p_{\ell} \right\rbrace \cdot \frac{\mathbb P(S=0,D=1|P=p_\ell)-\mathbb P(S=0,D=1|P=p_{\ell}-1)}{p_{\ell}-p_{\ell-1}} \right) \\
		& \hspace{20pt} + \mathbbm{1}\left\lbrace p \leq p_{1} \right\rbrace \cdot \left(1 - \dfrac{\mathbb P(S=1,D=1|P=p_1)}{p_1}\right) \\
		& \hspace{20pt} + \mathbbm{1}\left\lbrace p_{K} < p \right\rbrace \cdot \left(1 - \dfrac{\mathbb P(S=1,D=0|P=p_K)}{1 - p_K}\right).
	\end{align*}

	Under Assumptions \ref{RA}, \ref{Dist}, and \ref{MONS}, the above quantities are positive, and sum up to one. Indeed, for any $p \in \left[p_{\ell-1}, p_{\ell}\right]$ such that $\ell \in \left\lbrace 2, \ldots, K \right\rbrace$.
	\begin{eqnarray*}
		\tilde{\pi}_{OO|p}+\tilde{\pi}_{NO|p} &=& \frac{\mathbb P(S=1,D=1|P=p_\ell)-\mathbb P(S=1,D=1|P=p_{\ell-1})}{p_\ell-p_{\ell-1}},
	\end{eqnarray*}
	implying that
	\begin{eqnarray*}
		\tilde{\pi}_{OO|p}+\tilde{\pi}_{NO|p} + \tilde{\pi}_{NN|p} &=& \frac{\mathbb P(D=1|P=p_\ell)-\mathbb P(D=1|P=p_{\ell-1})}{p_\ell-p_{\ell-1}},\\
		&=& \frac{p_\ell-p_{\ell-1}}{p_\ell-p_{\ell-1}}=1.
	\end{eqnarray*}

	Now, define $f_{\left. \tilde{Y}^*_0 \right\vert \tilde{S}_0, \tilde{S}_1, \tilde{V}}\left(\left. y_{0} \right\vert k, p \right) = \frac{\partial \mathbb P(\tilde{Y}^*_0\leq y_0|k,\tilde{V}=p)}{\partial y_0}$ and $f_{\left. \tilde{Y}^*_1 \right\vert \tilde{S}_0, \tilde{S}_1, \tilde{V}}\left(\left. y_{1} \right\vert k, p \right) = \frac{\partial \mathbb P(\tilde{Y}^*_1\leq y_1|k,\tilde{V}=p)}{\partial y_1}$ for any $k \in \left\lbrace OO, NO, NN \right\rbrace$. Define
	\begin{eqnarray*}
		&&\mathbb P(\tilde{Y}_1^* \leq y_1, \tilde{S}_1=1| \tilde{V}=p)= \sum_{\ell = 2}^{K} \mathbbm{1}\left\lbrace p_{\ell - 1} < p \leq p_{\ell} \right\rbrace \nonumber\\
		&&\qquad \qquad \cdot \left( \frac{\mathbb P(Y\leq y_1,S=1,D=1|P=p_\ell) - \mathbb P(Y\leq y_1,S=1,D=1|P=p_{\ell-1})}{p_\ell - p_{\ell-1}} \right) \\
		&& \qquad \qquad + \mathbbm{1}\left\lbrace p \leq p_{1} \right\rbrace \cdot \dfrac{\mathbb P(Y\leq y_1,S=1,D=1|P=p_1)}{p_1} \\
		&& \qquad \qquad + \mathbbm{1}\left\lbrace p_{K} < p \right\rbrace \cdot \dfrac{\mathbb P(Y\leq y_1,S=1,D=0|P=p_K)}{1 - p_K}, \\
		&&\mathbb P(\tilde{Y}_0^* \leq y_0, \tilde{S}_0=1|\tilde{V}=p)= \sum_{\ell = 2}^{K} \mathbbm{1}\left\lbrace p_{\ell - 1} < p \leq p_{\ell} \right\rbrace \nonumber\\
		&&\qquad \qquad \cdot \left(- \frac{\mathbb P(Y\leq y_0,S=1,D=0|P=p_\ell) - \mathbb P(Y\leq y_0,S=1,D=0|P=p_{\ell-1})}{p_\ell - p_{\ell-1}}\right) \\
		&& \qquad \qquad + \mathbbm{1}\left\lbrace p \leq p_{1} \right\rbrace \cdot \dfrac{\mathbb P(Y\leq y_0,S=1,D=1|P=p_1)}{p_1} \\
		&& \qquad \qquad + \mathbbm{1}\left\lbrace p_{K} < p \right\rbrace \cdot \dfrac{\mathbb P(Y\leq y_0,S=1,D=0|P=p_K)}{1 - p_K},\text{ and }\\
	    && \mathbb P(\tilde{Y}_0^* \leq y_0|S_{0}=1,S_{1}=1,\tilde{V}=p) = \mathbb P(\tilde{Y}_0^* \leq y_0|\tilde{S}_0=1,\tilde{V}=p).
	\end{eqnarray*}
	Define $\tilde{Y}_1\sim F_{\tilde{Y}^*_1|\tilde{S}_1=1,\tilde{V}=p}$. Define also
	\begin{eqnarray*}
		\mathbb P(\tilde{Y}^*_1\leq y_1|OO, \tilde{V}=p)&=& \mathbb P\left(\tilde{Y}_1\leq y_1|\tilde{Y}_1\leq F^{-1}_{\tilde{Y}_1}\left(\frac{\tilde{\pi}_{OO|p}}{\tilde{\pi}_{OO|p}+\tilde{\pi}_{NO|p}}\right)\right),\\
		\mathbb P(\tilde{Y}^*_1\leq y_1|NO,\tilde{V}=p)&=& \mathbb P\left(\tilde{Y}_1\leq y_1|\tilde{Y}_1 > F^{-1}_{\tilde{Y}}\left(\frac{\tilde{\pi}_{OO|p}}{\tilde{\pi}_{OO|p}+\tilde{\pi}_{NO|p}}\right)\right),\\
		\mathbb P(\tilde{Y}^*_1\leq y_1|NN,\tilde{V}=p)&=& \mathbb P(\tilde{Y}_1^* \leq y_1, \tilde{S}_1=1| \tilde{V}=p),\\
		\mathbb P(\tilde{Y}^*_0\leq y_0|k,\tilde{V}=p)&=& \mathbb P(\tilde{Y}_0^* \leq y_0, \tilde{S}_0=1|\tilde{V}=p) \text{ for } k\in \left\{NO,NN\right\}.
		\end{eqnarray*}

	Notice that the lower bound in Proposition \ref{thm2} is attained by the distributions of $\left. \tilde{Y}_{0}^{*} \right\vert OO, \tilde{V}$ and $\left. \tilde{Y}_{0}^{*} \right\vert OO, \tilde{V}$ because
	\begin{equation*}
		\underline{\Delta}_{LATE}\left(\ell\right) = \mathbb{E}\left[\left. \tilde{Y}_{1}^{*} - \tilde{Y}_{0}^{*} \right\vert OO, p_{\ell - 1} < V \leq p_{\ell} \right]
	\end{equation*}
	for any $\ell \in \left\lbrace 2,\ldots,K\right\rbrace$.

	Finally, we show that the joint distribution of $(\tilde{Y}^*_0, \tilde{Y}^*_1,\tilde{S}_0, \tilde{S}_1, \tilde{V}, Z)$ induces the joint distribution on the data $(Y,S,D,Z)$. Let $\tilde{D} = \mathbbm{1}\left\lbrace \tilde{V} \leq P\left(Z\right) \right\rbrace$. For any $\left(y, \ell\right) \in \mathbb{R} \times \left\lbrace 2,\ldots,K\right\rbrace$,
	\begin{align*}
		& \mathbb{P} \left(\left. \tilde{Y} \leq y, \tilde{S} = 1, \tilde{D} = 1 \right\vert Z = z_{\ell} \right) = \mathbb{P} \left(\left. \tilde{Y}_{1}^{*} \leq y, \tilde{S}_{1} = 1, \tilde{V} \leq p_{\ell} \right\vert  Z = z_{\ell} \right) \\
		& \hspace{20pt} = \mathbb{P} \left(\tilde{Y}_{1}^{*} \leq y, \tilde{S}_{1} = 1, \tilde{V} \leq p_{\ell}\right) = \int_{0}^{p_{\ell}} \mathbb{P} \left(\left. \tilde{Y}_{1}^{*} \leq y, \tilde{S}_{1} = 1 \right\vert  \tilde{V} = v \right) \, \text{d} v \\
		& \hspace{20pt} = \bigints_{0}^{p_{\ell}} \left[\begin{array}{c}
			\sum_{k = 2}^{\ell} \mathbbm{1}\left\lbrace p_{k - 1} < v \leq p_{k} \right\rbrace \cdot  \frac{\mathbb P(Y\leq y_1,S=1,D=1|P=p_k) - \mathbb P(Y\leq y_1,S=1,D=1|P=p_{k-1})}{p_k - p_{k-1}}  \\ \\
			+ \mathbbm{1}\left\lbrace v \leq p_{1} \right\rbrace \cdot \dfrac{\mathbb P(Y\leq y_1,S=1,D=1|P=p_1)}{p_1}
		\end{array}\right] \, \text{d} v \\
		& \hspace{20pt} = \mathbb{P}\left(\left. Y\leq y, S = 1, D = 1 \right\vert P = p_{\ell}\right) = \mathbb{P}\left(\left. Y\leq y, S = 1, D = 1 \right\vert Z = z_{\ell}\right)
	\end{align*}
	and, analogously,
	\begin{align*}
		\mathbb{P} \left(\left. \tilde{Y} \leq y, \tilde{S} = 1, \tilde{D} = 0 \right\vert Z = z_{\ell} \right) & = \mathbb{P}\left(\left. Y\leq y, S = 1, D = 0 \right\vert Z = z_{\ell}\right), \\
		\mathbb{P} \left(\left. \tilde{S} = 0, \tilde{D} = 1 \right\vert Z = z_{\ell} \right) & = \mathbb{P}\left(\left. S = 0, D = 1 \right\vert Z = z_{\ell}\right), \\
		\mathbb{P} \left(\left. \tilde{S} = 0, \tilde{D} = 0 \right\vert Z = z_{\ell} \right) & = \mathbb{P}\left(\left. S = 0, D = 0 \right\vert Z = z_{\ell}\right).
	\end{align*}

	Similar reasoning holds for the upper bound, $\overline{\Delta}_{LATE}$. To attain any function $\delta \in \left(\underline{\Delta}_{LATE}, \overline{\Delta}_{LATE}\right)$, we can use convex combinations of the joint distributions that achieve the lower and upper bounds, where the weights of the convex combination may depend on the value $\ell \in \left\lbrace 2, \ldots, K \right\rbrace$.

	\section{Sharp testable implications for Assumptions \ref{RA}, \ref{Dist} and \ref{MONS}}\label{testable}
	Suppose that $Z$ contains at least one continuous instrument. Whenever Assumptions \ref{RA}, \ref{Dist} and \ref{MONS} hold, inequalities (\ref{test1}), (\ref{test2}) and (\ref{test3}) must hold, i.e.,
	\begin{eqnarray}
		&& 0 \leq \frac{\partial \mathbb E[\mathbbm{1}\left\{Y\in A\right\}SD|P(Z)=p]}{\partial p} \leq 1,\label{test1a}\\
		&& 0 \leq -\frac{\partial \mathbb E[\mathbbm{1}\left\{Y\in A\right\}S(1-D)|P(Z)=p]}{\partial p} \leq 1,\label{test2a}\\
		&&0 \leq \frac{\partial \mathbb P(S=1|P(Z)=p)}{\partial p} \leq 1 \label{test3a}
	\end{eqnarray}
	for all borel sets $A \subset \mathbb R$ and $p\in (0,1)$, where the last inequality holds because $$\mathbb P(NO|V=p)= \frac{\partial \mathbb P(S=1|P(Z)=p)}{\partial p}.$$
	In addition to the inequalities above, the following equalities must hold:
	\begin{eqnarray}
		\mathbb P(Y\in A, S=1, D=1|Z=z) &=&\mathbb P(Y\in A,S=1,D=1|P(Z)=P(z)),\label{test4a}\\
		\mathbb P(Y\in A,S=1,D=0|Z=z) &=&\mathbb P(Y\in A,S=1,D=0|P(Z)=P(z)), \label{test5a}\\
		\mathbb P(S=0, D=1|Z=z) &=&\mathbb P(S=0,D=1|P(Z)=P(z)),\label{test6a}\\
		\mathbb P(S=0,D=0|Z=z) &=&\mathbb P(S=0,D=0|P(Z)=P(z)). \label{test7a}
	\end{eqnarray}
	These equalities hold trivially when $P(z)$ is strictly monotone in $z$.
	\begin{proposition}
		Consider the model (\ref{seq1}).
		\begin{enumerate}
			\item [(i)] If Assumptions \ref{RA}, \ref{Dist} and \ref{MONS} hold, then inequalities (\ref{test1a}) to (\ref{test3a}) and equalities (\ref{test4a}) to (\ref{test7a})~hold.
			\item [(ii)] If inequalities (\ref{test1a}) to (\ref{test3a}) and equalities (\ref{test4a}) to (\ref{test7a}) hold, then there exists a vector $(\tilde{Y}^*_0,\tilde{Y}^*_1,\tilde{V}, \tilde{S}_0,\tilde{S}_1,,Z)$ that satisfies model (\ref{seq1}) and Assumptions \ref{RA}, \ref{Dist} and~\ref{MONS}.
		\end{enumerate}
	\end{proposition}
	These testable implications are identical to those in \cite{heckman2005structural} when there is no sample selection, i.e., $S=1$ almost surely.
	If the instrument $Z$ is binary, these testable implications become
	\begin{eqnarray*}
		&& 0 \leq \frac{\mathbb P(Y\in A, S=1, D=1|Z=1)-\mathbb P(Y\in A, S=1, D=1|Z=0)}{P(1)-P(0)}\leq 1,\\
		&& 0 \leq - \frac{(\mathbb P(Y\in A, S=1, D=0|Z=1)-\mathbb P(Y\in A, S=1, D=0|Z=0))}{P(1)-P(0)}\leq 1,\\
		&& 0 \leq \frac{\mathbb P(S=1|Z=1)-\mathbb P(S=1|Z=0)}{P(1)-P(0)}\leq 1.
	\end{eqnarray*}
	These latter inequalities generalize those in \cite{balke1997} and \cite{heckman2005structural}  to the sample selection case, and can therefore be tested using the procedures proposed by \cite{Machado2018}, \cite{MW2017}, \cite{Kitagawa2015}, \cite{Huber2015} or \cite{Laffers2017}.

	\begin{proof}
		\begin{enumerate}
			\item [(i)] Inequalities (\ref{test1a}) to (\ref{test3a}) have been shown in the main text. It remains to show equalities (\ref{test4a}) to (\ref{test7a}). We show (\ref{test4a}) and the proofs for the other equalities can be obtained similarly.
			\begin{eqnarray*}
				\mathbb P(Y\in A, S=1, D=1|Z=z) &=&\mathbb P(Y^*_1\in A,S_1=1,V\leq P(z)|Z=z),\\
				&=&\mathbb P(Y^*_1\in A,S_1=1,V\leq P(z)),\\
				&=&\mathbb P(Y^*_1\in A,S_1=1,V\leq P(z)|P(Z)=P(z)),\\
				&=&\mathbb P(Y^*_1\in A,S_1=1,V\leq P(Z)|P(Z)=P(z)),\\
				&=&\mathbb P(Y^*_1\in A,S_1=1,D=1|P(Z)=P(z)),\\
				&=&\mathbb P(Y\in A,S=1,D=1|P(Z)=P(z)),
			\end{eqnarray*}
			where the second and third equalities hold under Assumption \ref{RA}.
			\item [(ii)] Define $P(z)=\mathbb P(D=1|Z=z)$, and
			$\tilde{\pi}_{k|p,z}$ the probability of the stratum $k$ given $(\tilde{V}=p,Z=z)$.

			Define
\begin{eqnarray*}\mathbb P\left(\tilde{Y}^*_1 \leq y_1, \tilde{S}_1=1\vert \tilde{V}=p\right) =
	\left\{ \begin{array}{lcl}
		\frac{\partial \mathbb P\left(Y\leq y_1, S=1,D=1\vert P(Z)=p\right)}{\partial p}\ \text{ if }\ p\in [\underline{p}, \overline{p}]\\ \\
		\frac{\mathbb P\left(Y \leq y_1, S=1,D=1\vert P(Z)=\underline{p}\right)}{ \underline{p}}\ \text{ if }\ p < \underline{p} \\ \\
		\frac{\mathbb P\left(Y\leq y_1, S=1,D=0\vert P(Z)=\overline{p}\right)}{1-\overline{p}}\ \text{ if }\ p > \overline{p}
	\end{array} \right.
\end{eqnarray*}
\begin{eqnarray*}\mathbb P\left(\tilde{S}_1=0\vert \tilde{V}=p\right) =
	\left\{ \begin{array}{lcl}
		\frac{\partial \mathbb P\left(S=0,D=1\vert P(Z)=p\right)}{\partial p}\ \text{ if }\ p\in [\underline{p}, \overline{p}]\\ \\
		\frac{\mathbb P\left(S=0,D=1\vert P(Z)=\underline{p}\right)}{ \underline{p}}\ \text{ if }\ p < \underline{p} \\ \\
		\frac{\mathbb P\left(S=0,D=0\vert P(Z)=\overline{p}\right)}{1-\overline{p}}\ \text{ if }\ p > \overline{p}
	\end{array} \right.
\end{eqnarray*}
\begin{eqnarray*}\mathbb P\left(\tilde{Y}^*_0 \leq y_0, \tilde{S}_0=1\vert \tilde{V}=p\right) =
	\left\{ \begin{array}{lcl}
		-\frac{\partial \mathbb P\left(Y \leq y_0, S=1,D=0\vert P(Z)=p\right)}{\partial p}\ \text{ if }\ p\in [\underline{p}, \overline{p}]\\ \\
		\epsilon \frac{\mathbb P\left(Y \leq y_0, S=1,D=1\vert P(Z)=\underline{p}\right)}{ \underline{p}}\ \text{ if }\ p < \underline{p} \\ \\
		\frac{\mathbb P\left(Y\leq y_0, S=1,D=0\vert P(Z)=\overline{p}\right)}{1-\overline{p}}\ \text{ if }\ p > \overline{p}
	\end{array} \right.
\end{eqnarray*}
and
\begin{eqnarray*}\mathbb P\left(\tilde{S}_0=0\vert \tilde{V}=p\right) =
	\left\{ \begin{array}{lcl}
		-\frac{\partial \mathbb P\left(S=0,D=0\vert P(Z)=p\right)}{\partial p}\ \text{ if }\ p\in [\underline{p}, \overline{p}]\\ \\
		1-\epsilon \frac{\mathbb P\left(S=1,D=1\vert P(Z)=\underline{p}\right)}{ \underline{p}}\ \text{ if }\ p < \underline{p} \\ \\
		\frac{\mathbb P\left(S=0,D=0\vert P(Z)=\overline{p}\right)}{1-\overline{p}}\ \text{ if }\ p > \overline{p}
	\end{array} \right.
\end{eqnarray*}
where $\epsilon \in (0,1)$.\\
	Define the distribution on the strata:
	\begin{align*}
		\tilde{\pi}_{OO|p} & = \mathbb P(\tilde{S}_0=1\vert \tilde{V}=p) \\
		\tilde{\pi}_{NO|p} & =\mathbb P(\tilde{S}_1=1\vert \tilde{V}=p) -\mathbb P(\tilde{S}_0=1\vert \tilde{V}=p)  \\
		\tilde{\pi}_{NN|p} & = \mathbb P(\tilde{S}_1=0\vert \tilde{V}=p).
	\end{align*}
			Inequalities (\ref{test1a}) to (\ref{test3a}) imply that the above quantities are positive and sum up to one.
			Define
			\begin{equation*}
				\mathbb P(\tilde{Y}^*_1\leq y_1|k, \tilde{V}=p,Z=z)= \mathbb P(\tilde{Y}^*_1\leq y_1|\tilde{S}_1=1,\tilde{V}=p),\ \ k \in \left\lbrace OO, NO, NN \right\rbrace
			\end{equation*}
			and
			\begin{equation*}
				\mathbb P(\tilde{Y}^*_0\leq y_0|k,\tilde{V}=p,Z=z)= \mathbb P(\tilde{Y}^*_0 \leq y_0 \vert \tilde{S}_0=1, \tilde{V}=p),\ \ k\in \left\{OO,NO,NN\right\}.
			\end{equation*}
			Define the joint conditional distribution of $(\tilde{Y}^*_0,\tilde{Y}^*_1, \tilde{S}_0, \tilde{S}_1)$ given $(\tilde{V}=p,Z=z)$:
			\begin{eqnarray*}
				\mathbb P({\tilde{Y}^*_0\leq y_0,\tilde{Y}^*_1\leq y_1,(\tilde{S}_0,\tilde{S}_1)=k,|\tilde{V}=p,Z=z})&=& \mathbb P(\tilde{Y}^*_0\leq y_0|k,\tilde{V}=p,Z=z) \cdot \\
				&&\mathbb P(\tilde{Y}^*_1\leq y_1|k,\tilde{V}=p,Z=z) \cdot \tilde{\pi}_{k|p},\\
				&& k \in \left\{OO,NO,NN\right\},\\
				\mathbb P(\tilde{V}\leq p|Z=z)&=&p.
			\end{eqnarray*}
			Finally, define
			\begin{eqnarray}\label{seq1a}
				\left\{ \begin{array}{lcl}
					\tilde{Y}^*&=&\tilde{Y}^{*}_{1}\tilde{D}+\tilde{Y}^{*}_{0}(1-\tilde{D})\\ \\
					\tilde{D}&=&\mathbbm{1}\left\{\tilde{V}\leq P(Z)\right\} \\ \\
					\tilde{S}&=&\tilde{S}_1\tilde{D}+\tilde{S}_0(1-\tilde{D})\\ \\
					\tilde{Y}&=&\tilde{Y}^*\tilde{S}
				\end{array} \right.
			\end{eqnarray}
		\end{enumerate}
	We can show that $(\tilde{Y},\tilde{S}, \tilde{D},Z)$ has the same joint distribution as $(Y,S,D,Z)$.
		\begin{eqnarray*}
			\mathbb P(\tilde{Y}\leq y, \tilde{S}=1, \tilde{D}=1|Z=z) &=& \mathbb P(\tilde{Y}^*_1\leq y, \tilde{S}_1=1,\tilde{V}\leq P(z)|Z=z),\\
			&=& \mathbb P(\tilde{Y}^*_1\leq y, \tilde{S}_1=1|\tilde{V}\leq P(z),Z=z)\mathbb P(\tilde{V}\leq P(z)|Z=z),\\
			&=&  \mathbb P(\tilde{Y}^*_1\leq y, \tilde{S}_1=1|\tilde{V}\leq P(z),Z=z)P(z),\\
			&=& \left[ \int^{P(z)}_0 \mathbb P(\tilde{Y}^*_1\leq y, \tilde{S}_1=1|\tilde{V}=v,Z=z)\frac{f_{\tilde{V}|Z=z}(v)}{P(z)}dv \right] \cdot P(z),\\
			&=&  \int^{P(z)}_0 \mathbb P(\tilde{Y}^*_1\leq y,\tilde{S}_1=1\vert \tilde{V}=v,Z=z)dv,\\
			&=& \int^{P(z)}_0 \mathbb P(\tilde{Y}^*_1\leq y, \tilde{S}_1=1|\tilde{V}=v)dv,\\
			&=&  \int^{\underline{p}}_0 \frac{\mathbb P\left(Y \leq y_1, S=1,D=1\vert P(Z)=\underline{p}\right)}{ \underline{p}} dv\\
			&& + \int^{P(z)}_{\underline{p}} \frac{\partial \mathbb P\left(Y\leq y_1, S=1,D=1\vert P(Z)=p\right)}{\partial p} dv,\\
			&=& \mathbb P(Y\leq y, S=1,D=1|P(Z)=P(z))\\
			&=& \mathbb P(Y\leq y, S=1,D=1|Z=z).
		\end{eqnarray*}
		Similarly,
		\begin{eqnarray*}
			\mathbb P(\tilde{Y}\leq y, \tilde{S}=1, \tilde{D}=0|Z=z) &=& \mathbb P(Y\leq y, S=1,D=1|Z=z),\\
			\mathbb P(\tilde{S}=0, \tilde{D}=1|Z=z) &=& \mathbb P(S=0,D=1|Z=z),\\
			\mathbb P(\tilde{S}=0, \tilde{D}=0|Z=z) &=& \mathbb P(S=0,D=0|Z=z).
		\end{eqnarray*}
		Finally, by construction, Assumptions \ref{RA}, \ref{Dist} and~\ref{MONS} hold.
	\end{proof}

	\section{Numerical Illustration}\label{Sim}
	In this appendix, we highlight the feasibility and usefulness of the bounds proposed in Section \ref{Ident} by considering a numerical illustration of a simple structural model with endogenous treatment and sample selection. In Appendix \ref{nummain}, we present the bounds for $MTE^{OO}$ based on Propositions \ref{THMnoassumption} and \ref{thm1}. The illustration provides insights on the functioning of the bounds as well as the mechanisms driving how informative those bounds are. In Appendix \ref{numselected}, we discuss what is identified by the \textit{local instrumental variable} (LIV) estimand \citep{heckman1999} when directly applied to the selected sample and explain its limitations. Appendix \ref{numdetail} details the data-generating process used in this numerical illustration.

	\subsection{Illustrating the Bounds around $MTE^{OO}$}\label{nummain}
	Consider the following data generating process (DGP):
	\begin{eqnarray}\label{seqsim}
		\left\{ \begin{array}{lcl}
			Y&=&Y^{*}S\\ \\
			Y^{*}&=&Y^{*}_{1} D + Y^{*}_{0} (1-D)\\ \\
			S&=&\mathbbm{1}\left\{U_{S}\leq \delta_0 + \delta_1 D \right\}\\ \\
			D&=&\mathbbm{1}\left\{V\leq \Phi(Z)\right\}
		\end{array} \right.
	\end{eqnarray}
	we set
	\begin{eqnarray}\label{seqsim2}
		\left\{ \begin{array}{lcl}
			V &=& \Phi(\theta) \\ \\
			U_S&=&\frac{1}{\sqrt{2}}(\theta+\epsilon_S)\\ \\
			Y^{*}_{0} & = & T \cdot \beta_{0,1}\theta + \left(1 - T\right) \cdot \left(- \beta_{0,0}\theta \right)\\ \\
			Y^{*}_{1}&=& T \cdot \beta_{1,1}\theta + \left(1 - T\right) \cdot \left(- \beta_{1,0}\theta\right)
		\end{array} \right.
	\end{eqnarray}
	where $(\theta,\epsilon_S,Z, \xi)' \sim N(0,I)$, $T = \mathbbm{1}\left\lbrace \xi \geq 0 \right\rbrace$, $I$ is the identity matrix, $\Phi(.)$ is the standard normal CDF, and $\Phi^{-1}(\cdot)$ its inverse. The potential outcomes equations has  random coefficients in this illustration. Intuitively, there are two sets of individuals that might face different returns to treatment due to, for example, their gender or race. From a technical standpoint, this choice guarantees reasonable overlap for treated and untreated groups in the observed population over the support of the outcome conditional on $V$.

\begin{figure}[htbp]
		\begin{center}
			\subfigure[{Bounds as a function of the propensity score}]{
				\includegraphics[width = 2.8in]{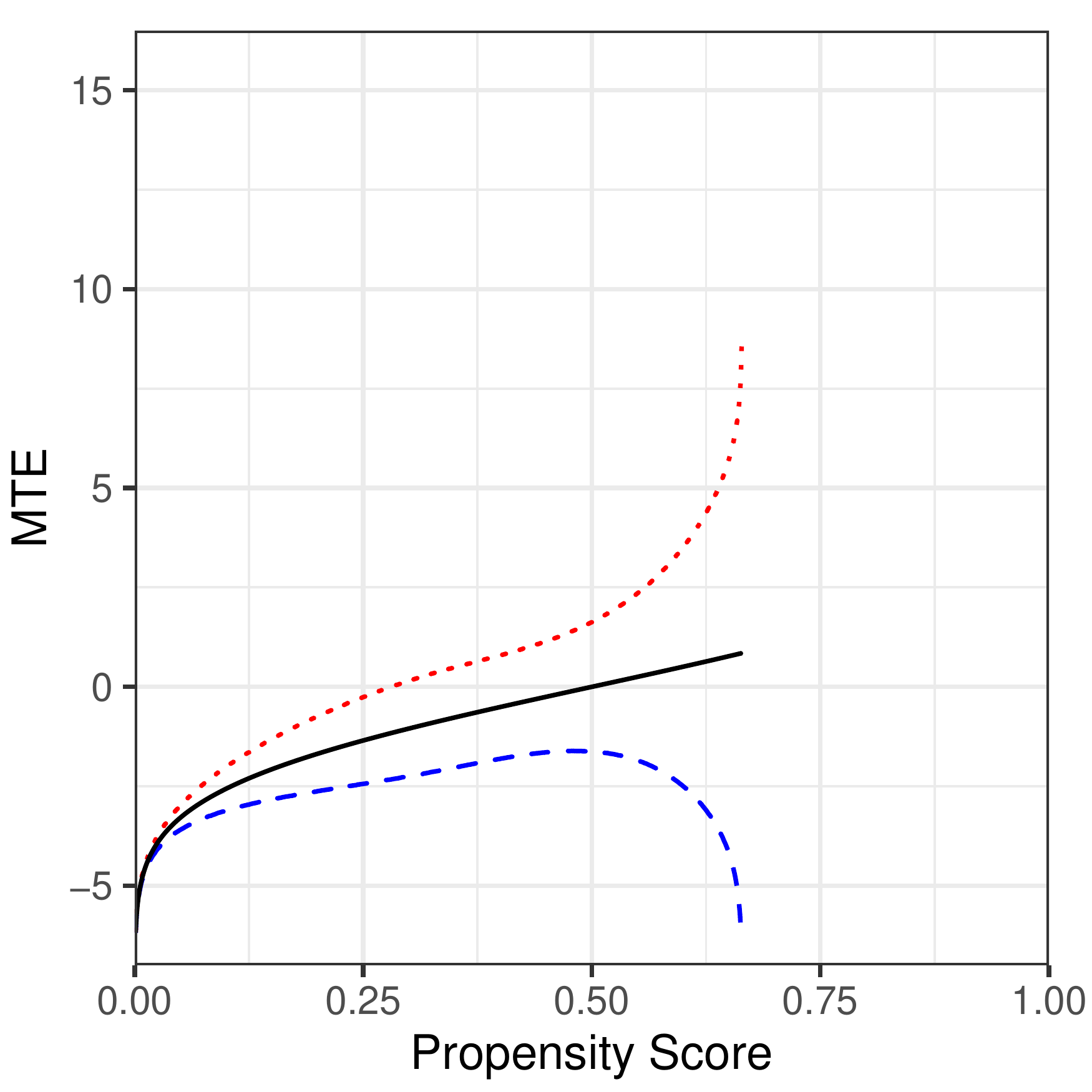}
				\label{fig:bounds1NOassumption}
			}
			\subfigure[{Bounds as a function of $\mathbb{P}\left[\left. S_0 = 1 \right\vert V = p \right] + \mathbb{P}\left[\left. S_1 = 1 \right\vert V = p \right] - 1$}]{
				\includegraphics[width = 2.8in]{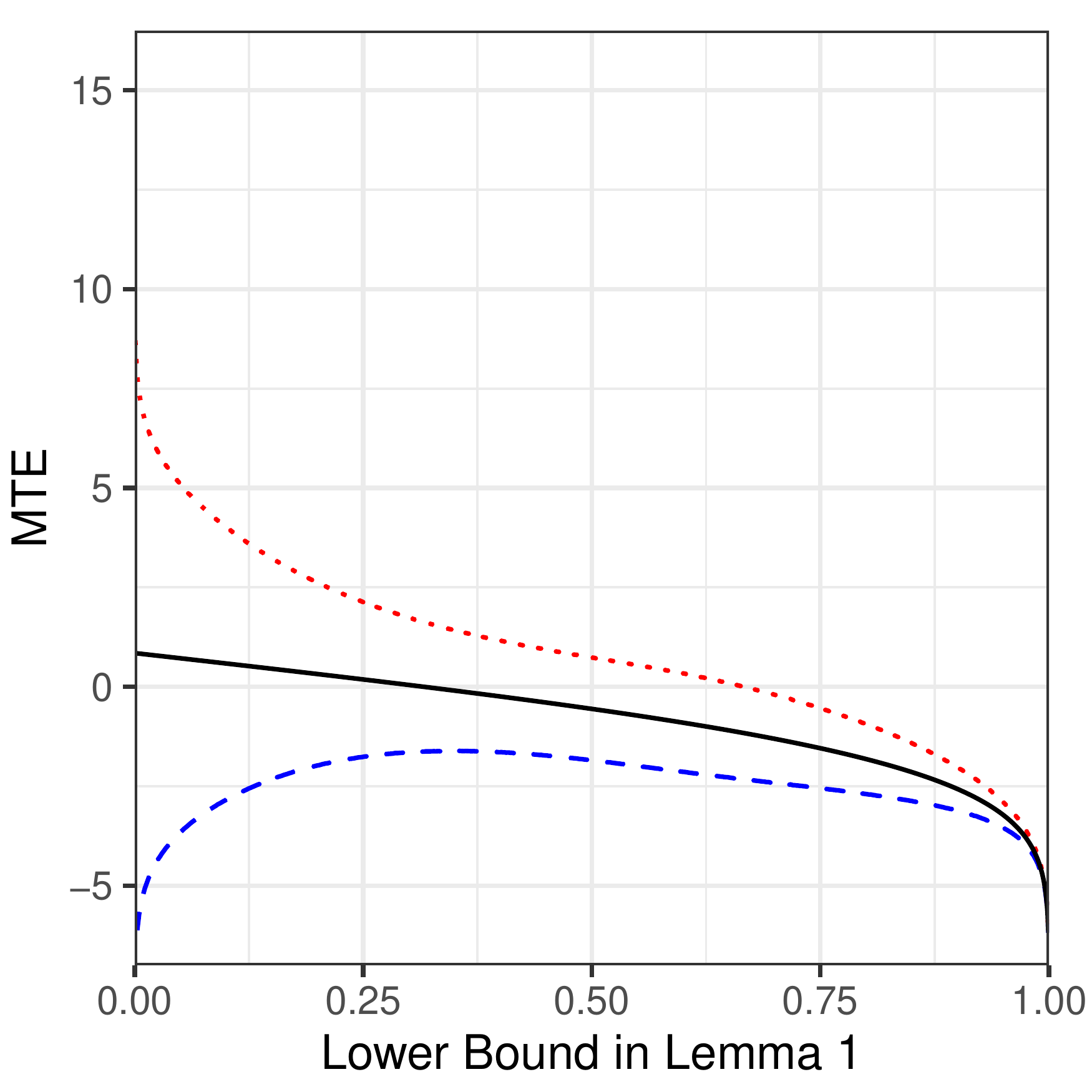}
				\label{fig:bounds2NOassumption}
			}
		\end{center}
		\footnotesize{Notes: The solid lines are the true values of the $MTE^{OO}$. The red dotted lines and the blue dashed lines are, respectively, the values of the upper and lower bounds around the $MTE^{OO}$ computed by numerical integration using 100,000 simulated points for each value of the propensity score.}
		\caption[]{Numerical Bounds based on Proposition \ref{THMnoassumption}}
		\label{figure:boundsNOassumption}
	\end{figure}

\begin{figure}[htbp]
		\begin{center}
			\subfigure[{Bounds as a function of the propensity score}]{
				\includegraphics[width = 2.8in]{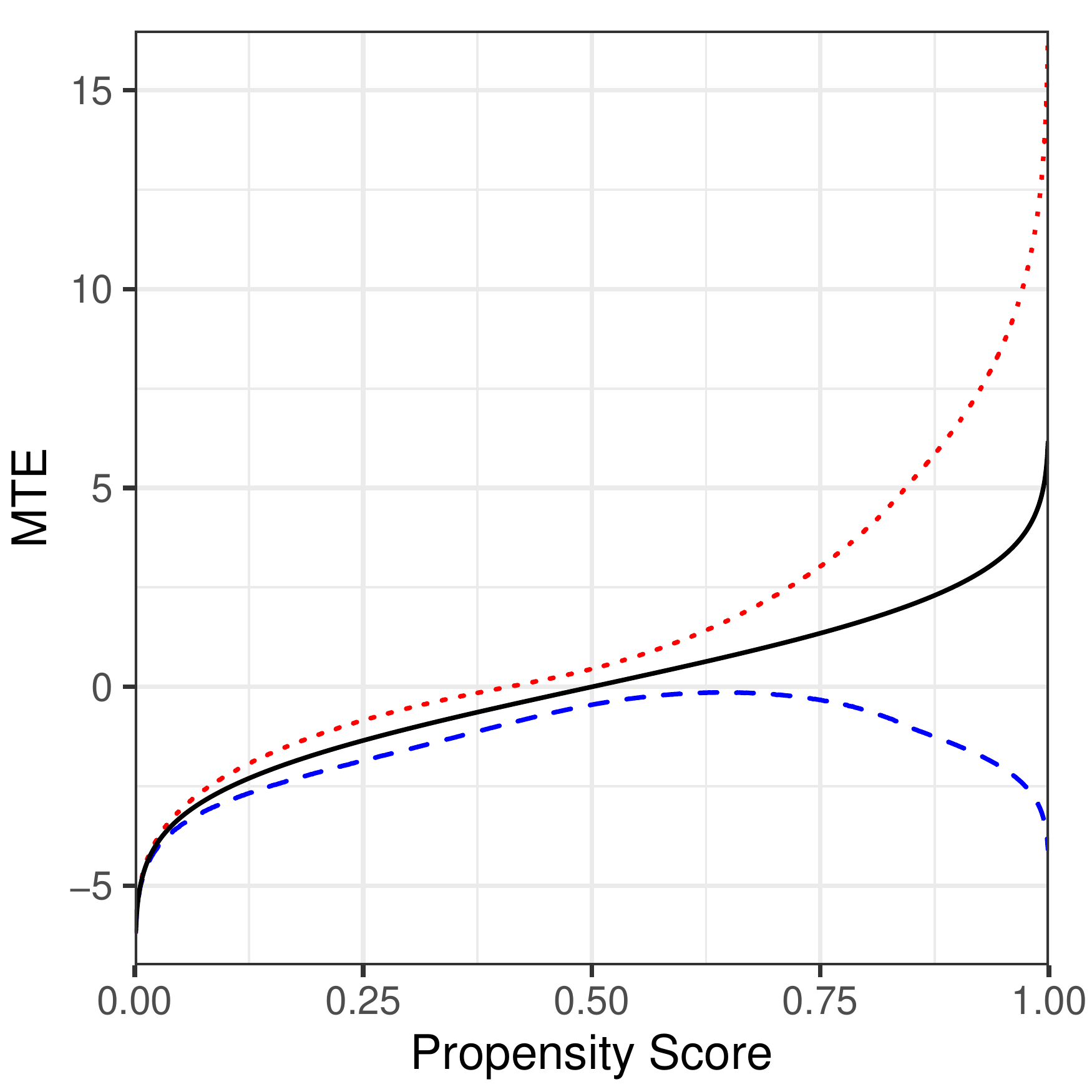}
				\label{fig:bounds1}
			}
			\subfigure[{Bounds as a function of $\alpha\left(p\right)$}]{
				\includegraphics[width = 2.8in]{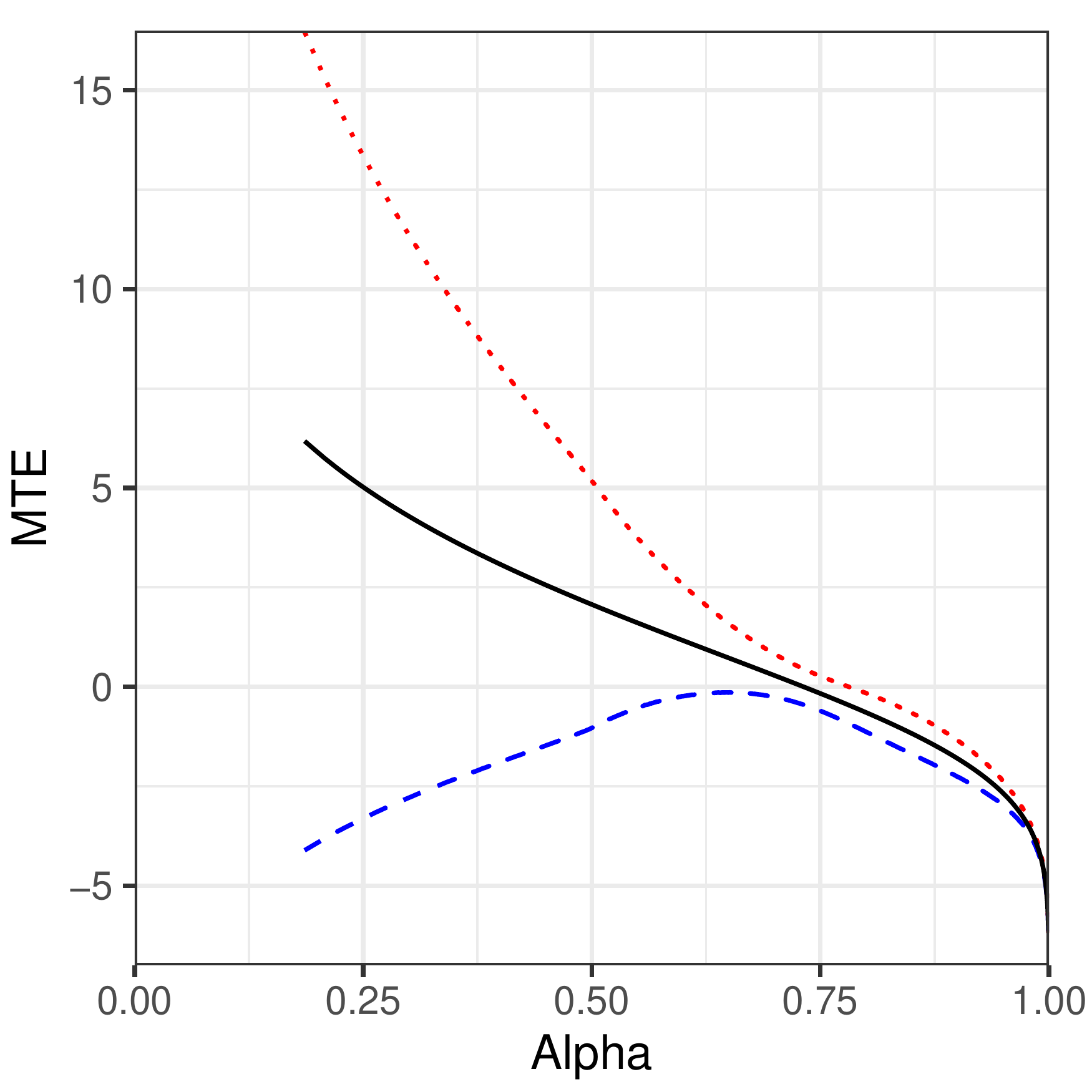}
				\label{fig:bounds2}
			}
		\end{center}
		\footnotesize{Notes: The solid lines are the true values of the $MTE^{OO}$. The red dotted lines and the blue dashed lines are, respectively, the values of the upper and lower bounds around the $MTE^{OO}$ computed by numerical integration using 1,000,000 simulated points for each value of the propensity score.}
		\caption[]{Numerical Bounds based on Proposition \ref{thm1}}
		\label{figure:bounds}
	\end{figure}

	We present the bounds for the $MTE^{OO}$ described in Proposition \ref{THMnoassumption} ($ \underline{\Delta}_{1},  \overline{\Delta}_{1}$) in Figure \ref{figure:boundsNOassumption}, and Proposition \ref{thm1} ($ \underline{\Delta}_{2},  \overline{\Delta}_{2}$) in Figure \ref{figure:bounds} for parameters $\delta_{0} = 0.1$, $\delta_{1} = 0.4$, $\beta_{0,0} = \beta_{0,1} = \beta_{1,0} = 1$ and $\beta_{1,1} = 5$.\footnote{Note that these bounds can be computed using numeric integration. See Appendix \ref{numdetail} for more details.}

	As can be seen on Subfigure \ref{fig:bounds1NOassumption}, the bounds based on Proposition \ref{THMnoassumption} are not very informative for a large part of the support of $V$. In this DGP, when the propensity score is small ($p$ is close to zero), $\upsilon^{\ell}$ --- the lower bound on the proportion of the always-observed $\left(\mathbb{P}\left[\left. S_0 = 1 \right\vert V = p \right] + \mathbb{P}\left[\left. S_{1} = 1 \right\vert V = p \right] - 1\right)$ --- approaches 1 and the $MTE^{OO}$ is almost point-identified as the bounds are close to each other. On the other hand, when $p$ is larger than 0.664, the lower bound on the proportion of the always-observed becomes exactly zero, the $MTE^{OO}$ is not identified and the bounds diverge. Nevertheless, the sign of $MTE^{OO}$ is identified for propensity scores smaller than 0.28.

	Subfigure \ref{fig:bounds2NOassumption} plots the identified interval for $MTE^{OO}$ against $\upsilon^{\ell}$ on the horizontal axis, emphasizing the important role of the lower bound on the proportion of the always-observed. As $\upsilon^{\ell}$ increases, the expectation of the observed outcomes conditional on $V=p$ becomes heavily composed by the always-observed group, leading to point identification of the $MTE^{OO}$ when it reaches one.

	Figure \ref{figure:bounds} plots the $MTE^{OO}$ and its bounds based on Proposition \ref{thm1}, i.e., with the addition of the monotonicity assumption. The bounds presented on Subfigure \ref{fig:bounds1} are in general informative. As discussed above, when the propensity score is small ($p$ is close to zero), the proportion of the always-observed $\left(\alpha\left(p\right)\right)$ approaches 1 and the $MTE^{OO}$ identified set is very tight. When $p$ is close to 1, the proportion of the always-observed decreases and the identified set around the $MTE^{OO}$ expands. The sign of the $MTE^{OO}$ is identified for $p<0.409$, illustrating that Assumption \ref{MONS} allow us to identify the sign of $MTE^{OO}$ in more cases than in Figure \ref{figure:boundsNOassumption}.

	Subfigure \ref{fig:bounds2} plots the same curves with $\alpha(p)$ on the horizontal axis, emphasizing the trimming proportion's importance. As $\alpha(p)\rightarrow 1$ , the observed expectation of the outcomes conditional on $V=p$ is fully composed by the always-observed group, leading to point identification of the $MTE^{OO}$. Moreover, under Assumption \ref{MONS}, $\alpha(p)$ never reaches zero, allowing us to non-trivially bound the $MTE^{OO}$ for all values of the propensity score.

\begin{figure}[htbp]
		\begin{center}
			\subfigure[{Proposition \ref{THMnoassumption}}]{
				\includegraphics[width = 2.8in]{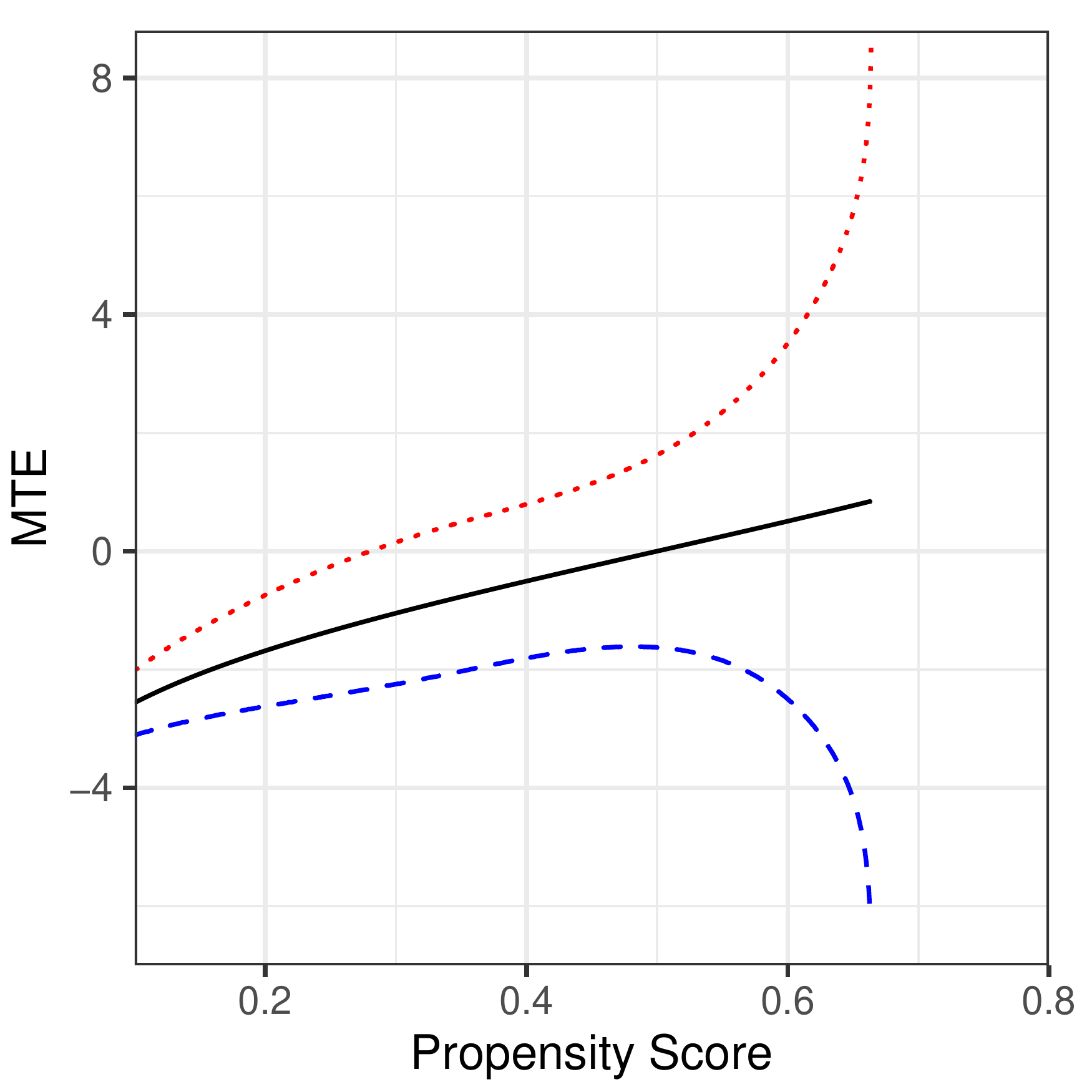}
				\label{zoom1}
			}
			\subfigure[{Proposition \ref{thm1}}]{
				\includegraphics[width = 2.8in]{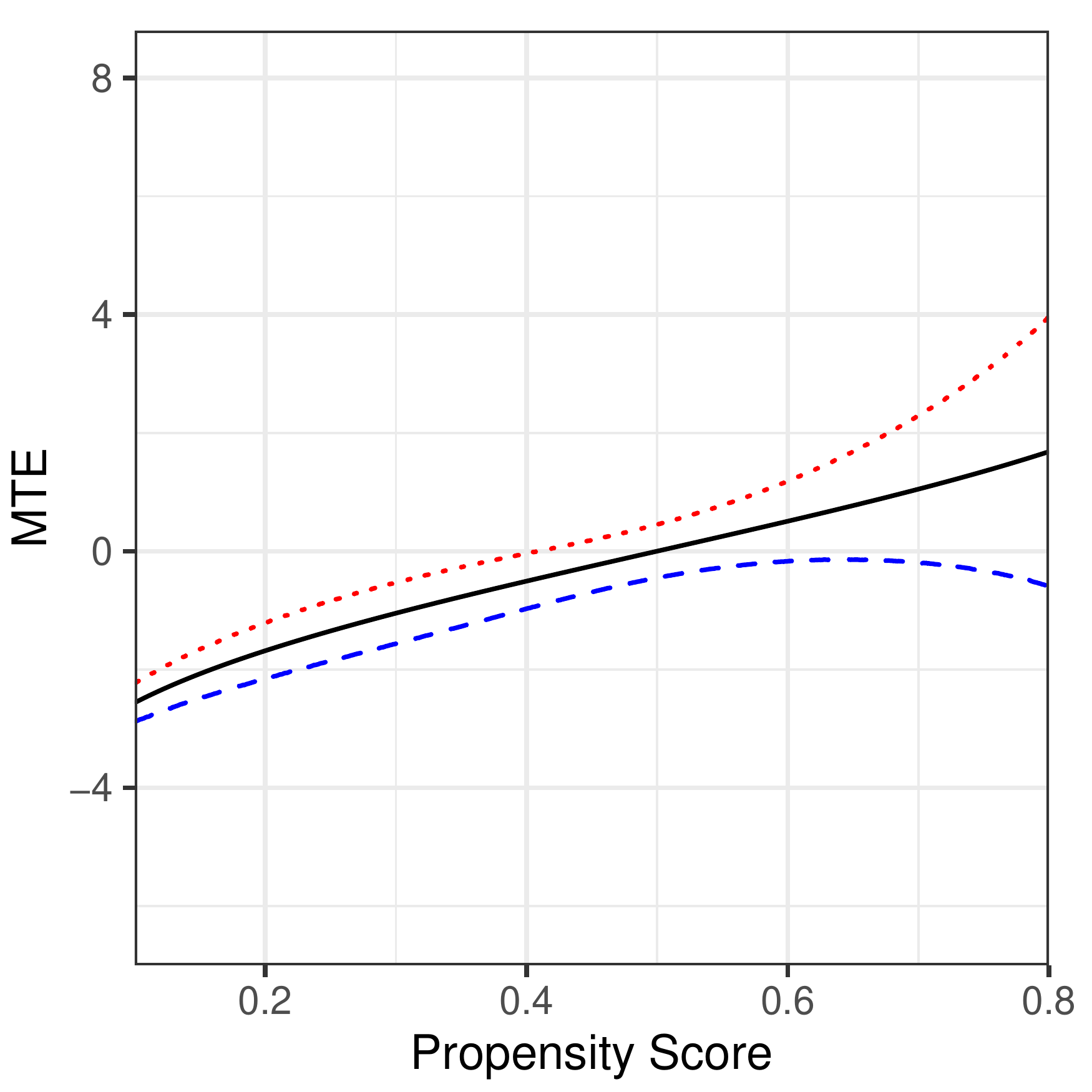}
				\label{zoom2}
			}
		\end{center}
		\footnotesize{Notes: The solid lines are the true values of the $MTE^{OO}$. The red dotted lines and the blue dashed lines are, respectively, the values of the upper and lower bounds around the $MTE^{OO}$ computed by numerical integration using 1,000,000 simulated points for each value of the propensity score.}
		\caption[]{Comparing Proposition \ref{THMnoassumption} and Proposition \ref{thm1}}
		\label{zoom}
	\end{figure}

	Figure \ref{zoom} is a zoomed version of Subfigures \ref{fig:bounds1NOassumption} and \ref{fig:bounds1}. Note that the bounds based on Proposition \ref{thm1} are much tighter than the bounds based on Proposition \ref{THMnoassumption}, especially for larger values of $p$. This is expected as the difference in trimming proportions in Proposition \ref{THMnoassumption} and  Proposition \ref{thm1} increases with the propensity score for this DGP.\footnote{Assumption \ref{DOM} holds with equality in this DGP, implying that the lower bound is equal to the true $MTE^{OO}$ for the case considered in Proposition \ref{thmDOM}.}

\subsection{Understanding the $LIV$ Estimand in the Selected Sample}\label{numselected}

A researcher may believe that sample selection is not a problem and decide to apply the \textit{local instrumental variable} (LIV) estimator \citep{heckman1999} directly to the selected sample ($S = 1$). Similarly to the issues generated by sample selection when trying to identify $ATE$ or $LATE$ parameters \citep{Heckman1979,lee2009training,chen2015bounds}, the LIV estimand applied directly to the selected sample will not identify our target parameter ($MTE^{OO}$). The LIV estimand in this case can be described as:
\begin{align}
LIV_{S=1}\left(p\right) & \coloneqq \dfrac{\partial \mathbb{E}\left[\left. Y \right\vert P\left(Z\right) = p, S = 1\right]}{\partial p} \nonumber \\
& = \dfrac{\partial}{\partial p} \left\lbrace \dfrac{\mathbb{E}\left[\left. Y \cdot S \right\vert P\left(Z\right) = p\right]}{\mathbb{E}\left[\left. S \right\vert P\left(Z\right) = p\right]} \right\rbrace \nonumber \\
& = \dfrac{1}{\mathbb{E}\left[\left. S \right\vert P\left(Z\right) = p\right]} \cdot \dfrac{\partial \mathbb{E}\left[\left. Y \cdot S \right\vert P\left(Z\right) = p\right]}{\partial p} - \dfrac{\mathbb{E}\left[\left. Y \cdot S \right\vert P\left(Z\right) = p\right]}{\left(\mathbb{E}\left[\left. S \right\vert P\left(Z\right) = p\right]\right)^{2}} \cdot \dfrac{\partial \mathbb{E}\left[\left. S \right\vert P\left(Z\right) = p\right]}{\partial p} \nonumber \\
& = \label{EQliv} \dfrac{\mathbb{E}\left[\left. Y_{1} - Y_{0} \right\vert V = p\right]}{\mathbb{E}\left[\left. S \right\vert P\left(Z\right) = p\right]} - \dfrac{\mathbb{E}\left[\left. Y \cdot S \right\vert P\left(Z\right) = p\right]}{\left(\mathbb{E}\left[\left. S \right\vert P\left(Z\right) = p\right]\right)^{2}} \cdot \mathbb{E}\left[\left. S_{1} - S_{0} \right\vert V = p\right]
\end{align}
for any $p \in \mathcal{P}$ such that $\mathbb{E}\left[\left. S \right\vert P\left(Z\right) = p\right] \neq 0$. As Equation \eqref{EQliv} illustrates, the $LIV$ estimand captures a linear combination of the marginal treatment effect on the non-censored outcome variable $\left(\mathbb{E}\left[\left. Y_{1} - Y_{0} \right\vert V = p\right]\right)$ and of the marginal treatment effect on the sample selection $\left(\mathbb{E}\left[\left. S_{1} - S_{0} \right\vert V = p\right]\right)$. Since the weights on this linear combination can be negative, the LIV estimand does not identify an interpretable treatment effect parameter when applied directly to the selected sample. For instance, observe that in the job training example and in the health insurance example (Section \ref{empirical}), the weight on the marginal treatment effect on the sample selection $\left(-\dfrac{\mathbb{E}\left[\left. Y \cdot S \right\vert P\left(Z\right) = p\right]}{\left(\mathbb{E}\left[\left. S \right\vert P\left(Z\right) = p\right]\right)^{2}}\right)$ must be negative, since $Y$ and $S$ are non-negative. Note also that, even when there is no differential sample selection $\left( \mathbb{E}\left[\left. S_{1} - S_{0} \right\vert V = p\right] = 0\right)$, the $LIV$ estimand still does not have a clear interpretation. In this case, it will capture the marginal treatment effect on the non-censored outcome variable magnified by the probability of being selected into the sample.

\begin{figure}[htbp]
	\begin{center}
		\subfigure[{LIV Estimand and the $MTE^{OO}$}]{
			\includegraphics[width = 2.8in]{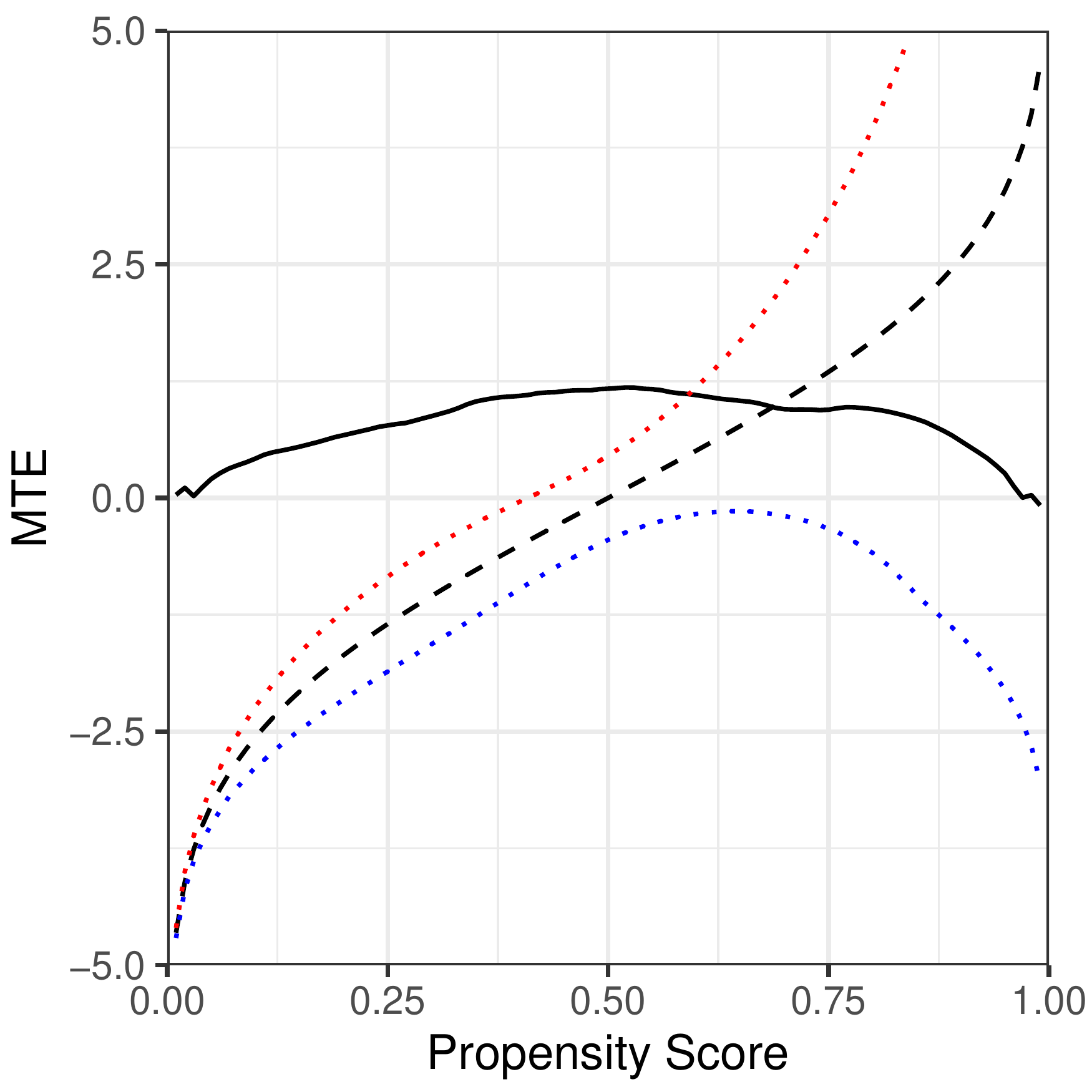}
			\label{LIV1}
		}
		\subfigure[{LIV Estimand's Bias}]{
			\includegraphics[width = 2.8in]{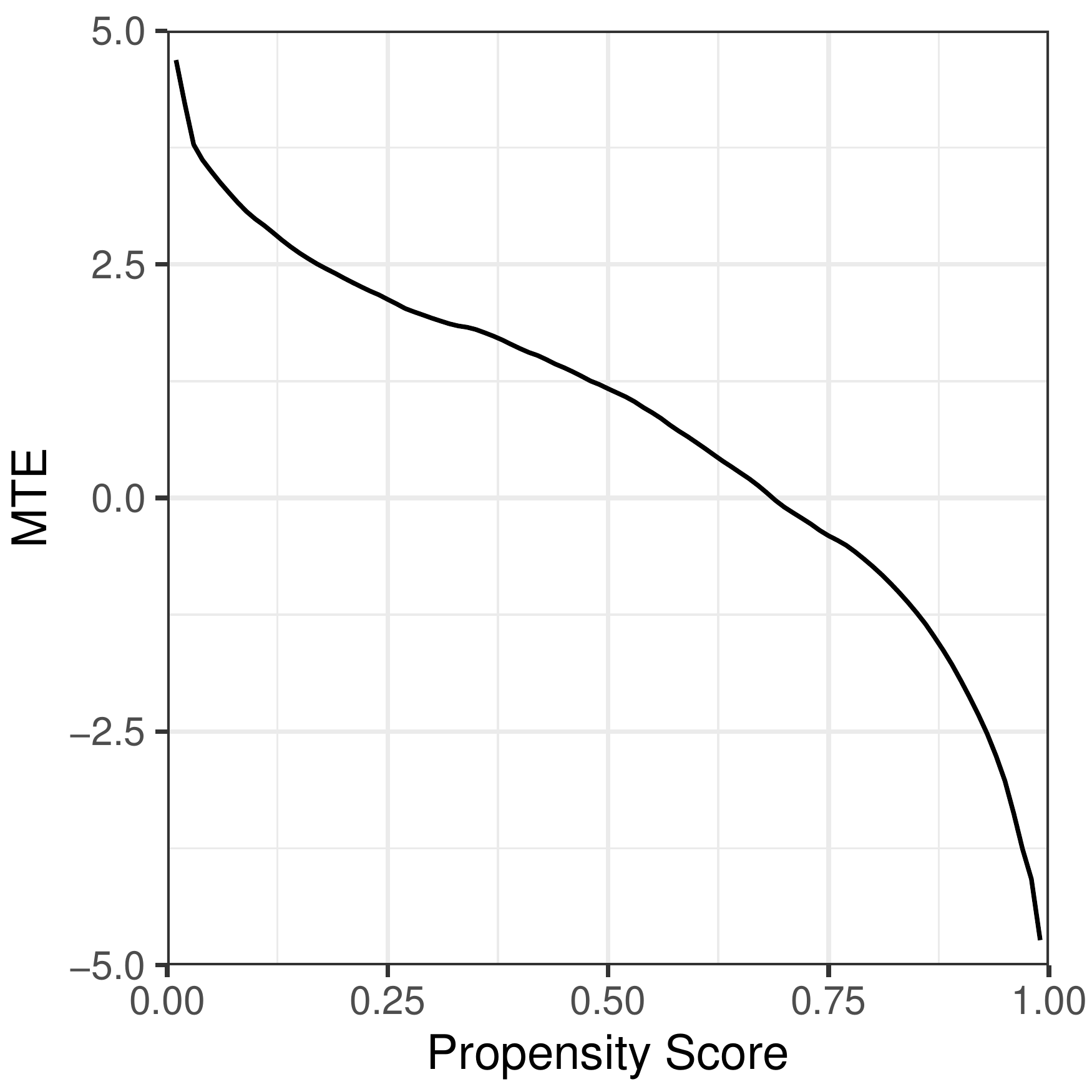}
			\label{LIV2}
		}
	\end{center}
	\footnotesize{Notes: In Subfigure \ref{LIV1}, the solid line is the LIV estimand (Equation \eqref{EQliv}) computed by numerical integration using 600,000 simulated points while the dashed line is the true value of the $MTE^{OO}$. The dotted lines are the upper and lower bounds around the $MTE^{OO}$ from Subfigure \ref{fig:bounds1}. In Subfigure \ref{LIV2}, the solid line is the LIV estimand's bias in comparison with the target parameter ($MTE^{OO}$).}
	\caption[]{Understanding the $LIV$ Estimand in the Selected Sample}
	\label{LIV}
\end{figure}

Figure \ref{LIV} illustrates the differences between the $LIV$ estimand and our target parameter --- marginal treatment effect for the always-observed subpopulation ($MTE^{OO}$). In Subfigure \ref{LIV1}, the dashed line is the $MTE^{OO}$ function explained in Appendix \ref{nummain}, while the solid line is the $LIV$ estimand described in Equation \eqref{EQliv}. Note that the target parameter is monotone and highly heterogeneous, assuming large positive and negative values. However, the $LIV$ estimand is non-monotone and only assumes small positive values, hiding important aspects related to the heterogeneity captured by the $MTE^{OO}$. As a consequence, the $LIV$ estimand's bias $\left(LIV_{S=1}\left(p\right) - MTE^{OO}\left(p\right)\right)$ is large as illustrated in Subfigure \ref{LIV2}. To have a better understanding of the relevance of the $LIV$ estimand's bias, we can compare this estimand against the bounds based on Proposition \ref{thm1}. These bounds are depicted in Subfigure \ref{LIV1} as dotted lines. Note that $LIV$ estimand's bias is so large that this estimand is outside the bounds based on Proposition \ref{thm1} for most values of the propensity score.

Since the $LIV$ estimand applied directly to the selected sample does not capture an intepretable treatment effect parameter and would be biased as a measure for the $MTE^{OO}$, we do not recommend using it when there is sample selection. When faced with selection-into-treatment and sample selection, the researcher should use a methodology that addresses both challenges.

	\subsection{Details on the Numerical Illustration}\label{numdetail}
	This appendix lists the relevant densities, expectations and objects of interest implied by the DGP used in Appendices \ref{Sim} and \ref{mcsimulation}. We have that
	\begin{align*}
		&\mathbb{P}\left[\left. S_0 = 1, S_1 = 1 \right\vert V = p \right] = \Phi \left(\delta_0 \sqrt{2} - \Phi^{-1}\left(p\right) \right), \\
		&\mathbb{P}\left[\left. S_0 = 1 \right\vert V = p \right]  = \Phi \left(\delta_0 \sqrt{2} - \Phi^{-1}\left(p\right) \right), \\
		&\mathbb{P}\left[\left. S_1 = 1 \right\vert V = p \right]  = \Phi \left(\delta_0 \sqrt{2} + \delta_1 \sqrt{2} - \Phi^{-1}\left(p\right) \right), \\
		&\alpha\left(p, \upsilon^{\ell}\right) = \max \left\lbrace 1 + \dfrac{\Phi \left(\delta_0 \sqrt{2} - \Phi^{-1}\left(p\right) \right) - 1}{\Phi \left(\delta_0 \sqrt{2} + \delta_1 \sqrt{2} - \Phi^{-1}\left(p\right) \right)}, 0 \right\rbrace, \\
		&\beta\left(p, \upsilon^{\ell}\right)  = \max \left\lbrace 1 + \dfrac{\Phi \left(\delta_0 \sqrt{2} + \delta_1 \sqrt{2} - \Phi^{-1}\left(p\right) \right) - 1}{\Phi \left(\delta_0 \sqrt{2} - \Phi^{-1}\left(p\right) \right)}, 0 \right\rbrace, \\
		&\alpha\left(p\right) = \dfrac{\Phi \left(\delta_0 \sqrt{2} - \Phi^{-1}\left(p\right) \right)}{\Phi \left(\delta_0 \sqrt{2} + \delta_1 \sqrt{2} - \Phi^{-1}\left(p\right) \right)}, \\
		&E\left[\left.Y^{*}_{1}-Y^{*}_{0} \right\vert T=1, S_0 = 1, S_1 = 1, V = p \right] = (\beta_{1,1} - \beta_{0,1}) \cdot \Phi^{-1}(p), \\
		&E\left[\left.Y^{*}_{1}-Y^{*}_{0} \right\vert T=0, S_0 = 1, S_1 = 1, V = p \right] = (\beta_{0,0} - \beta_{1,0}) \cdot \Phi^{-1}(p), \\
		&E\left[\left.Y^{*}_{1}-Y^{*}_{0} \right\vert S_0 = 1, S_1 = 1, V = p \right] = \left(\beta_{1,1} - \beta_{1,0} - \beta_{0,1} + \beta_{0,0} \right) \cdot \dfrac{\Phi^{-1}(p)}{2}, \\
		&P\left[\left.Y^{*}_{0} \leq y \right\vert T = 1, S_0 = 1, V = p \right] = \Phi\left( y - \beta_{0,1} \Phi^{-1} \left(p\right) \right),  \\
		&P\left[\left.Y^{*}_{1} \leq y \right\vert T = 1, S_1 = 1, V = p \right] = \Phi\left( y - \beta_{1,1} \Phi^{-1} \left(p\right) \right),  \\
		&P\left[\left.Y^{*}_{0} \leq y \right\vert T = 0, S_0 = 1, V = p \right] = \Phi\left( y + \beta_{0,0} \Phi^{-1} \left(p\right) \right), \\
		&P\left[\left.Y^{*}_{1} \leq y \right\vert T = 0, S_1 = 1, V = p \right] = \Phi\left( y + \beta_{1,0} \Phi^{-1} \left(p\right) \right), \\
		&P\left[\left.Y^{*}_{0} \leq y \right\vert S_0 = 1, V = p \right] = \dfrac{1}{2} \Phi\left( y - \beta_{0,1} \Phi^{-1} \left(p\right) \right) + \dfrac{1}{2} \Phi\left( y + \beta_{0,0} \Phi^{-1} \left(p\right) \right), \\
		&P\left[\left.Y^{*}_{1} \leq y \right\vert S_1 = 1, V = p \right]  = \dfrac{1}{2} \Phi\left( y - \beta_{1,1} \Phi^{-1} \left(p\right) \right) + \dfrac{1}{2} \Phi\left( y + \beta_{1,0} \Phi^{-1} \left(p\right) \right).
	\end{align*}


\section{ Two Economic Models Satisfying Assumptions \ref{RA}-\ref{DOM}}\label{econmodel}
\subsection{Model 1: Job Training Program}
To better understand the intuition behind the stochastic dominance assumption, we provide a simple economic model of a worker deciding to enroll in a job training program that satisfies assumptions \ref{RA}-\ref{DOM}.

Let $Y_{0}^{*}$ be the untreated wage whose support satisfy $\mathcal{Y}_{0}^{*} = \left[0, c_{0} \cdot \overline{y} \right)$, where $c_{0} \in \left(1, +\infty\right)$ and $\overline{y} \in \mathbb{R}_{++}$. Let $\tilde{V} \sim F_{\tilde{V}}$ be the training program's effect on wages whose support satisfy $\mathcal{V} = \left[0, \overline{v}\right]$, where $\overline{v} \in \mathbb{R}_{++}$, and whose cumulative distribution function is strictly increasing in its support. To simplify our argument, assume that $\tilde{V} \independent Y_{0}^{*}$. Define $Y_{1}^{*} \coloneqq Y_{0}^{*} + \tilde{V}$ as the treated wage. Let $Z$ be the training program's cash allowance whose support is given by $\mathcal{Z} = \mathbb{R}_{+}$ and whose cumulative distribution function is continuous. To further simplify our argument, assume that the support of the joint distribution of $\tilde{V}$ and $Z$ satisfy $\text{supp}\left(\tilde{V},Z\right) = \mathcal{Z} \times \mathcal{V}$. The worker enrolls in the training program if its total benefits are larger than its non-random and homogeneous cost, $c_{D} \in \mathbb{R}_{++}$. Formally, define $D \coloneqq \mathbbm{1}\left\lbrace Z + \tilde{V} \geq c_{D} \right\rbrace = \mathbbm{1}\left\lbrace F_{-\tilde{V}}\left(Z - c_{D}\right) \geq F_{-\tilde{V}}\left(- \tilde{V}\right) \right\rbrace = \mathbbm{1}\left\lbrace P\left(Z \right) \geq V \right\rbrace$, where $P\left(Z \right) = F_{-\tilde{V}}\left(Z - c_{D}\right)$ and $V = F_{-\tilde{V}}\left(- \tilde{V}\right)$. Moreover, the worker is hired if her productivity is high enough, i.e., $S_{d} \coloneqq \mathbbm{1}\left\lbrace Y_{d}^{*} \geq \overline{y} \right\rbrace$ for any $d \in \left\lbrace 0, 1 \right\rbrace$. Finally, assume that the cash allowance is randomly chosen so that Assumption \ref{RA} holds by design.

Now, we check whether Assumptions \ref{continuity}-\ref{DOM} hold in this economic model. First, observe that Assumption \ref{MONS} holds because $\mathcal{V} = \mathbb{R}_{+}$ and $S_{1} = \mathbbm{1}\left\lbrace Y_{0}^{*} + \tilde{V} \geq \overline{y} \right\rbrace$. Assumption \ref{continuity} holds because $F_{\tilde{V}}$ is a strictly increasing function and $Z$ is a absolutely continuous random variable. For Assumption \ref{positive}, note that $\text{supp}\left(\tilde{V},Z\right) = \left[0, \overline{v}\right] \times \mathbb{R}_{+}$ and $c_{D} > 0$ imply $\mathbb{P}\left[Z + \tilde{V} \geq c_{D}\right] > 0$ and $\mathbb{P}\left[Z + \tilde{V} < c_{D}\right] > 0$, and because $S_{1} \geq S_{0}$, $\mathcal{Y}_{0}^{*} = \left[0, c_{0} \cdot \overline{y} \right)$, $c_{0} > 1$ and $\tilde{V} \independent Y_{0}^{*}$ imply that $\mathbb{P}\left[\left. S_{0} = 1, S_{1} = 1 \right\vert V = p\right] = \mathbb{P}\left[\left. S_{0} = 1 \right\vert V = p\right] = \mathbb{P}\left[\left. Y_{0}^{*} \geq \overline{y} \right\vert V = p\right] = \mathbb{P}\left[Y_{0}^{*} \geq \overline{y} \right] > 0$ for any $p \in \mathcal{P}$. Assumption \ref{finite} holds because $\sup \left\vert \mathcal{Y}_{0}^{*} \right\vert = c_{0} \cdot \overline{y} < + \infty$ and $\sup \left\vert \mathcal{Y}_{1}^{*} \right\vert = c_{0} \cdot \overline{y} + \overline{v} < + \infty$, while Assumption \ref{Dist} holds by definition.

Finally, we discuss whether the stochastic dominance assumption (Assumption \ref{DOM}) holds. Since $\tilde{V} \independent Y_{0}^{*}$, we have, for any $y \in \mathbb{R}_{+}$ and $p \in \mathcal{P}$, that
\begin{align*}
\mathbb{P}\left[\left.Y_{1}^{*} \leq y \right\vert V = p, S_0 = 1, S_1 = 1 \right] & = \prob\left[\left.Y_{1}^{*} \leq y \right\vert F_{-\tilde{V}}\left(-\tilde{V}\right) = p, Y_{0}^{*} \geq \overline{y} \right] \\
& = \mathbb{P}\left[\left.Y_{0}^{*} - F_{-\tilde{V}}^{-1}\left(p\right) \leq y \right\vert F_{-\tilde{V}}\left(-\tilde{V}\right) = p, Y_{0}^{*} \geq \overline{y} \right] \\
& = \mathbb{P}\left[\left.Y_{0}^{*} \leq y + F_{-\tilde{V}}^{-1}\left(p\right) \right\vert Y_{0}^{*} \geq \overline{y} \right]
\end{align*}
and
$$\mathbb{P}\left[\left.Y_{1}^{*} \leq y \right\vert V = p, S_0 = 0, S_1 = 1 \right] = \mathbb{P}\left[\left.Y_{0}^{*} \leq y + F_{-\tilde{V}}^{-1}\left(p\right) \right\vert  \overline{y} + F_{-\tilde{V}}^{-1}\left(p\right) \leq Y_{0}^{*} < \overline{y} \right].$$ As a consequence, we have that
\begin{enumerate}
	\item if $y \leq \overline{y}$, then $\mathbb{P}\left[\left.Y_{1}^{*} \leq y \right\vert V = p, S_0 = 1, S_1 = 1 \right] = 0$ and $\mathbb{P}\left[\left.Y_{1}^{*} \leq y \right\vert V = p, S_0 = 0, S_1 = 1 \right] = 0$;

	\item if $y > \overline{y}$ and $ y + F_{-\tilde{V}}^{-1}\left(p\right) \leq \overline{y}$, then $\mathbb{P}\left[\left.Y_{1}^{*} \leq y \right\vert V = p, S_0 = 1, S_1 = 1 \right] = 0$ and \linebreak $\mathbb{P}\left[\left.Y_{1}^{*} \leq y \right\vert V = p, S_0 = 0, S_1 = 1 \right] \in \left(0, 1 \right]$; and

	\item if $y > \overline{y}$ and $ y + F_{-\tilde{V}}^{-1}\left(p\right) > \overline{y}$, then $\mathbb{P}\left[\left.Y_{1}^{*} \leq y \right\vert V = p, S_0 = 1, S_1 = 1 \right] \in \left(0, 1 \right] $ and \linebreak $\mathbb{P}\left[\left.Y_{1}^{*} \leq y \right\vert V = p, S_0 = 0, S_1 = 1 \right] = 1$.
\end{enumerate}
We can, then, conclude that Assumption \ref{DOM} holds in this simple economic model.

\subsection{Model 2: Managed Care and Ambulatory Expenditures}\label{apx:stoch_dom}
Suppose that an individual's decision to choose a managed care plan (HMO or PPO) is based on a cost-benefit analysis, such that she decides to enroll in a managed care plan if its cost $V$ is less than its benefit $P(Z)$, i.e., $D=\mathbbm{1}\{V < P(Z)\}$. The individual's potential ambulatory expenditure when she enrolls in a managed care plan is $Y^*_1$, and her potential ambulatory expenditure when she enrolls in a fee-for-service plan is $Y^*_0$, where $Y^*_d \geq 0$ with $\mathbb P(Y^*_d =0)>0$ for all $d$. We observe the individual's ambulatory expenditure when $Y^*=Y^*_1 D + Y^*_0 (1-D)$ is positive, i.e., $S=\mathbbm{1}\{Y^* > 0\}$, and the observed ambulatory expenditure is $Y=Y^*S$. Suppose now that $Y^*_1=Y^*_0+\varepsilon$, where $\varepsilon\ \indep\  Y^*_0 \vert V$, and $\varepsilon \vert V=p \sim \mathcal U_{[0,1/p]}$. We are going to show that Assumptions \ref{MONS} and \ref{DOM} hold in this framework. By definition, $S_0=\mathbbm{1}\{Y^*_0 > 0\}$ and $S_1=\mathbbm{1}\{Y^*_1 > 0\}=\mathbbm{1}\{Y^*_0+\varepsilon > 0\}$. We can see that $\{S_0=1\}$ implies $\{S_1=1\}$ since $\varepsilon \geq 0$. Therefore, Assumption \ref{MONS} holds. We now show that $$\mathbb P(Y^*_1 \leq y \vert V=p, S_0=1, S_1=1) \leq \mathbb P(Y^*_1 \leq y \vert V=p, S_0=0, S_1=1).$$
We have
\begin{eqnarray*}
\mathbb P(Y^*_1 \leq y \vert V=p, S_0=0, S_1=1) &=& \mathbb P(Y^*_1 \leq y \vert V=p, Y^*_0=0, Y^*_1>0),\\
&=& \mathbb P(\varepsilon \leq y \vert V=p, \varepsilon>0)\ \text{ since }\  \varepsilon\ \indep\  Y^*_0 \vert V,\\
&=& \left\{ \begin{array}{lcl}
		p.y\ \text{ if }\ y \geq 0\\ \\
		0 \ \text{ if }\ y < 0
	\end{array} \right.
\end{eqnarray*}
and
\begin{eqnarray*}
\mathbb P(Y^*_1 \leq y \vert V=p, S_0=1, S_1=1) &=& \int_{y_0>0}\mathbb P(Y^*_1 \leq y \vert V=p, Y^*_0>0, Y^*_0=y_0, Y^*_1>0) \cdot \\
&& \qquad \qquad \qquad  \qquad \qquad \qquad f_{Y^*_0 \vert V=p, Y^*_0>0, Y^*_1>0}(y_0) dy_0,\\
&=& \int_{y_0>0}\mathbb P(\varepsilon \leq y-y_0 \vert V=p, \varepsilon> - y_0) \cdot \\
&& \qquad \qquad \qquad  f_{Y^*_0 \vert V=p, Y^*_0>0, Y^*_1>0}(y_0) dy_0\ \text{ since }\  \varepsilon\ \indep\  Y^*_0 \vert V,\\
&=& \int_{y_0>0}\mathbb P(\varepsilon \leq y-y_0 \vert V=p) \cdot  f_{Y^*_0 \vert V=p, Y^*_0>0, Y^*_1>0}(y_0) dy_0 \\
\end{eqnarray*}
and
\begin{eqnarray*}
\mathbb P(\varepsilon \leq y-y_0 \vert V=p) &=& \left\{ \begin{array}{lcl}
		p(y-y_0)\ \text{ if }\ y > y_0\\ \\
		0 \ \text{ if }\ y \leq y_0
	\end{array} \right.\\
	&\leq& \left\{ \begin{array}{lcl}
		p.y\ \text{ if }\ y \geq 0 \text{ and } y_0>0\\ \\
		0 \ \text{ if }\ y < 0
	\end{array} \right.
\end{eqnarray*}
Therefore,
\begin{eqnarray*}
\mathbb P(Y^*_1 \leq y \vert V=p, S_0=1, S_1=1)
	&\leq& \left\{ \begin{array}{lcl}
		\int_{y_0>0} p.y f_{Y^*_0 \vert V=p, Y^*_0>0, Y^*_1>0}(y_0) dy_0\ \text{ if }\ y \geq 0\\ \\
		0 \ \text{ if }\ y < 0
	\end{array} \right.\\
	&=& \left\{ \begin{array}{lcl}
		p.y\ \text{ if }\ y \geq 0\\ \\
		0 \ \text{ if }\ y < 0
	\end{array} \right.\\
	&=& \mathbb P(Y^*_1 \leq y \vert V=p, S_0=0, S_1=1).
\end{eqnarray*}
Hence, Assumption \ref{DOM} holds.


\setcounter{table}{0}
\renewcommand\thetable{F.\arabic{table}}

\setcounter{figure}{0}
\renewcommand\thefigure{F.\arabic{figure}}
\section{Estimation Details}\label{App_est}

This appendix presents the details on how to estimate the bounds proposed in Proposition \ref{thm1}, building upon the discussion on Section \ref{Inf} on the main text. For brevity, we focus on the bounds identified under monotonicity of sample selection in the treatment (Assumptions \ref{RA}-\ref{MONS}), as it is the most relevant (and feasible) case empirically. Estimators for the bounds proposed in Propositions \ref{THMnoassumption} and \ref{thmDOM} are natural extensions of the estimator discussed here.

The following subsections (\ref{estpscore}-\ref{estMTE}) present the details on a suggested non(semi)parametric estimation approach that could be implemented with a large and informative dataset.\footnote{Appendix \ref{mcsimulation} presents a Monte Carlo Simulation that evaluates the small sample properties of this estimator.} Naturally, parametric alternatives are available as long as the researcher is willing to specify functional forms for the probabilities described below, imposing some structure to the model. This can be an useful alternative which greatly simplifies estimation, especially when conditioning on covariates. For concreteness, we present the details of the particular parametric estimation procedure implemented in the empirical application (Section \ref{empirical}) at Appendix \ref{empirical_app}.

Recall from Section \ref{Inf} that we need estimates for:
\begin{equation*}
	\tilde{Y}_d| S=1,D=d,P(Z)=p\ \sim F_{\tilde{Y}_d|S=1,D=d,P(Z)=p}(y) = \frac{\frac{\partial \mathbb P\left[Y\leq y, S=1,D=d|P(Z)=p\right]}{\partial p}}{\frac{\partial \mathbb P\left[S=1,D=d|P(Z)=p\right]}{\partial p}}
\end{equation*}
for any $d \in \left\lbrace 0, 1 \right\rbrace$ and
\begin{equation*}
	\alpha\left(p\right) = - \dfrac{\frac{\partial \mathbb P\left[S=1,D=0|P(Z)=p\right]}{\partial p}}{\frac{\partial \mathbb P\left[S=1,D=1|P(Z)=p\right]}{\partial p}}.
\end{equation*}
Consequently, we need to estimate:
\begin{align*}
	\Gamma_{1}\left(p, y\right) & \coloneqq \frac{\partial \mathbb P\left[\left. Y \leq y, S=1,D=1 \right\vert P(Z)=p\right]}{\partial p}, &  \pi_{1}\left(p\right) & \coloneqq \frac{\partial \mathbb P\left[\left. S=1,D=1 \right\vert P(Z)=p\right]}{\partial p}, \\
	\Gamma_{0}\left(p, y\right) & \coloneqq - \frac{\partial \mathbb P\left[\left. Y \leq y, S=1,D=0 \right\vert P(Z)=p\right]}{\partial p}, &  \pi_{0}\left(p\right) & \coloneqq - \frac{\partial \mathbb P\left[\left. S=1,D=0 \right\vert P(Z)=p\right]}{\partial p}.
\end{align*} Furthermore, the estimation of the propensity score $P(Z)$ is necessary to obtain the moments of the conditional distribution of the observed outcome.

\subsection{Estimating the Propensity Score $P(Z)$}\label{estpscore}

The procedures proposed by \cite{carneiro2009estimating} to estimate $P(z)=\mathbbm{P}(D=1|Z=z)$ apply directly, since treatment status $D$ is observed for all individuals, and are summarized here. We suggest to model the probability as a partially linear additive regression model to improve precision of the estimates while avoiding the curse of dimensionality: $\mathbbm{P}\left[D=1|Z=z\right]=z^{pc}\vartheta+\sum_{j=1}^{d}\varphi_{j}(z_{j}^{c}),$ where $z$ is composed by nonparametric ($z^{c}$) and parametric ($z^{pc}$) components, $z^{c}$ is a continuous random vector of dimension $d$, $\vartheta$ is a vector of unknown parameters and $\varphi_{j}(\cdot)$ are unknown functions.

Let $\{p_{\kappa}: \kappa=1,2...\}$ be the basis for smooth functions that we will use to approximate $\varphi_{j}(\cdot)$ more closely as the number of approximating functions increases. For a given $\kappa>0$, define $P_{\kappa}(z)=[z^{pc}, p_{1}(z_{1}^{c}), \dots, p_{\kappa}(z_{1}^{c}),\dots,p_{1}(z_{d}^{c}), \dots, p_{\kappa}(z_{d}^{c})]^{\prime}$. Then, using \cite{carneiro2009estimating} notation, we have that $\tilde{\mathbbm{P}}(Z_{i}) = P_{\kappa}(Z_{i})^{\prime}\hat{\theta}_{\kappa}$, where $\hat{\theta}_{\kappa} = \left[\sum_{i=1}^{n}P_{\kappa}(Z_{i})P_{\kappa}(Z_{i})^{\prime}\right]^{-1}\left[\sum_{i=1}^{n}P_{\kappa}(Z_{i})D_{i}\right]$. As discussed in \cite{carneiro2009estimating}, the estimated probabilities might fall outside of the $[0,1]$ interval in finite samples. As a consequence, it is preferable to use the trimmed version, $\hat{P}_{i}\equiv\hat{\mathbbm{P}}(Z_{i})=\tilde{\mathbbm{P}}(Z_{i})+(1-\lambda-\tilde{\mathbbm{P}}(Z_{i}))\mathbbm{1}(\tilde{\mathbbm{P}}(Z_{i})>1)+(\lambda-\tilde{\mathbbm{P}}(Z_{i}))\mathbbm{1}(\tilde{\mathbbm{P}}(Z_{i})<0),$ for a suitably small positive $\lambda$. Alternatively, a typical conditional probability estimator of $P(\cdot)$ based on a logit or probit model could be used, so that the fitted probability is always  between 0 and 1.

\subsection{Estimating $\pi_{1}\left(p\right)$, $\pi_{0}\left(p\right)$ and $\alpha\left(p\right)$}\label{subPi}

To estimate $\pi_{1}\left(p\right)$ and $\pi_{0}\left(p\right)$, consider the local polynomial estimators \citep{fan1996local}:
\begin{align*}
	\hat{\pi}_{1}(p) & \coloneqq e_{2}\argmin\limits_{c_{0},c_{1},c_{2}}\sum_{i=1}^n\left[S_{i}D_{i}-c_{0}-c_{1}(\hat{P}_{i}-p)-c_{2}(\hat{P}_{i}-p)^{2}\right]^2K\left(\frac{\hat{P}_{i}-p}{h}\right), \\
	\hat{\pi}_{0}(p) & \coloneqq -e_{2}\argmin\limits_{c_{0},c_{1},c_{2}}\sum_{i=1}^n\left[S_{i}(1-D_{i})-c_{0}-c_{1}(\hat{P}_{i}-p)-c_{2}(\hat{P}_{i}-p)^{2}\right]^2K\left(\frac{\hat{P}_{i}-p}{h}\right)
\end{align*}
where $e_{g}$ is a conformable row vector of zeros with $g$-th element equal to one, $K(\cdot)$ is a kernel function and $h$ is a bandwidth.
When implementing these estimators, we recommend using recent developments in \cite{doi:10.1080/01621459.2017.1285776,2018arXiv180801398C} for optimal coverage error bandwidth and kernel selection methods that are nonparametric robust bias-corrected (RBC). We suggest implementing the RBC version of these estimators with optimal bandwidth choice as conveniently implemented in the software \textit{R}, using the package \textit{nprobust} \citep{2019arXiv190600198C}.

To estimate $\alpha\left(p\right)$, we simply take its sample analog: $\hat{\alpha}\left(p\right) \coloneqq \dfrac{\hat{\pi}_{0}\left(p\right)}{\hat{\pi}_{1}\left(p\right)}.$

\subsection{Estimating $\Gamma_{1}\left(p, y\right)$ and $\Gamma_{0}\left(p, y\right)$}\label{subGamma}
In order to estimate $\Gamma_{1}\left(p, y\right)$ and $\Gamma_{0}\left(p, y\right)$, we choose a grid for the outcome variable $\left(\left\lbrace y_{1}, \ldots, y_{K_{n}} \right\rbrace\right)$ and estimate the conditional density of the outcome for each bin in the grid, leading to the following local polynomial regression \citep{fan1996local},
\begin{small}
	\begin{align*}
	\hat{\gamma}_{1}(p,k) & \coloneqq e_{2}\argmin\limits_{c_{0},c_{1},c_{2}}\sum_{i=1}^{n} \left[ \mathbbm{1}\left\{y_{k-1} \leq Y_{i} \leq y_{k}\right\} S_{i}D_{i}-c_{0}-c_{1}(\hat{P}_{i}-p)-c_{2}(\hat{P}_{i}-p)^{2}\right]^{2}K\left(\frac{\hat{P}_{i}-p}{h}\right),\\
	\hat{\gamma}_{0}(p,k) & \coloneqq -e_{2}\argmin\limits_{c_{0},c_{1},c_{2}}\sum_{i=1}^{n} \left[\mathbbm{1}\left\{y_{k-1} \leq Y_{i}\leq y_{k} \right\} S_{i}(1-D_{i})-c_{0}-c_{1}(\hat{P}_{i}-p)-c_{2}(\hat{P}_{i}-p)^{2}\right]^{2} K\left(\frac{\hat{P}_{i}-p}{h}\right)
\end{align*}
\end{small}
for any $k \in \left\{2,\ldots,K_{n} \right\}$. Similarly to the estimation of $\hat{\pi}_{0}$ and $\hat{\pi}_{1}$, we implement the optimal bandwidth selector and RBC procedures mentioned previously.

For simplicity, let $\hat{f}_{d}(p,k)=\frac{\hat{\gamma}_{d}(p,k)}{\hat{\pi}_{d}(p)}$ for any $k \in \left\lbrace 2,\ldots,K_{n} \right\rbrace$. Natural estimators for $F_{1}\left(p, y\right) \coloneqq \frac{\Gamma_{1}\left(p, y\right)}{\pi_{1}(p)}$ and $F_{0}\left(p, y\right) \coloneqq \frac{\Gamma_{0}\left(p, y\right)}{\pi_{0}(p)}$ are given by $\hat{F}_{1}\left(p, y_{k}\right) \coloneqq \sum_{j = 2}^{k} \hat{f}_{1}(p, j)$ and $\hat{F}_{0}\left(p, y_{k}\right) \coloneqq \sum_{j = 2}^{k} \hat{f}_{0}(p, j)$.

The estimation of $\hat{\gamma}_{d}(p, k)$ is a crucial step and can be adversely affected by several features of the population DGP and the available data. For example, these estimators will perform well in situations for which the available data about the observed outcome covers the whole range of possible values of $Y$ for the values of $p$ being considered among both treated and untreated individuals. One can mitigate the challenges to feasibility of the estimator by choosing wider bins $[y_{k - 1} \leq Y_{i} \leq y_{k}]$, at the cost of obtaining a coarse description of the distribution of $Y^{*}_{d}|V=p$. That can be particularly harmful in the region around the trimming points, and should be considered carefully.

\subsection{Estimating $MTE^{OO}(p)$ Bounds}\label{estMTE}
The estimators for the bounds $LB_{2}(p)$ and $UB_{2}(p)$  can be obtained as
\begin{align}
	\widehat{LB}_{2}(p) & \coloneqq \label{estLBapp} \sum_{k = 2}^{K_{n}} \overline{y}_{k} \cdot \mathbbm{1}\left\lbrace \hat{F}_{1}\left(p, y_{k}\right) \leq \hat{\alpha}\left(p\right) \right\rbrace \cdot \dfrac{\hat{f}_{1}(p, k)}{\hat{\alpha}\left(p\right)} \\
	\widehat{UB}_{2}(p) & \coloneqq \label{estUBapp} \sum_{k = 2}^{K_{n}} \overline{y}_{k} \cdot \mathbbm{1}\left\lbrace 1 - \hat{F}_{1}\left(p, y_{k}\right) < \hat{\alpha}\left(p\right) \right\rbrace \cdot \dfrac{\hat{f}_{1}(p, k)}{\hat{\alpha}\left(p\right)},
\end{align}
where $\overline{y}_{k}$ is the center point of each bin $[y_{k-1}, y_{k}]$  for any $k \in \left\lbrace 2,\ldots,K_{n} \right\rbrace$. Moreover, we can estimate $\mathbb E\left[\tilde{Y}_0|S=1,D=0,P(Z)=p\right]$ in Proposition \ref{thm1} using $\hat{\Xi}_{OO, 0}(p) \coloneqq \sum_{k = 2}^{K_{n}} \overline{y}_{k} \cdot \hat{f}_{0}(p, k).$

Note that the estimators for $\widehat{LB}_{2}(p), \widehat{UB}_{2}(p)$ and $\hat{\Xi}_{OO, 0}(p)$ rely on the estimates for densities and trimming points obtained in previous steps, not using on the original data for the observed outcomes.

Naturally, the estimated $MTE^{OO}$ bounds can, then, be obtained by  $\hat{\underline{\Delta}}_{2}\left(p\right) \coloneqq  \widehat{LB}_{2}(p)-\hat{\Xi}_{OO, 0}(p)$ and $\hat{\overline{\Delta}}_{2}\left(p\right) \coloneqq \widehat{UB}_{2}(p)-\hat{\Xi}_{OO, 0}(p)$.

Analyzing the inference procedures and asymptotic properties of the proposed estimators is beyond the scope of this paper and an exciting area for future work.

\setcounter{table}{0}
\renewcommand\thetable{D.\arabic{table}}

\setcounter{figure}{0}
\renewcommand\thefigure{D.\arabic{figure}}
\section{Monte Carlo Simulation}\label{mcsimulation}

In this appendix, we use the DGP described in Appendix \ref{Sim} to  produce Monte Carlo simulations using the estimator proposed in the main text. We analyze two sets of parameters: (i) $\delta_{0} = 0.75$, $\delta_{1} = 1.5$, $\beta_{00} = \beta_{01} = \beta_{10} = 0.1$, $\beta_{11} = 0.2$ and (ii) $\delta_{0} = 0.2$, $\delta_{1} = 2.0$, $\beta_{00} = \beta_{01} = \beta_{10} = 0.1$, $\beta_{11} = 0.2$. The first set of parameters is ideal for the proposed estimator in the sense that the effective sample size is large since the trimming proportion $\alpha\left(p\right)$ is never small and the sample selection problem is not severe. The second set of parameters intentionally decreases the trimming proportion $\alpha\left(p\right)$, reducing the effective sample size and worsening the sample selection problem. We find that our estimator performs adequately in both DGPs when the sample size is equal to $n = 10,000$.

Based on the nonparametric estimation procedure described in the Appendix \ref{App_est}, we need to specify the propensity score estimator, the grid points for the observed outcome variable $\left(\left\lbrace y_{1}, \ldots, y_{K_{n}} \right\rbrace\right)$ and the evaluation points for the unobserved characteristic $V$. We estimate the propensity score with a logit estimator whose index is linear in the instrument $Z$, implying that the propensity score estimator is misspecified. For the grid points $\left(\left\lbrace y_{1}, \ldots, y_{K_{n}} \right\rbrace\right)$, we choose the sample percentiles 0.0, 0.1, 0.2, 0.3, 0.4, 0.5, 0.6, 0.7, 0.8, 0.9, 1.0, implying that $K_{n} = 11$. For the evaluation points of the the unobserved characteristic $V$, we choose $p \in \left\lbrace 0.1, 0.2, 0.3, 0.4, 0.5, 0.6, 0.7, 0.8, 0.9 \right\rbrace$.

In this simulation, we focus on the performance of six estimators: $\hat{\alpha}\left(p\right)$, $\hat{\Xi}_{OO,0}\left(p\right)$, $\widehat{LB}_{2}\left(p\right)$, $\widehat{UB}_{2}\left(p\right)$, $\hat{\underline{\Delta}}_2\left(p\right)$ and $\hat{\overline{\Delta}}_{2}\left(p\right)$. Table \ref{true} reports the true value of their estimands for the first and second sets of parameters in Panel A and B, respectively. The important distinction between Panels A and B is the value of $\alpha\left(p\right)$.

\begin{table}[p]
	\centering
	\captionsetup{justification=centering}
	\rotatebox{90}{\begin{varwidth}{\textheight}\centering

			\parbox{16cm}{\caption{True Value of the Estimands \label{true}}}
			\begin{lrbox}{\tablebox}
				\begin{tabular}{lccccccccc}
					\hline
					\hline

					& p = 0.1 & p = 0.2 & p = 0.3 & p = 0.4 & p = 0.5 & p = 0.6 & p = 0.7 & p = 0.8 & p = 0.9 \\
					& (1) &  (2) &  (3) & (4) & (5) & (6) & (7) & (8) & (9) \\

					\hline

					& \multicolumn{9}{c}{Panel A: $\delta_{0} = 0.75$, $\delta_{1} = 1.5$, $\beta_{00} = \beta_{01} = \beta_{10} = 0.1$, $\beta_{11} = 0.2$} \\

					$\alpha\left(p\right)$ & 0.99 & 0.97 & 0.94 & 0.91 & 0.86 & 0.79 & 0.71 & 0.59 & 0.42 \\
					$\Xi_{OO,0}\left(p\right)$ & 0.00 & 0.00 & 0.00 & 0.00 & 0.00 & 0.00 & 0.00 & 0.00 & 0.00 \\
					$LB_{1}\left(p\right)$ & -0.09 & -0.11 & -0.15 & -0.2 & -0.27 & -0.35 & -0.46 & -0.62 & -0.88 \\
					$UB_{1}\left(p\right)$ & -0.04 & 0.03 & 0.10 & 0.17 & 0.26 & 0.37 & 0.51 & 0.70 & 1.00 \\
					$\underline{\Delta}_2\left(p\right)$ & -0.09 & -0.11 & -0.15 & -0.2 & -0.27 & -0.35 & -0.46 & -0.62 & -0.88 \\
					$\overline{\Delta}_{2}\left(p\right)$ & -0.04 & 0.03 & 0.10 & 0.17 & 0.26 & 0.37 & 0.51 & 0.70 & 1.00 \\
					$MTE^{OO}\left(p\right)$ & -0.06 & -0.04 & -0.03 & -0.01 & 0.00 & 0.01 & 0.03 & 0.04 & 0.06 \\

					\\

					& \multicolumn{9}{c}{Panel B: $\delta_{0} = 0.2$, $\delta_{1} = 2.0$, $\beta_{00} = \beta_{01} = \beta_{10} = 0.1$, $\beta_{11} = 0.2$} \\

					$\alpha\left(p\right)$ & 0.94 & 0.87 & 0.79 & 0.7 & 0.61 & 0.51 & 0.41 & 0.29 & 0.16 \\
					$\Xi_{OO,0}\left(p\right)$ & 0.00 & 0.00 & 0.00 & 0.00 & 0.00 & 0.00 & 0.00 & 0.00 & 0.00 \\
					$LB_{1}\left(p\right)$ & -0.19 & -0.29 & -0.39 & -0.5 & -0.63 & -0.77 & -0.93 & -1.14 & -1.47 \\
					$UB_{1}\left(p\right)$ & 0.06 & 0.2 & 0.34 & 0.48 & 0.63 & 0.79 & 0.98 & 1.23 & 1.60 \\
					$\underline{\Delta}_2\left(p\right)$ & -0.19 & -0.29 & -0.39 & -0.5 & -0.63 & -0.77 & -0.93 & -1.14 & -1.47 \\
					$\overline{\Delta}_{2}\left(p\right)$ & 0.06 & 0.2 & 0.34 & 0.48 & 0.63 & 0.79 & 0.98 & 1.23 & 1.60 \\
					$MTE^{OO}\left(p\right)$ & -0.06 & -0.04 & -0.03 & -0.01 & 0.00 & 0.01 & 0.03 & 0.04 & 0.06 \\

					\hline

				\end{tabular}
			\end{lrbox}
			\usebox{\tablebox}\\
			\settowidth{\tableboxwidth}{\usebox{\tablebox}} \parbox{\tableboxwidth}{\footnotesize{Note: We define $\Xi_{OO,0}\left(p\right) \coloneqq \mathbb{E}\left[\left. Y_{0}^{*} \right\vert S_{0} = 1, S_{1} = 1, V = p\right]$. The true values of the estimands are computed by numerical integration using 100,000 simulated points for each value of the propensity score.}
			}
	\end{varwidth}}
\end{table}

Table \ref{tableAB} reports the average bias, while Table \ref{tableASB} presents the mean squared error (MSE) after normalizing it by the sample size $\left(n = 10,000\right)$. For the first set of parameters (Panel A), the estimators' average bias and MSE is smaller for intermediate values of the propensity score. In this DGP, the treatment is determined by $D = \mathbbm{1}\left\{V\leq \Phi(Z)\right\}$, implying that the data becomes sparser at low (high) values of the propensity score for the (un)treated group. As a consequence, performance is worse when the propensity score is either small or large. Moreover, when the propensity score is large, the sample selection problem reduces the effective sample size, worsening the estimator's performance. Despite those challenges, the estimator's average bias and MSE are reasonably small for both set of parameters.

\begin{table}[p]
	\centering
	\captionsetup{justification=centering}
	\rotatebox{90}{\begin{varwidth}{\textheight}\centering

			\parbox{16cm}{\caption{Average Bias: $\mathbb{E}\left[\hat{\theta}\right] - \theta_{0}$ \label{tableAB}}}
			\begin{lrbox}{\tablebox}
				\begin{tabular}{lccccccccc}
					\hline
					\hline

					& p = 0.1 & p = 0.2 & p = 0.3 & p = 0.4 & p = 0.5 & p = 0.6 & p = 0.7 & p = 0.8 & p = 0.9 \\
					& (1) &  (2) &  (3) & (4) & (5) & (6) & (7) & (8) & (9) \\

					\hline

					& \multicolumn{9}{c}{Panel A: $\delta_{0} = 0.75$, $\delta_{1} = 1.5$, $\beta_{00} = \beta_{01} = \beta_{10} = 0.1$, $\beta_{11} = 0.2$} \\

					$\hat{\alpha}\left(p\right)$ & -0.02 & 0.01 & 0.01 & -0.01 & -0.01 & -0.02 & -0.02 & 0.00 & 0.01  \\
					& (0.07) & (0.03) & (0.03) & (0.03) & (0.03) & (0.02) & (0.02) & (0.03) & (0.05) \\
					$\hat{\Xi}_{OO,0}\left(p\right)$ & -0.02 & -0.02 & -0.01 & 0 & -0.01 & -0.01 & -0.01 & -0.01 & -0.01 \\
					& (0.32) & (0.14) & (0.08) & (0.07) & (0.06) & (0.06) & (0.07) & (0.08) & (0.18) \\
					$\widehat{LB}_{1}\left(p\right)$ & 0.04 & 0.06 & -0.07 & -0.10 & -0.12 & -0.15 & -0.16 & -0.14 & -0.07 \\
					& (0.22) & (0.15) & (0.08) & (0.09) & (0.07) & (0.08) & (0.09) & (0.17) & (0.41) \\
					$\widehat{UB}_{1}\left(p\right)$ & 0.14 & 0.04 & -0.02 & 0.04 & 0.09 & 0.13 & 0.14 & 0.13 & 0.15 \\
					& (0.18) & (0.08) & (0.09) & (0.16) & (0.10) & (0.10) & (0.11) & (0.19) & (0.52) \\
					$\hat{\underline{\Delta}}_2\left(p\right)$ & 0.07 & 0.08 & -0.07 & -0.09 & -0.11 & -0.14 & -0.15 & -0.14 & -0.06 \\
					& (0.38) & (0.2) & (0.11) & (0.11) & (0.10) & (0.10) & (0.11) & (0.19) & (0.45) \\
					$\hat{\overline{\Delta}}_{2}\left(p\right)$ & 0.17 & 0.05 & -0.01 & 0.04 & 0.10 & 0.14 & 0.15 & 0.14 & 0.16 \\
					& (0.36) & (0.16) & (0.12) & (0.18) & (0.12) & (0.12) & (0.13) & (0.21) & (0.54) \\

					\\

					& \multicolumn{9}{c}{Panel B: $\delta_{0} = 0.2$, $\delta_{1} = 2.0$, $\beta_{00} = \beta_{01} = \beta_{10} = 0.1$, $\beta_{11} = 0.2$} \\

					$\hat{\alpha}\left(p\right)$ & 0.00 & 0.04 & 0.01 & -0.01 & -0.01 & -0.01 & -0.01 & 0.01 & 0.02 \\
					& (0.09) & (0.05) & (0.03) & (0.03) & (0.03) & (0.02) & (0.02) & (0.02) & (0.03) \\
					$\hat{\Xi}_{OO,0}\left(p\right)$ & 0.01 & 0.00 & 0.00 & -0.01 & -0.01 & -0.01 & -0.01 & -0.01 & 0.00 \\
					& (0.29) & (0.13) & (0.07) & (0.07) & (0.07) & (0.08) & (0.09) & (0.11) & (0.26) \\
					$\widehat{LB}_{1}\left(p\right)$ & 0.07 & 0.02 & -0.06 & -0.09 & -0.11 & -0.14 & -0.15 & -0.12 & 0.48 \\
					& (0.25) & (0.13) & (0.08) & (0.08) & (0.09) & (0.09) & (0.11) & (0.25) & (0.76) \\
					$\widehat{UB}_{1}\left(p\right)$ & 0.08 & -0.02 & 0.11 & 0.15 & 0.18 & 0.22 & 0.26 & 0.35 & 0.67 \\
					& (0.22) & (0.17) & (0.10) & (0.09) & (0.09) & (0.10) & (0.14) & (0.28) & (1.01) \\
					$\hat{\underline{\Delta}}_2\left(p\right)$ & 0.06 & 0.02 & -0.06 & -0.09 & -0.10 & -0.13 & -0.14 & -0.11 & 0.48 \\
					& (0.39) & (0.18) & (0.11) & (0.10) & (0.11) & (0.11) & (0.13) & (0.26) & (0.81) \\
					$\hat{\overline{\Delta}}_{2}\left(p\right)$ & 0.07 & -0.03 & 0.12 & 0.16 & 0.19 & 0.23 & 0.28 & 0.36 & 0.67 \\
					& (0.36) & (0.2) & (0.12) & (0.11) & (0.12) & (0.12) & (0.16) & (0.3) & (1.04) \\

					\hline

				\end{tabular}
			\end{lrbox}
			\usebox{\tablebox}\\
			\settowidth{\tableboxwidth}{\usebox{\tablebox}} \parbox{\tableboxwidth}{\footnotesize{Note: We define $\theta_{0}$ as the true population value of the estimand, $\hat{\theta}$ as the estimator of $\theta_{0}$ and $\Xi_{OO,0}\left(p\right) \coloneqq \mathbb{E}\left[\left. Y_{0}^{*} \right\vert S_{0} = 1, S_{1} = 1, V = p\right]$. The results are based on 1,000 Monte Carlo repetitions.}
			}
	\end{varwidth}}
\end{table}

\begin{table}[p]
	\centering
	\captionsetup{justification=centering}
	\rotatebox{90}{\begin{varwidth}{\textheight}\centering

			\parbox{16cm}{\caption{Mean Squared Error: $\mathbb{E}\left[n \cdot \left(\hat{\theta} - \theta_{0}\right)^{2}\right]$ \label{tableASB}}}
			\begin{lrbox}{\tablebox}
				\begin{tabular}{lccccccccc}
					\hline
					\hline

					& p = 0.1 & p = 0.2 & p = 0.3 & p = 0.4 & p = 0.5 & p = 0.6 & p = 0.7 & p = 0.8 & p = 0.9 \\
					& (1) &  (2) &  (3) & (4) & (5) & (6) & (7) & (8) & (9) \\

					\hline

					& \multicolumn{9}{c}{Panel A: $\delta_{0} = 0.75$, $\delta_{1} = 1.5$, $\beta_{00} = \beta_{01} = \beta_{10} = 0.1$, $\beta_{11} = 0.2$} \\

					$\hat{\alpha}\left(p\right)$ & 46.54 & 10.07 & 8.72 & 7.02 & 7.69 & 8.89 & 9.52 & 7.39 & 27.1 \\
					& (153.91) & (17.2) & (11.09) & (9.97) & (10.81) & (10.62) & (11.11) & (10.32) & (41.41) \\
					$\hat{\Xi}_{OO,0}\left(p\right)$ & 1005.74 & 185.55 & 57.02 & 44.87 & 41.4 & 40.3 & 43.53 & 60.12 & 311.93 \\
					& (1351.09) & (259.3) & (83.15) & (61.98) & (61.76) & (60.61) & (60.6) & (79.85) & (479.29) \\
					$\widehat{LB}_{1}\left(p\right)$ & 500.89 & 275.72 & 121.81 & 168.21 & 194.74 & 283.95 & 343.43 & 502.21 & 1723.3 \\
					& (673.6) & (221.19) & (155.71) & (193.96) & (182.51) & (241.64) & (312.75) & (651.88) & (2463.16) \\
					$\widehat{UB}_{1}\left(p\right)$ & 536.17 & 81.04 & 91.79 & 282.78 & 187.07 & 259.52 & 321.22 & 548.54 & 2956.32 \\
					& (1047.25) & (195.52) & (170.81) & (211.73) & (250.02) & (248.53) & (321.19) & (694.02) & (4377.82) \\
					$\hat{\underline{\Delta}}_2\left(p\right)$ & 1517.37 & 463.07 & 157.49 & 203.19 & 219.32 & 292.6 & 347.73 & 533.7 & 2075.69 \\
					& (2219.03) & (557.22) & (208.91) & (258.66) & (244.93) & (302.02) & (377.33) & (684.14) & (2894.41) \\
					$\hat{\overline{\Delta}}_{2}\left(p\right)$ & 1539.34 & 289.92 & 141.24 & 328.43 & 238.21 & 321.69 & 392.97 & 635.88 & 3180.7 \\
					& (2464.02) & (507.75) & (242.44) & (311.5) & (318.06) & (335.83) & (411.57) & (830.93) & (4596.81) \\

					\\

					& \multicolumn{9}{c}{Panel B: $\delta_{0} = 0.2$, $\delta_{1} = 2.0$, $\beta_{00} = \beta_{01} = \beta_{10} = 0.1$, $\beta_{11} = 0.2$} \\

					$\hat{\alpha}\left(p\right)$ & 87.38 & 47.18 & 10.25 & 8.52 & 7.62 & 5.71 & 4.27 & 4.73 & 14.29 \\
					& (203.68) & (53.35) & (15) & (12.27) & (11.64) & (8.22) & (6.3) & (6.59) & (20.35) \\
					$\hat{\Xi}_{OO,0}\left(p\right)$ & 832.73 & 157.01 & 55.6 & 48.6 & 54.85 & 59.32 & 74.1 & 115.4 & 666.71 \\
					& (1192.08) & (226.59) & (81.12) & (71.15) & (80.75) & (86.91) & (119.01) & (171.63) & (979.21) \\
					$\widehat{LB}_{1}\left(p\right)$ & 667.96 & 177.78 & 96.89 & 147.49 & 200.67 & 264.66 & 350.65 & 761.64 & 8061.64 \\
					& (740.58) & (332.85) & (123.89) & (170.56) & (217.53) & (269.91) & (416.65) & (1136.95) & (8254.23) \\
					$\widehat{UB}_{1}\left(p\right)$ & 558.54 & 279.09 & 223.32 & 315.69 & 413.11 & 579.47 & 878.94 & 2019.95 & 14733.92 \\
					& (1225.8) & (242.89) & (242.07) & (288.42) & (366.06) & (490.23) & (779.26) & (2477.32) & (34135.5) \\
					$\hat{\underline{\Delta}}_2\left(p\right)$ & 1538.87 & 326.63 & 150.56 & 181.28 & 229.65 & 285.19 & 370.49 & 818 & 8804.48 \\
					& (2203.9) & (504.4) & (213.22) & (231.91) & (286.43) & (347.38) & (478.26) & (1159.35) & (9722.85) \\
					$\hat{\overline{\Delta}}_{2}\left(p\right)$ & 1337.34 & 409.26 & 281.66 & 379.82 & 491.03 & 674.47 & 1008.06 & 2186 & 15299.42 \\
					& (2259.27) & (517.54) & (331.17) & (387.45) & (476.41) & (620.89) & (972.28) & (2637.36) & (34933.81) \\

					\hline

				\end{tabular}
			\end{lrbox}
			\usebox{\tablebox}\\
			\settowidth{\tableboxwidth}{\usebox{\tablebox}} \parbox{\tableboxwidth}{\footnotesize{Note: We define $\theta_{0}$ as the true population value of the estimand, $\hat{\theta}$ as the estimator of $\theta_{0}$ and $\Xi_{OO,0}\left(p\right) \coloneqq \mathbb{E}\left[\left. Y_{0}^{*} \right\vert S_{0} = 1, S_{1} = 1, V = p\right]$. The results are based on 1,000 Monte Carlo repetitions. The sample size in each generate data set is $n = 10,000$.}
			}
	\end{varwidth}}
\end{table}

\pagebreak

\section{Empirical Illustration: Additional Details and Results}\label{empirical_app}
In this appendix, we present the parametric specification and estimation details for the results in Section \ref{empirical}.

As mentioned in the main text, to properly account for covariates, we calculate bounds for $MTE^{OO}\left(p, x\right)$ based on parametric estimates for the functions $\mathbb{P}\left[\left.D=1\right\vert X = x, Z = z\right]$, $\mathbb{P}\left[\left. S = 1, D = d \right\vert X = x, P\left(Z\right) = p \right]$, and $\mathbb{P}\left[\left. y_{k-1} \leq Y < y_{k}, S = 1, D = d \right\vert X = x, P\left(Z\right) = p \right]$ for $d=\{0,1\}$, $p \in \left[0, 1 \right]$, $k \in \left\lbrace 1,\ldots,K \right\rbrace$ and covariate value $x$ observed in the sample.

In particular, the probabilities above are modeled as logit functions that depend on a linear index of the covariates, instruments (for the propensity score), and a quadratic function of the propensity score (for the last two equations), while using $K=20$ grid points ($y_{k}$) for the outcome variable in the last case. Let $\Lambda(\cdot)$ and $\lambda(\cdot)$ be the logistic distribution's CDF and PDF respectively, then:
\begin{align*}
&\mathbb{P}\left[\left.D=1\right\vert X=x,Z=z\right]=\Lambda(x\psi_{X}^{d}+z\psi_{Z})\\
&\mathbb{P}\left[\left. S = 1, D = d \right\vert X = x, P\left(Z\right) = p \right]=\Lambda(\delta_{1}^{d}p+\delta_{2}^{d}p^{2}+x\delta_{X}^{d}),\\
&\mathbb{P}\left[\left. y_{k-1} \leq Y < y_{k}, S = 1, D = d \right\vert X = x, P\left(Z\right) = p \right]=\Lambda(\theta_{1}^{d}p+\theta_{2}^{d}p^{2}+x\theta_{X}^{d}).
\end{align*}
The coefficients in the index function can then be estimated using a standard logit estimator, where the estimated $\hat{p}_{i}=\Lambda(x_{i}\hat{\psi}_{X}^{d}+z_{i}\hat{\psi}_{Z})$ replace $p_{i}$ in the estimation for the last two terms. To enforce the common support assumption, we trim the bottom 1\% and top 1\% of the overlapping estimated propensity score distribution.

Hence, the estimates for the partial derivatives of these functions, $\pi_{d}(p,x)$ and $\gamma_{d}(p,x,k)$, can be obtained by plugging in the relevant estimates:
\begin{align*}
	\hat{\pi}_{d}(p,x)=\lambda(\hat{\delta}_{1}^{d}p+\hat{\delta}_{2}^{d}p^{2}+x\hat{\delta}_{X}^{d})(\hat{\delta}_{1}^{d}+2\hat{\delta}_{2}^{d}p)(-1)^{1-d},\\
	\hat{\gamma}_{d}(p,x,k)=\lambda(\hat{\theta}_{1}^{d}p+\hat{\theta}_{2}^{d}p^{2}+x\hat{\theta}_{X}^{d})(\hat{\theta}_{1}^{d}+2\hat{\theta}_{2}^{d}p)(-1)^{1-d}.
\end{align*}

The values for $\alpha(p,x)$ and $\beta(p,x)$ depend on the assumptions being considered. For the case in which the sample selection mechanism is unrestricted we can obtain $\alpha(p,\upsilon^{\ell},x)$ and $\beta(p,\upsilon^{\ell}, x)$ by plugging in $\hat{\pi}_{d}(p,x)$ in the formulas in Lemma \ref{BFbounds}. In the leading case of ``monotonicity of sample selection in the treatment'' (Assumptions \ref{RA}-\ref{MONS}), $\hat{\alpha}\left(p, x\right) =\dfrac{\hat{\pi}_{0}\left(p,x\right)}{\hat{\pi}_{1}\left(p,x\right)}$.

 After estimating $\alpha\left(p, x\right)$, $\beta\left(p, x\right)$,
 the bounds around $MTE^{OO}\left(p, x\right)$ for each covariate value $x$, can be obtained. Below we focus on the ``monotonicity of sample selection in the treatment'' case, but the other situations can be handled analogously.
 For simplicity, let $\hat{f}_{d}(p,x,k)=\frac{\hat{\gamma}_{d}(p,x,k)}{\hat{\pi}_{d}(p,x)}$. Then, estimates of the bounds in Proposition \ref{thm1} are given by,
 \begin{align}
 	\widehat{LB}_{2}(p,x) & \coloneqq \sum_{k = 2}^{K_{n}} \overline{y}_{k} \cdot \mathbbm{1}\left\lbrace \hat{F}_{1}\left(p,x,k\right) \leq \hat{\alpha}\left(p,x\right) \right\rbrace \cdot \dfrac{\hat{f}_{1}(p,x,k)}{\hat{\alpha}\left(p,x\right)} \\
 	\widehat{UB}_{2}(p,x) & \coloneqq \sum_{k = 2}^{K_{n}} \overline{y}_{k} \cdot \mathbbm{1}\left\lbrace 1 - \hat{F}_{1}\left(p,x,k\right) < \hat{\alpha}\left(p,x\right) \right\rbrace \cdot \dfrac{\hat{f}_{1}(p,x,k)}{\hat{\alpha}\left(p,x\right)},
 \end{align}
 where $\overline{y}_{k}$ is the center point of each bin $[y_{k-1}, y_{k}]$  for any $k \in \left\lbrace 2,\ldots,K_{n} \right\rbrace$, and $\hat{F}_{1}\left(p,x,k\right) \coloneqq \sum_{j = 2}^{k} \hat{f}_{1}(p,x,j)$ and $\hat{F}_{0}\left(p,x,k\right) \coloneqq \sum_{j = 2}^{k} \hat{f}_{0}(p,x,j)$ for any $k \in \left\lbrace 2,\ldots,K_{n} \right\rbrace$. Moreover, we can estimate $\mathbb E\left[\tilde{Y}_0|S=1,D=0,X=x,P(Z)=p\right]$ in Proposition \ref{thm1} using $\hat{\Xi}_{OO, 0}(p,x) \coloneqq \sum_{k = 2}^{K_{n}} \overline{y}_{k} \cdot \hat{f}_{0}(p,x,k)$.

 Finally, the $MTE^{OO}(p,x)$ estimated bounds are given by $\hat{\underline{\Delta}}_{2}\left(p\right) \coloneqq  \widehat{LB}_{2}(p)-\hat{\Xi}_{OO, 0}(p)$ and $\hat{\overline{\Delta}}_{2}\left(p\right) \coloneqq \widehat{UB}_{2}(p)-\hat{\Xi}_{OO, 0}(p)$.

  To summarize the treatment effects, we average the bounds for $MTE^{OO}\left(p, x\right)$ across the sample using observed covariates values. By averaging with respect to the observed density of the covariates, we compute bounds around the summary measure of the conditional $MTE$ for the always-observed subgroup: $$SCMTE^{OO}\left(p\right) \coloneqq \int \mathbb{E}\left[\left. Y_{1}^{*} - Y_{0}^{*} \right\vert V = p, S_{0} = 1, S_{1} = 1, X = x^{\prime} \right] \, \text{d}F_{X}\left(x^{\prime}\right).$$ If $X \indep \left(V = p, S_{0} = 1, S_{1} = 1\right)$ holds (i.e., the covariates are exogenous), then
  \begin{align*}
  SCMTE^{OO}\left(p\right) & = \int \mathbb{E}\left[\left. Y_{1}^{*} - Y_{0}^{*} \right\vert V = p, S_{0} = 1, S_{1} = 1, X = x^{\prime} \right] \, \text{d}F_{\left. X \right\vert V = p, S_{0} = 1, S_{1} = 1}\left(x^{\prime}\right) \\
  & = \mathbb{E}\left[\left. Y_{1}^{*} - Y_{0}^{*} \right\vert V = p, S_{0} = 1, S_{1} = 1\right] \\
  & = MTE^{OO}\left(p\right),
  \end{align*}
  implying that the summary bounds are valid for the unconditional $MTE$ function for the always-observed subgroup. Importantly, \cite{Deb2006} assumed that the covariates are fully exogenous, implying that $X \indep \left(V = p, S_{0} = 1, S_{1} = 1\right)$ holds. \cite{carneiro2009estimating} used a similar exogeneity assumption.


\section{Extension: Bounds for the distributional marginal treatment effect (DMTE)}\label{DMTE}
In this appendix, we extend Proposition \ref{thm1} to uniformly and sharply bound the distributional marginal treatment effect for the always-observed sub-population. As discussed by \cite{carneiro2009estimating}, policy-makers may care about distributional effects of a policy instead of the average effects, as individuals could respond differently to it depending on their position in the outcome distribution. For example, people at the bottom of the income distribution may not have the same response to a policy that increases college accessibility as people at the top of the income distribution. The distributional marginal treatment effect for the always-observed for any set $A$ and any $p \in \left[0, 1\right]$ is defined as
\begin{align*}
	DMTE^{OO}(A;p) \coloneqq \mathbb P\left[Y^*_1\in A| S_{0}=1,S_{1}=1, V=p\right]- \mathbb P\left[Y^*_0 \in A | S_{0}=1,S_{1}=1, V=p\right].
\end{align*}

\cite{carneiro2009estimating} show point identification results for $P\left[Y^*_1\in A| V=p\right]- \mathbb P\left[Y^*_0 \in A | V=p\right]$ when there is no sample selection. However, in the presence of sample selection, the $DMTE^{OO}\left(A; p \right)$ is only partially identified.

Combining Equation \eqref{eqmix} and Corollary 1.2 by \cite{horowitz1995}, we obtain sharp bounds on the $DMTE^{OO}(A;p)$ for any set $A$.
\begin{proposition}\label{prop1}
	For any set $A$, bounds on the $DMTE^{OO}(A;p)$ under Assumptions \ref{RA}-\ref{MONS} are given by
	\begin{equation*}
		\underline{\Delta}_{2}^{A}\left(p\right) \leq DMTE^{OO}(A;p) \leq  \overline{\Delta}_{2}^{A}\left(p\right)
	\end{equation*}
	for any $p \in \mathcal{P}$, where $\underline{\Delta}_{2}^{A}\coloneqq \mathcal{P} \rightarrow \mathbb{R}$ and $\overline{\Delta}_{2}^{A}\coloneqq \mathcal{P} \rightarrow \mathbb{R}$ are given by
	\begin{align*}
		\underline{\Delta}_{2}^{A}\left(p\right) &\coloneqq \max \left\lbrace 0,\frac{\mathbb P\left[Y^*_1 \in A | S_1 =1, V=p\right]-(1-\alpha(p))}{\alpha(p)} \right\rbrace -\mathbb P\left[Y^*_0\in A| S_{0}=1,S_{1}=1, V=p\right],\\
		\overline{\Delta}_{2}^{A}\left(p\right) &\coloneqq \min \left\lbrace 1,\frac{\mathbb P\left[Y^*_1 \in A | S_1 =1, V=p\right]}{\alpha(p)} \right\rbrace -\mathbb P\left[Y^*_0\in A| S_{0}=1,S_{1}=1, V=p\right]
	\end{align*}
	for any $p \in \mathcal{P}$. Moreover, these bounds are sharp.
\end{proposition}

\end{document}